\documentclass[runningheads,openany]{svmult}

\usepackage{palatino}

\usepackage{euler}

\usepackage{helvet}

\usepackage{graphicx}  

\usepackage{physprbb}  

\usepackage{proc}  

\usepackage{epsfig}

\usepackage{amssymb}

\makeindex             

\usepackage{amsmath}   
\usepackage{multicol}
\usepackage{bm}

\let\tocauthor\authorrunning

\begin{document}

\frontmatter


\thispagestyle{empty}
\parindent=0pt

{\Large\sc Blejske delavnice iz fizike \hfill Letnik~6, \v{s}t. 2}

\smallskip

{\large\sc Bled Workshops in Physics \hfill Vol.~6, No.~2}

\smallskip

\hrule

\hrule

\hrule

\vspace{0.5mm}

\hrule

\medskip
{\sc ISSN 1580--4992}

\vfill

\bigskip\bigskip
\begin{center}

{\bfseries 
{\Large  Proceedings to the $8^\textrm{th}$ Workshop}\\
{\Huge What Comes Beyond the Standard Models\\}
\bigskip
{\Large Bled, July 19--29, 2005}\\
\bigskip
}

\vspace{5mm}

\vfill

{\bfseries\large
Edited by

\vspace{5mm}
Norma Manko\v c Bor\v stnik

\smallskip

Holger Bech Nielsen

\smallskip

Colin D. Froggatt

\smallskip

Dragan Lukman

\bigskip


\vspace{12pt}

\vspace{3mm}

\vrule height 1pt depth 0pt width 54 mm}

\vspace*{3cm}

{\large {\sc  DMFA -- zalo\v{z}ni\v{s}tvo} \\[6pt]
{\sc Ljubljana, december 2005}}
\end{center}
\newpage

\thispagestyle{empty}
\parindent=0pt
\begin{flushright}
{\parskip 6pt
{\bfseries\large
                  The 8th Workshop \textit{What Comes Beyond  
                  the Standard Models}, 19.-- 29. July 2005, Bled}

\bigskip\bigskip

{\bfseries\large was organized by}

{\parindent8pt
\textit{Department of Physics, Faculty of Mathematics and Physics,
University of Ljubljana}

}

\bigskip

{\bfseries\large and sponsored by}

{\parindent8pt
\textit{Slovenian Research Agency}

\textit{Department of Physics, Faculty of Mathematics and Physics,
University of Ljubljana}


\textit{Society of Mathematicians, Physicists and Astronomers
of Slovenia}}}
\bigskip
\medskip

{\bfseries\large Organizing Committee}

\medskip

{\parindent9pt
\textit{Norma Manko\v c Bor\v stnik}

\textit{Colin D. Froggatt}

\textit{Holger Bech Nielsen}}

\end{flushright}


\setcounter{tocdepth}{0}

\tableofcontents

\cleardoublepage

\chapter*{Preface}
\addcontentsline{toc}{chapter}{Preface}

The series of workshops on "What Comes Beyond the Standard Model?" started 
in 1998 with the idea of organizing a real workshop, in which participants 
would spend most of the time in discussions, confronting different 
approaches and ideas. The picturesque town of Bled by the lake of the 
same name, surrounded by beautiful mountains and offering pleasant walks, 
was chosen to stimulate the discussions.

The idea was successful and has developed into an annual workshop, which is taking place 
every year since 1998.  
Very open-minded and fruitful discussions have become the trade-mark of 
our workshop, producing several published works. It takes place in 
the house of Plemelj, which belongs to the Society of Mathematicians, 
Physicists and Astronomers of Slovenia.

In this eight workshop, which took place from 19 to 29 of July 2005 at Bled, 
Slovenia, we have tried to answer some of the open questions which the 
Standard models leave unanswered, like:
 
\begin{itemize}
\item  Why has Nature made a choice of four (noticeable) dimensions? While all 
the others, if existing, are hidden?  And what are the properties of 
space-time in the hidden dimensions? 
\item  How could Nature make the decision about the breaking of symmetries 
down to the noticeable ones, if coming from some higher dimension $d$?
\item  Why is the metric of space-time Minkowskian and how is the choice 
of metric connected with the evolution of our universe(s)? 
\item  Why do massless fields exist at all? Where does the weak scale 
come from?
\item  Why do only left-handed fermions carry the weak charge? Why does 
the weak charge break parity?
\item  What is the origin of Higgs fields? Where does the Higgs mass come from?
\item  Where does the small hierarchy come from? (Or why are some Yukawa 
couplings so small and where do they come from?) 
\item  Where do the generations come from? 
\item  Can all known elementary particles be understood as different states of 
only one particle, with a unique internal space of spins and charges?
\item  How can all gauge fields (including gravity) be unified and quantized?
\item  How can different geometries and boundary conditions influence conservation laws?
\item  Does noncommutativity of coordinate manifest in Nature?
\item  Can one make the Dirac see working for fermions and bosons?
\item  What is our universe made out of (besides the baryonic matter)?
\item  What is the role of symmetries in Nature?
\end{itemize}
We have discussed these and other questions for ten days. Some results of 
this efforts appear in these Proceedings. Some of the ideas are treated 
in a very preliminary way.
Some ideas still wait to be discussed (maybe in the next workshop) 
and understood better before appearing in the next proceedings
of the Bled workshops.
The discussion will certainly continue next year, 
again at Bled, again in the house 
of Josip Plemelj.

Physics and mathematics are to our understanding both a part of Nature. 
To have ideas how to try to understand Nature, physicists need besides the knowledge also the intuition, 
inspiration, imagination and much more. Accordingly it is not surprising that there are also poets among us. 
One of the poems of Astri Kleppe can be found at the end of this Proceedings.

The organizers are grateful to all the participants for the lively discussions 
and the good working atmosphere.\\[2 cm]

\parbox[b]{40mm}{%
                 \textit{Norma Manko\v c Bor\v stnik}\\
                 \textit{Holger Bech Nielsen}\\
                 \textit{Colin Froggatt}\\
                 \textit{Dragan Lukman} }
\qquad\qquad\qquad\qquad\qquad\qquad\quad
\textit{Ljubljana, December 2005}

\newpage


\cleardoublepage


\mainmatter

\parindent=20pt

\setcounter{page}{1}


\title{Can MPP Together with Weinberg-Salam Higgs Provide %
Cosmological Inflation?}
\author{D.L. Bennett${}^1$ and H.B. Nielsen${}^2$}
\institute{%
${}^1$ Brookes Institute for Advanced Studies,
B\o gevej 6, 2900 Hellerup, Denmark\\
${}^2$ The Niels Bohr Institute, Blegdamsvej 17, 
2100 Copenhagen {\O}, Denmark
}

\titlerunning{Can MPP Together with Weinberg-Salam Higgs \ldots}
\authorrunning{D.L. Bennett and H.B. Nielsen}
\maketitle

\begin{abstract}
We investigate the possibility of producing inflation with the use of walls 
rather than the customary inflaton field(s). In the usual picture one needs to
fine-tune in such a way as to have at least 60 e-foldings. In the alternative 
picture with walls that we consider here the ``only'' fine-tuning required is
the assumption of a multiply degenerate vacuum alias the multiple point
principle (MPP).   
\end{abstract}

\section{Introduction}

One of the major problems with inflation models in general is that 
one needs the inflation to go on for very many e-foldings - of the order of 
sixty - so that the universe can expand by a needed factor of the order of 
$\exp 60$ while the scalar field causing this inflation remains at roughly the 
same field value, or at least does not fall down to the present vacuum state
value (see for example \cite{DBkolbturner}).
Accepting the prejudice that the field value should  not be 
larger than the the Planck scale, a normal inflation with an inflaton field 
would require a rather flat effective potential over a large range of field 
values but with the restriction that the present vacuum field value
should not be further away in 
field value space than about a Planck energy. This would seemingly suggest a 
a rather unnatural essentially  theta-function-like effective potential.
Of course one can with finetuning just postulate that the inflaton effective
potential has whatever strange shape may be needed, but it would be nicer 
if we could instead use some finetuning principle that could also be useful
in other contexts.

Now we have for some time worked on the idea of {\em unifying} the apparently
needed finetunings 
in the Standard Model (or perhaps some  model behind the Standard model)  
into a single  
finetuning principle that postulates the coexistence of 
{\em many vacua or phases with the same 
cosmological 
constant namely zero.} This is what we called the multiple point principle
(MPP)\cite{bennett1,bennett2} which states that observed
coupling constant values correspond to a {\em maximally} degenerate vacuum.
 
The point of the present article is to attempt to find some way to use the 
finetuning to the degenerate vacua (which is what happens if MPP is assumed) 
to replace the finetuning otherwise needed
to get 
the rather 
theta-function-like behavior of the  inflaton effective potential. 
In fact it is not so 
unlikely that such a replacement of one finetuning by another one could work.
We might think about it in the following way: 

The essence of what is needed is some field variable $\phi$ - 
that could be present say
as a scalar degree of freedom in any little neighborhood in space - 
that can simulate the inflaton field. This variable $\phi$ should now be 
associated 
with an effective potential - it could be a potential for a Higgs field - 
having roughly the properties of the inflaton 
effective potential that are needed needed for a good fitting of inflation. 
That is to say 
we should have an effective potential as a function of this scalar field $\phi$
which would behave like the needed theta function. 
Note that if  we restrict ourselves to using {\em only the Standard Model} we 
would have to use as $\phi$ the only scalar field available in the 
Standard Model namely
the Higgs field.  
Now the characteristics of the 
many degenrate vacua effective potential for some Higgs field say is that 
there can, without a need for a volume-proportional energy, be several 
vacua or phases present in the same spatial  region\cite{hbncdf}. The energy needed for 
this is {\em only proportional to an area} and not to a volume since all
the different vacua have energy density zero. In fact the 
states that we here refer to consist of two distinct phases separated from 
one another by a system of walls. Essentially we envision simulating the 
theta function by having an effective potential that as a function of 
a degree of freedom corresponding to the distance between phase-delineating 
walls become
``saturated'' at some constant value when the distance exceeds some threshold.
This essentially means that instead of the usual scenario with inflation
in three spatial directions, we instead have lack inflation along one spatial 
direction. As a consequence, inflation is restricted to taking place along
(within)the wall system that make up the boundary between the two
(or many more than two in principle) phases.    

We speculate about whether there could be, even during an ongoing 
Hubble expansion, a network 
of walls that could keep on adjusting itself so as to keep both the
energy density 
and the negative pressure constant so that inflation
could be simulated. Such a possibility could come about  
if the walls expand locally {\em more} than the regions between the 
walls
with the result that the average distance between the walls does not have to 
increase at the same rate under a Hubble expansion as 
would be the case for
isotropic geometrical expansion which would lead to an increase in the distance
between the walls
by the same factor as time goes on as the expansion along the walls. 
With this (anisotropic) expansion only {\em along} walls 
we can imagine that the walls must curl up in order to cope with that 
the average expansion of the whole of space filled with many walls lying 
in some 
complicated way is less than the expansion of the walls locally. Such a curling
up of walls complicates estimating what really will go on. Hence the 
possibility for some stabilization of the local density and the revelation of
crude features 
of the wall network is not totally excluded. Most important is that the
walls are not driven by forces caused by volume energy density differences 
- as would generically  be the case - because we have assumed the degeneracy 
of the vacua (our MPP assumption). It is for this reason that the assumption
of the multiple point principle leads to  wall motion (i.e., field values in 
the transition region between phases) that is much less 
strongly 
driven and in this way mimicks a locally flat effective potential for the 
inflaton field which is needed phenomenologically to get slow roll.
That is to say that the 
stability and lack of strong forces on the walls allow the walls to develop 
much more slowly than would be the case without the finetuning provided by 
the multiple point 
principle (MPP), and this alone could make a period of inflation simulated by 
the presence of walls last  longer - roll away slower so to speak - 
than without this MPP. It is in this sense that our finetuning by multiple 
point prinicple can provide a delay in the disappearance of the 
inflation-causing features - in our case the walls - that is similiar in effect
to having flatness in the effective potential for the inflaton near the point 
where inflation goes on.  
The role of this flatness (the theta function plateau) is of course to
keep the inflation going over a long time. 

So the possibility that 
the inflation era could be simulated by an era of walls cannot be excluded. 
Moreover we can even speculate that the assumption of  
our multiple point principle may even  
be helpful for getting a more natural way of achieving the slow roll.
If we are successful in achieving slow roll in this way, it  
will of cause only be as a consequence of a finetuning assumption in the form
of the
multiple point principle instead of the finetuning provided by the clever 
adjustment of the effective potential that yields slow roll in the usual 
simple inflation scenario 
with just a scalar field.
However, we have in other works argued that our multiple point principle
is useful in providing a explanation for other - perhaps all or at least many -
finetuning enigmas in physics. For example it is widely accepted that
finetuning is needed in order to account for the extreme smallness of the
cosmological constant 
compared to the huge value which would a priori be 
expected if one took the Planck units to be the fundamental ones.

A possible model for a system of walls strictly separating two different
phases that nevertheless both permeate all of space is provided by structures 
formed for appropriate parameter values by membranes consisting of amphiphilic 
molecules. These are the socalled bicontinuous phases - sometimes called the
plumber's nightmare - that for given temperature are observed for sufficiently 
small values of chemical potentials.

In the scenario of the ``plumber's nightmare'', the labyrinth of 
``pipe'' walls separating the two phases 
proliferates in such a way that the volume between pipe walls remains constant 
on average while the ``nightmare'' as a whole expands.

\section{Can it work?}

It is not yet clear that one can replace an inflation period 
with a  simple scalar field taking a constant value by a situation 
with an  effective potential that depends on field degrees of freedom 
corresponding to the distance between walls that separate two (or more)
degenerate vacua having the same (vanishing) energy density.
Of course if the cosmological constant for the vacua corresponding to
the bottoms of the potentials 
were not essentially zero (we take
the tiny cosmological constant of today as being essentially zero, because we 
are interested in the early era of inflation when the inflation was enormous 
compared to what could be achieved by today's cosmological constant) we could 
still get inflation, but here our idea is to obtain the inflation due to the 
walls, 
which means due to the scalar field potential in the transition regions - 
the walls - between the two phases. If we can ignore the effects of
the field gradients which are of course non-zero in the wall regions too, 
there 
would be an inflation effect coming simply from the walls. 
It would just be reduced 
by the ratio of the wall volume to the full volume one could say compared to 
the inflation that would be obtained by having the inflaton field just sitting 
on the maximum of the effective potential between the two minima corresponding 
to the two vacua. The gradient in the direction perpendicular to the walls 
will then give a contribution to the energy momentum tensor which depends on 
the direction and when averaged over directions contribute more energy 
relative to pressure than the cosmological constant or equivalently constant 
scalar field terms.   

Let us for our purposes optimistically imagine that during the main 
era of inflation we have the situation  
that, compared to the spaces in between the walls, the walls themselves 
get bigger and bigger areas.
Then the wall density would seem at first quite likely to increase with time
and the walls might even curl up or even collide and interact with each 
other.
In this scenario the energy density might even seem to have the possibility 
of increasing.
Really however the overall expansion will diminish the energy density because 
the negative pressure 
is not high enough to compensate. 

Now we may ask which effects will come to dominate 
when the density is respectively high or low. If the density is low one would 
say that each piece of wall will be isolated in first approximation and the 
expansion of the wall area will go on while the distance to the next wall 
hardly will change significantly percentwise. That suggests that in the low 
density case we would expect that the local expansion of the walls forcing 
them to 
curl up will dominate. This in turn  will cause the density to become higher. 

On the 
other hand, if the density is high, then the network of walls makes up a kind 
of complicated 
matter. Adjacent walls are not isolated from one another; they now ``feel''
one another and it
must be justifiable  to use the 
approximation that the main effect of the Hubble expansion can be assumed to 
be the effect on this matter. therefore the expansion will be isotropic (and
not just along the walls) provided the 
``network of walls is on the average isotropic. In this case the effect of total 
expansion dominates. That would cause the density of walls to fall as under 
simple Hubble expansion in which the single wall plays no important role.

Using these two estimates we then see that the density will tend to be driven 
to some intermediate density as the inflation-like Hubble expansion goes on.
If the density becomes too low it will tend to grow, and if too high it will 
fall. If such a stabilization indeed takes place the density will reach 
an intrmediate stable level. 

Once the stable density has been reached the Hubble constant will become 
constant and in this sense we will then have obtained a situation that even in 
the mathematical form for the expansion - exponential form - will simulate 
the usual simple 
inflation for a constant scalar inflaton field. 
The constant Hubble expansion constant referred to above is really the average 
Hubble expansion - denote it by $H_{av}$ - of the bulk system of walls as well
as the space in between. $H_{av}$ is to be distinguished from the Hubble 
expansion {\em along} the walls - let us denote this as $H_{wall}$. Recall that we expect that $H_{av}<H_{wall}$ - also at the stable wall density value.  

Then denoting the the ratio of the total volume to the part of it taken up
by the walls by $\xi$ 
\begin{equation}
\xi = \frac{``\hbox{full \quad volume''}}{\hbox{``wall-volume''}}
\end{equation}
we have that the Hubble expansion in the wall-regions $H_{wall}$ is bigger than 
outside by a factor $\xi$,
\begin{equation}
H_{wall} = \sqrt{\xi} H_{av}.
\end{equation}

 
\section{ What is the stabile  wall density?}

If indeed such a speculated stabilization takes place, we may make a 
dimensionality argument for what the stability density will be. 

Let us take the picture that the walls are topologically stabilized and 
therefore
unable to decay everywhere except where they collide with each other. 
Roughly 
we expect the walls to be about to cross in a fraction  $1/\xi^2$ of the total 
space or in a fraction $1/\xi$ of the wall-space. We need an estimate as to how 
fast this wall interaction region gets turned into being one or the other of 
the 
vacuum phases by the decay of its energy into other types of particles. 
This is 
what is usually supposed to happen during reheating in normal inflation 
models - namely that pairs of non-inflaton particles are produced during 
the decay of the scalar particle involved in the inflation. We shall also 
in our picture imagine that such a reheating-like mechanism is going on. For 
dimensional reasons we expect the typical energy scale for the  
field $\phi$ at the wall to give the scale for the decay rate. This wall-scale 
is say the
maximal value of the effective potential for $\phi$ in the interval between 
the two minima. This scale we could call $V_{max}$ or $V_{wall}$ because 
it is also the typical value of the effective potential $V_{eff}$ in the 
wall-volume. We must admit though that strictly speaking the scale that 
would be most relevant for the reheating decay rate would
be the mass scale of the $\phi$-particle which decays or some effective 
mass scale denoting the energy of a state for this particle as a bound state 
in the
surrounding fields. A rough estimate for this mass scale would be the 
square root of the second derivative of the effective potential, 
mass $\approx\sqrt{V_{eff}''}$.
As an estimate for this we might take an expression involving 
the difference $\phi_{vac1} - \phi_{vac2}$ between the $\phi$-values 
for the say two minima, $\phi_{vac1}$ and $\phi_{vac2}$. In fact we estimate 
$V_{eff}'' \approx V_{wall}/(\phi_{vac1}-\phi_{vac2})^2$. Therefore
the better mass scale to use would be $ \sqrt{V_{wall}}/|\phi_{vac1} - 
\phi_{vac2}|$. Let us take the decay rate - i.e. the inverse of the time scale 
for the decay of the field to its locally topologigally stable situation -
to be:
\begin{equation}
\hbox{``decay rate ''} \approx n \frac{\sqrt{V_{wall}} } 
{|\phi_{vac1} - \phi_{vac2}|}.
\end{equation} 
Here $n$ is a dimensionless quantity that is essentailly the number of decay
channels weighted by the appropriate products of 
coupling constants.

To have stability we now have to have that the amount of walls produced 
percentwise from the Hubble expansion of the walls 
(which is roughly $H_{wall}$)
should  be in balance with the destruction rate which percentwise 
becomes 
$$n  \frac{\sqrt{V_{wall}} } 
{|\phi_{vac1} - \phi_{vac2}|\xi}.$$ 
I.e. we have
\begin{equation}
H_{wall} \approx n  \frac{\sqrt{V_{wall}} } 
{|\phi_{vac1} - \phi_{vac2}|\xi}.
\end{equation} 
If we took all the coupling constants to be of order unity and included only
two-particle decays with all the selection rules ignored, the number $n$  
of decay 
channels would of course be the square of the number of particle species 
into which the $\phi$-particle could decay. Denoting this number by $n_s$ 
(where s
stands for ``species''), we get $n = n_s^2$.  

If in the philosophy of taking the scale for the scalar field to be of the 
Planck energy scale we take $|\phi_{vac1} - \phi_{vac2}| \approx 1$ 
(in Planck units), then we get in these Planck units $H_{wall} \approx n 
\sqrt{V_{wall}}/\xi$,
or using $ H_{wall} \approx \sqrt{V_{wall}}$ as is true in Planck units,
i. e. we get $\xi \approx n$.


\section{How did the present Universe come about?}

Our wall-dominated model for slow roll must of course contain the ultimate
demise of these walls 
because a wall-dominated Universe is not what we see today. A first glance
this might seem to be a problem since MPP says that the different vacua are 
degenerate and therefore there is not a dominant or preferred phase that can
squeeze all other phases out of existence and thereby eliminate the walls
that delineate these weaker phases.

However, while MPP claims degeneracy of multiple vacua primordially and during 
inflation,
it does not stipulate that there is a symmetry between the different degenerate
vacua. Different phases (i.e., degenerate vacua) could for example differ in
having heavier or lighter particles in which case the addition of heat
at the end of the inflationary era would reveal the assymmetries of the phases
e.g., the phase with lightest degrees of freedom would due to the $-TS$ term in
the free energy grow in spatial extent at the expense of phases with heavier 
degrees of freedom. Concurrent with the disappearence of these phases with 
heavier degrees of freedom the walls
delineating them would also disappear. But during the inflation period in 
which the 
temperature is effectively zero the  phase assymmetries are hidden and the 
conditions for having delineating walls separating effectively symmetric 
(and degenerate) phases are plausibly good.

\section{Conclusion}

We propose that the inflationary era of the Universe was dominated by a 
network of walls separating degenerate vacua. Assuming such a multidegenerate vacuum 
is equivalent to assuming the validity of our multiple point principle (MPP).
The assumption of MPP amounts to a finetuning to this multiply degenerate 
vacuum and as such is to be regarded as an alternative to the usual fintuning
necessary for having a constant value of the inflaton potential for a scalar 
field
for of the order of 60 e-foldings.

It is conjectured that this inflation era wall density is stabilized such 
that for too low a density Hubble expansion occurs predominantly {\em along} 
the
walls (and essentially not within the phase volume delineated by the walls).
This would lead to a corrective increase in wall density to the stable value.
If the wall density were to exceed the stable density value, the walls are 
no longer isolated from one another with the result that the presumeably
isotropic wall network materia is subject to an isotropic (bulk) Hubble
expansion. Such an expansion of this bulk materia would tend to reduce the 
wall density until the stabile wall density is attained

While phases separated by the walls are degenerate (and assumed to all have 
a vanishing energy density) they need not be symmetric. Such assymmetris
while being effectively hidden during inflation can be manifested during 
reheating at the end of the era of inflation. So two phases could for example
differ in having lighter and heavier degrees of freedom respectively. This
assymmetry would during reheating be manifested as a difference in the free 
energy of the two phases so that only one phase would survive. The other phase
would be obliterated together with the wall network system that separated the 
two phases during the inflationary era. Disappearence of the walls is of 
course required by the phenomenology of our present day Universe.


\def\MBBCprg#1{\medskip{\bf #1}}     
\def\MBBCpgt{{\scriptstyle\rm PGT}}  
\def\MBBCLeff{\hbox{$\mit\Lambda_{\hspace{.6pt}\rm eff}$}}
\def\MBBCba#1{\begin{array}{#1}}          \def\MBBCea{\end{array}}
\def\MBBClab#1{\label{eq:#1}}             
\def\MBBCeq#1{(\ref{eq:#1})}
\title{Conserved Charges in 3d Gravity With Torsion}
\author{M.\ Blagojevi\'c${}^{1,2}$ and B. Cvetkovi\'c${}^1$} 
\institute{%
$^1$ Institute of Physics, P.O.Box 57, 11001 Belgrade, Serbia\\
$^2$ Department of Physics, Univ. of Ljubljana,
1000 Ljubljana, Slovenia}
\titlerunning{Conserved Charges in 3d Gravity With Torsion}
\authorrunning{M.\ Blagojevi\'c and B. Cvetkovi\'c}
\maketitle

\begin{abstract}
We review some new developments in three-dimensional gravity based
on Rie\-mann-Cartan geometry. In particular, we discuss the structure
of asymptotic symmetry, and clarify its fundamental role in
understanding the gravitational conservation laws.
\end{abstract}

\section{Introduction}

Although general relativity (GR) successfully describes all the known
observational data, such fundamental issues as the nature of
classical singularities and the problem of quantization remain
without answer. Faced with such difficulties, one is naturally led to
consider technically simpler models that share the same conceptual
features with GR. A particularly useful model of this type is
three-dimensional (3d) gravity \cite{mbbc1}. Among many interesting
results achieved in the last twenty years, we would like to mention
(a) the asymptotic conformal symmetry of 3d gravity, (b) the
Chern-Simons formulation, (c) the existence of the black hole
solution, and (d) understanding of the black hole entropy
\cite{mbbc2,mbbc3,mbbc4,mbbc5}.

Following a widely spread belief that GR is the most reliably
approach to describe the gravitational phenomena, 3d gravity has been
studied mainly in the realm of Riemannian geometry. However, there is
a more general conception of gravity, based on Riemann-Cartan
geometry \cite{mbbc6}, in which both the curvature and the torsion are
used to describe the gravitational dyna\-mics. In this review, we
focus our attention on some new developments in 3d gravi\-ty, in
the realm of Riemann-Cartan geometry \cite{mbbc7,mbbc8,mbbc9,mbbc10,mbbc11}. In
particular, we show that the symmetry of anti-de Sitter asymptotic
conditions is described by two independent Virasoro algebras with
different central charges, in contrast to GR, and discuss the
importance of the asymptotic structure for the concept of conserved
charges---energy and angular momentum \cite{mbbc10}.

\section{Basic dynamical features}
\setcounter{equation}{0}

Theory of gravity with torsion can be formulated as Poincar\'e gauge
theory (PGT), with an underlying geometric structure described by
Riemann-Cartan space \cite{mbbc6}.

\MBBCprg{PGT in brief.} Basic gravitational variables in PGT are the
triad field $b^i$ and the Lorentz connection $A^{ij}=-A^{ji}$
(1-forms). The corresponding field strengths are the torsion and the
curvature: $T^i= db ^i+A^i{_m}\wedge b^m$,
$R^{ij}=dA^{ij}+A^i{_m}\wedge A^{mj}$ (2-forms). Gauge symmetries of
the theory are local translations and local Lorentz rotations,
parametrized by $\xi^\mu$ and $\varepsilon^{ij}$.

In 3D, we can simplify the notation by introducing
$$
A^{ij}=-\varepsilon^{ijk}\omega_k\, ,\qquad  R^{ij}=-\varepsilon^{ijk}R_k\, ,\qquad
\varepsilon^{ij}=-\varepsilon^{ijk}\theta_k\, .
$$
In local coordinates $x^\mu$, we have $b^i=b^i{_\mu}dx^\mu$,
$\omega^i=\omega^i{_\mu}dx^\mu$. The field strengths take the form
\begin{eqnarray}
&&T^i= db^i+\varepsilon^i{}_{jk}\omega^j\wedge b^k
     =\frac{1}{2}T^i{}_{\mu\nu}dx^\mu\wedge dx^\nu\, ,         \nonumber\\
&&R^i=d\omega^i+\frac{1}{2}\,\varepsilon^i{}_{jk}\omega^j\wedge\omega^k
     =\frac{1}{2}R^i{}_{\mu\nu}dx^\mu\wedge dx^\nu\, ,         \MBBClab{2.1}
\end{eqnarray}
and gauge transformations are given as
\begin{eqnarray}
\delta_0 b^i{_\mu}&=& -\varepsilon^i{}_{jk}b^j{}_{\mu}\theta^k-(\partial_\mu\xi^\rho)b^i{_\rho}
     -\xi^\rho\partial_\rho b^i{}_\mu\equiv\delta_\MBBCpgt b^i{}_\mu\, ,      \nonumber\\
\delta_0\omega^i{_\mu}&=& -\nabla_\mu\theta^i-(\partial_\mu\xi^\rho)\omega^i{_\rho}
     -\xi^\rho\partial_\rho\omega^i{}_\mu\equiv\delta_\MBBCpgt\omega^i{}_\mu\, ,    \MBBClab{2.2}
\end{eqnarray}
where $\nabla_\mu\theta^i=\partial_\mu\theta^i+\varepsilon^i{}_{jk}\omega^j{_\mu}\theta^k$ is the
covariant derivative of $\theta^i$.

To clarify the geometric meaning of PGT, we introduce the metric
tensor as a bilinear combination of the triad fields:
$$
g=\eta_{ij}b^i\otimes b^j\equiv g_{\mu\nu}dx^\mu\otimes dx^\nu\, ,
\qquad \eta_{ij}=(+,-,-)\, .
$$
Although metric and connection are in general independent geometric
objects, in PGT they are related to each other by the {\it metricity
condition\/}: $\nabla g=0$. Consequently, the geometric structure of
PGT is described by {\it Riemann-Cartan geometry\/}. Using the
metricity condition, one can derive the useful identity
\begin{equation}
\omega^i={\widetilde{\omega}}^i+K^i\, ,                                       \MBBClab{2.3}
\end{equation}
where ${\widetilde{\omega}}^i$ is Riemannian connection,
$K_{ijk}=-\frac{1}{2}(T_{ijk}-T_{kij}+T_{jki})$ is the contortion,
and $K^i$ is defined by $K^{ij}{_m}b^m\equiv K^{ij}=-\varepsilon^{ijk}K_k$.

\MBBCprg{Topological action.} General gravitational dynamics is defined
by Lagrangians which are at most quadratic in field strengths.
Omitting the quadratic terms, Mielke and Baekler proposed a {\it
topological\/} model for 3D gravity \cite{mbbc7}, defined by the action
\begin{subequations}\MBBClab{2.4}
\begin{equation}
I=aI_1+\Lambda I_2+\alpha_3I_3+\alpha_4I_4+I_M\, ,                      \MBBClab{2.4a}
\end{equation}
where
\begin{eqnarray}
&&I_1\equiv 2\int b^i\wedge R_i\, ,                             \nonumber\\
&&I_2\equiv-\frac{1}{3}\,\int\varepsilon_{ijk}b^i\wedge b^j\wedge b^k\,,\nonumber\\
&&I_3\equiv\int\left(\omega^i\wedge d\omega_i
  +\frac{1}{3}\varepsilon_{ijk}\omega^i\wedge\omega^j\wedge\omega^k\right)\,,\nonumber\\
&&I_4\equiv\int b^i\wedge T_i\, ,                          \MBBClab{2.4b}
\end{eqnarray}
\end{subequations}
and $I_M$ is a matter contribution. The first term, with $a=1/16\pi
G$, is the usual Einstein-Cartan action, the second term is a
cosmological term, $I_3$ is the Chern-Simons action for the Lorentz
connection, and $I_4$ is a torsion counterpart of $I_1$. The
Mielke-Baekler model is a natural generalization of Riemannian GR
with a cosmological constant ({GR$_\Lambda$}).

\MBBCprg{Field equations.} Variation of the action with respect to $b^i$
and $\omega^i$ yields the gravitational field equations. In order to
understand the canonical structure of the theory in the asymptotic
region, it is sufficient to consider the field equations in vacuum
(for isolated gravitational systems, gravitational sources can be
practically ignored in the asymptotic region). In the sector
$\alpha_3\alpha_4-a^2\ne 0$, these equations take the simple form
\begin{eqnarray}
2T^i=p\varepsilon^i{}_{jk}\,b^j\wedge b^k\, ,\qquad
2R^i=q\varepsilon^i{}_{jk}\,b^j\wedge b^k\, ,                      \MBBClab{2.5}
\end{eqnarray}
where
$$
p=\frac{\alpha_3\Lambda+\alpha_4 a}{\alpha_3\alpha_4-a^2}\, ,\qquad
q=-\frac{(\alpha_4)^2+a\Lambda}{\alpha_3\alpha_4-a^2}\, .
$$
Thus, the vacuum solution is characterized by constant torsion and
constant curvature. For $p=0$, the vacuum geometry is Riemannian
($T^i=0$), while for $q=0$, it becomes teleparallel ($R^i=0$).

In Riemann-Cartan spacetime, one can use the identity (2.3) to
express the curvature $R^i(\omega)$ in terms of its Riemannian piece
${\tilde R}^i\equiv R^i({\widetilde{\omega}})$ and the contortion:
$R^i(\omega)={\tilde R}^i+ \nabla K^i-\frac{1}{2}\varepsilon^{imn}K_m\wedge K_n$.
This result, combined with the field equations \MBBCeq{2.5}, leads to
\begin{equation}
2{\tilde R}^i=\MBBCLeff\,\varepsilon^i{}_{jk}\,b^j\wedge b^k\, ,
\qquad \MBBCLeff\equiv q-\frac{1}{4}p^2\, ,                    \MBBClab{2.6}
\end{equation}
where $\MBBCLeff$ is the effective cosmological constant. Consequently,
our spacetime is maximally symmetric: for $\MBBCLeff<0$ ($\MBBCLeff\ge 0$),
the spacetime manifold is anti-de Sitter (de Sitter/Minkowski). In
what follows, our attention will be focused on the model \MBBCeq{2.4}
with $\alpha_3\alpha_4-a^2\ne 0$, and with negative $\MBBCLeff$ (anti-de Sitter
sector):
\begin{equation}
\MBBCLeff\equiv-\frac{1}{\ell^2}<0 \, .                        \MBBClab{2.7}
\end{equation}

\section{The black hole solution}
\setcounter{equation}{0}

For $\MBBCLeff<0$, equation \MBBCeq{2.6} has a well known solution for the
metric --- the BTZ black hole. Using the static coordinates
$x^\mu=(t,r,\varphi)$ with $0\le\varphi<2\pi$, the black hole metric is
given by
\begin{eqnarray}
&&ds^2=N^2dt^2-N^{-2}dr^2-r^2(d\varphi+N_\varphi dt)^2\, ,     \nonumber\\
&&N^2=\left(-8Gm+\frac{r^2}{\ell^2}+\frac{16G^2J^2}{r^2}\right)\, ,
  \qquad N_\varphi=\frac{4GJ}{r^2}\, ,                       \MBBClab{3.1}
\end{eqnarray}
with $m\ge 0$, $\ell m\ge \vert J\vert$. If $r_+$ and $r_-$ are the
zeros of $N^2$, then we have: $m=(r_+^2+r_-^2)/8G\ell^2$,
$J=r_+r_-/4G\ell$. The relation between the parameters $m,J$, and the
black hole energy and angular momentum, will be clarified later.
Since the triad field corresponding to \MBBCeq{3.1} is determined only up
to a local Lorentz transformation, we can choose $b^i$ to have the
simple form:
\begin{subequations}\MBBClab{3.2}
\begin{equation}
b^0=Ndt\, ,\qquad b^1=N^{-1}dr\, ,\qquad
b^2=r\left(d\varphi+N_\varphi dt\right)\, .                    \MBBClab{3.2a}
\end{equation}
To find the connection, we combine the relation $K^i=(p/2)b^i$, which
follows from the first field equation in \MBBCeq{2.5}, with the identity
\MBBCeq{2.3}. This yields
\begin{equation}
\omega^i={\widetilde{\omega}}^i+\frac{p}{2}\,b^i\, ,                          \MBBClab{3.2b}
\end{equation}
where the Riemannian connection ${\widetilde{\omega}}^i$ is defined by
$d{\widetilde{\omega}}^i+\varepsilon^i{}_{jk}{\widetilde{\omega}}^j b^k=0$:
\begin{equation}
{\widetilde{\omega}}^0=-Nd\varphi\, ,\qquad {\widetilde{\omega}}^1=N^{-1}N_\varphi dr\, ,\qquad
{\widetilde{\omega}}^2=-\frac{r}{\ell^2}dt-rN_\varphi d\varphi\, .             \MBBClab{3.2c}
\end{equation}
\end{subequations}
Equations \MBBCeq{3.2} define the BTZ black hole in {\it
Riemann--Cartan\/} spacetime \cite{mbbc8,mbbc9}. As a constant curvature
spacetime, the black hole is locally isometric to the AdS solution
(AdS$_3$), obtained formally from \MBBCeq{3.2} by the replacement $J=0$,
$8Gm=-1$.

\section{Asymptotic conditions}
\setcounter{equation}{0}

For isolated gravitational systems, matter is absent from the
asymptotic region. In spite of that, it can influence global
properties of spacetime through the asymptotic conditions, the
symmetries of which are closely related to the gravitational
conserved charges \cite{mbbc10}.

\MBBCprg{AdS asymptotics.} For $\MBBCLeff<0$, maximally symmetric AdS
solution has the role analogous to the role of Minkowski space in the
$\MBBCLeff=0$ case. Following the analogy, we could choose that all the
fields approach the single AdS$_3$ configuration at large distances.
The asymptotic symmetry would be the global AdS symmetry $SO(2,2)$,
the action of which leaves the AdS$_3$ configuration invariant.
However, this choice would exclude the important black hole solution.
This motivates us to introduce the {\it asymptotic AdS
configurations\/}, determined by the following requirements:
\begin{itemize}
\item[(a)] the asymptotic conditions include the black hole
configuration,
\item[(b)] they are invariant under the action of the AdS group,
and 
\item[(c)] the asymptotic symmetries have well defined canonical
generators.
\end{itemize}

The asymptotics of the triad field $b^i{_\mu}$ that satisfies (a) and
(b) reads:
\begin{subequations}\MBBClab{4.1}
\begin{equation}
b^i{_\mu}= \left( \MBBCba{ccc}
   \displaystyle\frac{r}{\ell}+{\cal O}_1   & O_4  & O_1   \\
   {\cal O}_2 & \displaystyle\frac{\ell}{r}+{\cal O}_3  & O_2   \\
   {\cal O}_1 & {\cal O}_4                     & r+{\cal O}_1
                  \MBBCea
          \right)   \, .                                   \MBBClab{4.1a}
\end{equation}
Here, for any ${\cal O}_n=c/r^n$, we assume that $c$ is not a constant,
but a function of $t$ and $\varphi$, $c=c(t,\varphi)$, which is the
simplest way to ensure the global $SO(2,2)$ invariance. This
assumption is of crucial importance for highly non-trivial structure
of the resulting asymptotic symmetry.

The asymptotic form of $\omega^i{_\mu}$ is defined in accordance with
\MBBCeq{3.2b}:
\begin{equation}
\omega^i{_\mu}=\left( \MBBCba{ccc}
   \displaystyle\frac{pr}{2\ell}+{\cal O}_1 & {\cal O}_4 &\displaystyle-\frac{r}{\ell}+{\cal O}_1 \\
   {\cal O}_2 & \displaystyle\frac{p\ell}{2r}+{\cal O}_3 & {\cal O}_2                   \\
   \displaystyle-\frac{r}{\ell^2}+{\cal O}_1 & {\cal O}_4 & \displaystyle \frac{pr}{2}+{\cal O}_1
                  \MBBCea
                \right)  \, .                              \MBBClab{4.1b}
\end{equation}
\end{subequations}
A verification of the third condition (c) is left for the next section.

\MBBCprg{Asymptotic parameters.} Having chosen the asymptotic conditions,
we now wish to find the subset of gauge transformations \MBBCeq{2.2} that
respect these conditions. They are defined by restricting the
original gauge parameters in accordance with \MBBCeq{4.1}, which yields
\begin{subequations}\MBBClab{4.2}
\begin{eqnarray}
&&\xi^0=\ell\left[ T
  +\frac{1}{2}\left(\frac{\partial^2 T}{\partial t^2}\right)
              \frac{\ell^4}{r^2}\right] +{\cal O}_4\, , \qquad
  \xi^1=-\ell\left(\frac{\partial T}{\partial t}\right)r+{\cal O}_1\, ,   \nonumber\\
&&\xi^2=S-\frac{1}{2}\left(\frac{\partial^2 S}{\partial\varphi^2}\right)
              \frac{\ell^2}{r^2}+{\cal O}_4\, ,                 \MBBClab{4.2a}
\end{eqnarray}
and
\begin{eqnarray}
&&\theta^0=-\frac{\ell^2}{r}\partial_0\partial_2T+{\cal O}_3\, ,\qquad
  \theta^1=\partial_2 T+{\cal O}_2\, ,                                  \nonumber\\
&&\theta^2=\frac{\ell^3}{r}\partial_0^2T+{\cal O}_3\, .                 \MBBClab{4.2b}
\end{eqnarray}
\end{subequations}
The functions $T$ and $S$ are such that $\partial_\pm(T\mp S)=0$, with
$x^\pm\equiv x^0/\ell \pm x^2$, which implies
$$
T+S=g(x^+)\, , \qquad T-S=h(x^-)\, ,
$$
where $g$ and $h$ are two arbitrary, periodic functions.

The commutator algebra of the Poincar\'e gauge transformations
\MBBCeq{2.2} is closed: $[\delta_0',\delta_0'']=\delta_0'''$, where
$\delta_0'=\delta_0(\xi',\theta')$ and so on. Using the related composition law
with the restricted parameters \MBBCeq{4.2}, and keeping only the lowest
order terms, one finds the relation
\begin{eqnarray}
&&T'''=T'\partial_2 S''+S'\partial_2 T''-T''\partial_2 S'-S''\partial_2 T'\, , \nonumber\\
&&S'''=S'\partial_2 S''+T'\partial_2 T''-S''\partial_2 T'-T''\partial_2 ST'\, ,\MBBClab{4.3}
\end{eqnarray}
which is expected to be the composition law for $(T,S)$. To verify
this assumption, we separate the parameters \MBBCeq{4.2} into two pieces:
the leading terms containing $T$ and $S$ define a $(T,S)$
transformation, while the rest defines the residual (pure gauge)
transformation. The PGT commutator algebra implies that the
commutator of two $(T,S)$ transformations produces not only a $(T,S)$
transformations, but also an additional pure gauge transformation.
This result motivates us to introduce an improved definition of the
asymptotic symmetry: it is the symmetry defined by the parameters
\MBBCeq{4.2}, modulo pure gauge transformations. As we shall see in the
next section, this symmetry coincides with the conformal symmetry.

\section{Canonical generators and conserved charges}
\setcounter{equation}{0}

In this section, we continue our study of the asymptotic symmetries
and conservation laws in the canonical formalism \cite{mbbc10}.

\MBBCprg{Hamiltonian and constraints.} Introducing the canonical momenta
$(\pi_i{^\mu},\Pi_i{^\mu})$, corresponding to the Lagrangian variables
$(b^i{_\mu},\omega^i{_\mu})$, we find that the primary constraints of the
theory \MBBCeq{2.4} are of the form:
\begin{eqnarray}
&&\phi_i{^0}\equiv\pi_i{^0}\approx 0\, ,\hspace{82.5pt}
   \Phi_i{^0}\equiv\Pi_i{^0}\approx 0\, ,                  \nonumber\\
&&\phi_i{^\alpha}\equiv\pi_i{^\alpha}
  -\alpha_4\varepsilon^{0\alpha\beta}b_{i\beta}\approx 0\, ,\qquad
   \Phi_i{^\alpha}\equiv\Pi_i{^\alpha}
  -\varepsilon^{0\alpha\beta}(2ab_{i\beta}+\alpha_3\omega_{i\beta})\approx 0\, .       \nonumber
\end{eqnarray}
Up to an irrelevant divergence, the total Hamiltonian reads
\begin{eqnarray}
&&{\cal H}_T=b^i{_0}{\cal H}_i+\omega^i{_0}{\cal K}_i
        +u^i{_0}\pi_i{^0}+v^i{_0}\Pi_i{^0} \, ,          \MBBClab{5.1}\\
&&{\cal H}_i=-\varepsilon^{0\alpha\beta}\left(aR_{i\alpha\beta}+\alpha_4T_{i\alpha\beta}
            -\Lambda \varepsilon_{ijk}b^j{_\alpha}b^k{_\beta}\right)           \nonumber\\
&&\hspace{141pt}-\nabla_\beta\phi_i{^\beta}
  +\varepsilon_{imn}b^m{_\beta}\left(p\phi^{n\beta}+q\Phi^{n\beta}\right)\,,\nonumber\\
&&{\cal K}_i=-\varepsilon^{0\alpha\beta}\left(aT_{i\alpha\beta}+\alpha_3R_{i\alpha\beta}
                   +\alpha_4\varepsilon_{imn}b^m{_\alpha}b^n{_\beta}\right)
      -\nabla_\beta\Phi_i{^\beta}-\varepsilon_{imn}b^m{_\beta}\phi^{n\beta}\, .\nonumber
\end{eqnarray}
The constraints ($\pi_i{^0},\Pi_i{^0},{\cal H}_i,{\cal K}_i$) are first class,
($\phi_i{^\alpha},\Phi_i{^\alpha}$) are second class.

\MBBCprg{Canonical generators.} Applying the general Castellani's
algorithm \cite{mbbc6}, we find the canonical gauge generator of the
theory:
\begin{eqnarray}
&& G=-G_1-G_2\, ,                                          \nonumber\\
&&G_1\equiv\dot\xi^\rho\left(b^i{_\rho}\pi_i{^0}+\omega^i{_\rho}\Pi_i{^0}\right)
  +\xi^\rho\left[b^i{_\rho}{\cal H}_i +\omega^i{_\rho}{\cal K}_i
  +(\partial_\rho b^i{_0})\pi_i{^0}+(\partial_\rho\omega^i{_0})\Pi_i{^0}\right]\,,\nonumber\\
&&G_2\equiv\dot\theta^i\Pi_i{^0}
  +\theta^i\left[{\cal K}_i-\varepsilon_{ijk}\left( b^j{_0}\pi^{k0}
  +\omega^j{_0}\,\Pi^{k0}\right)\right]\, .                   \MBBClab{5.2}
\end{eqnarray}
Here, the time derivatives $\dot b^i{_\mu}$ and $\dot\omega^i{_\mu}$ are
shorts for $u^i{_\mu}$ and $v^i{_\mu}$, respectively, and the
integration symbol $\int d^2x$ is omitted in order to simplify the
notation. The transformation law of the fields, defined by
$\delta_0\phi\equiv \{\phi\,,G\}$, is in complete agreement with the
gauge transformations \MBBCeq{2.2} {\it on shell\/}.

\MBBCprg{Asymptotics of the phase space.} The behaviour of momentum
variables at large distances is defined by the following general
principle: the expressions that vanish on-shell should have an
arbitrarily fast asymptotic decrease, as no solution of the field
equations is thereby lost. Applied to the primary constraints, this
principle gives the asymptotic behaviour of $\pi_i{^\mu}$ and
$\Pi_i{^\mu}$. The same principle can be also applied to the secondary
constraints and the true equations of motion.

\MBBCprg{The improved generator.} The canonical generator acts on
dynamical variables via the Poisson bracket operation, which is
defined in terms of functional derivatives. In general, $G$ does not
have well defined functional derivatives, but the problem can be
corrected by adding suitable surface terms \cite{mbbc6}. The improved
canonical generator ${\tilde G}$ reads:
\begin{equation}
{\tilde G}=G+\Gamma\, ,\qquad
\Gamma\equiv-\int_0^{2\pi}d\varphi
         \left(\xi^0{\cal E}^1+\xi^2{\cal M}^1\right)\, ,            \MBBClab{5.3}
\end{equation}
where
\begin{eqnarray}
&&{\cal E}^\alpha\equiv
   2\varepsilon^{0\alpha\beta}\left[\left(a+\frac{\alpha_3p}{2}\right)\omega^0{}_\beta
  +\left(\alpha_4+\frac{ap}{2}\right)b^0{}_\beta+\frac{a}{\ell}b^2{}_\beta
  +\frac{\alpha_3}{\ell}\omega^2{}_\beta\right]b^0{}_0\, ,           \nonumber\\
&&{\cal M}^\alpha\equiv
  -2\varepsilon^{0\alpha\beta}\left[\left(a+\frac{\alpha_3p}{2}\right)\omega^2{}_\beta
  +\left(\alpha_4+\frac{ap}{2}\right)b^2{}_\beta+\frac{a}{\ell}b^0{}_\beta
  +\frac{\alpha_3}{\ell}\omega^0{}_\beta\right]b^2{}_2\, .           \nonumber
\end{eqnarray}
The adopted asymptotic conditions guarantee differentiability and
finiteness of ${\tilde G}$. Moreover, ${\tilde G}$ is also conserved.

The value of the improved generator ${\tilde G}$ defines the {\it
gravitational charge\/}. Since ${\tilde G}\approx \Gamma$, the charge is
completely determined by the boundary term $\Gamma$. Note that $\Gamma$
depends on $T$ and $S$, but not on pure gauge parameters. In other
words, $(T,S)$ transformations define non-vanishing charges, while
the charges corresponding to pure gauge transformations vanish.

\MBBCprg{Energy and angular momentum.} For $\xi^2=0$, ${\tilde G}$ reduces to
the time translation generator, while for $\xi^0=0$ we obtain the
spatial rotation generator. The corresponding surface terms,
calculated for $\xi^0=1$ and $\xi^2=1$, respectively, have the
meaning of energy and angular momentum:
\begin{equation}
E=\int_0^{2\pi}d\varphi\,{\cal E}^1 \, ,\qquad
M=\int_0^{2\pi}d\varphi\,{\cal M}^1 \, .                          \MBBClab{5.4}
\end{equation}
Energy and angular momentum are conserved gravitational charges.

Using the above results, one can calculate energy and angular
momentum of the black hole configuration \MBBCeq{3.2} \cite{mbbc8,mbbc9}:
\begin{equation}
E= m+\frac{\alpha_3}{a}\left(\frac{pm}{2}-\frac{J}{\ell^2}\right)\, ,
\qquad M= J+\frac{\alpha_3}{a}\left(\frac{pJ}{2}-m\right)\, .  \MBBClab{5.5}
\end{equation}
These expressions differ from those in Riemannian {GR$_\Lambda$}, where
$\alpha_3=0$. Moreover, $E$ and $M$ are the only independent black hole
charges.

\MBBCprg{Canonical algebra.} The Poisson bracket algebra of the improved
ge\-ne\-rators contains essential informations on the asymptotic
symmetry structure. In the notation $G'\equiv G[T',S']$, $G''\equiv
G[T'',S'']$, and so on, the Poisson bracket algebra is found to have
the form
\begin{subequations}\MBBClab{5.6}
\begin{equation}
\left\{{\tilde G}'',\,{\tilde G}'\right\} ={\tilde G}''' + C''' \,.             \MBBClab{5.6a}
\end{equation}
where the parameters $T'''$, $S'''$  are determined by the
composition rules  \MBBCeq{4.3}, and $C'''$ is the
{\it central term\/} of the canonical algebra:
\begin{eqnarray}
C'''=&&~(2a+\alpha_3p)\ell\int_0^{2\pi}d\varphi
       \left(\partial_2S''\partial_2^2T'-\partial_2S'\partial_2^2T''\right)    \nonumber\\
&&~-2\alpha_3\int_0^{2\pi}d\varphi
       \left(\partial_2T''\partial_2^2T'+\partial_2S''\partial_2^2S'\right)\, .\MBBClab{5.6b}
\end{eqnarray}
\end{subequations}
Expressed in terms of the Fourier modes, the canonical algebra
\MBBCeq{5.6} takes a more familiar form---the form of two independent
Virasoro algebras with classical central charges:
\begin{subequations}
\begin{eqnarray}
&&\left\{L_n,L_m\right\}=-i(n-m)L_{n+m}
                 -\frac{c}{12}in^3\delta_{n,-m}\, ,            \nonumber\\
&&\left\{{\bar L}_n,{\bar L}_m\right\}=-i(n-m){\bar L}_{m+n}
                 -\frac{\bar c}{12}in^3\delta_{n,-m}\, ,       \nonumber\\
&&\{L_n,\bar{L}_m\}=0\, .                                  \MBBClab{5.7a}
\end{eqnarray}
The central charges have the form:
\begin{equation}
c=\frac{3\ell}{2G}+24\pi\alpha_3\left(\frac{p\ell}{2}+1\right)\, ,\qquad
\bar{c}=\frac{3\ell}{2G}
        +24\pi\alpha_3\left(\frac{p\ell}{2}-1\right)\, .       \MBBClab{5.7b}
\end{equation}
\end{subequations}
Asymptotically, the gravitational dynamics is characterized by the
conformal symmetry with two {\it different\/} central charges, in
contrast to Riemannian {GR$_\Lambda$}, where $c=\bar c=3\ell/2G$. As a
consequence, the entropy of the black hole \MBBCeq{3.2} differs from the
corresponding Riemannian result \cite{mbbc12}:
\begin{equation}
S=\frac{2\pi r_+}{4G}
  +4\pi^2\alpha_3\left(pr_+ -2\frac{r_-}{\ell}\right)\, .
\end{equation}

\section{Concluding remarks}

\begin{itemize}

\item 3d gravity with torsion, defined by the action \MBBCeq{2.4},
is based on an underlying Riemann-Cartan geometry of spacetime.
\item The theory possesses the black hole solution \MBBCeq{3.2}, a
generalization of the Riemannian BTZ black hole. Energy and angular
momentum of the black hole differ from the corresponding Riemannian
expressions in {GR$_\Lambda$}.
\item The AdS asymptotic conditions \MBBCeq{4.1} imply the conformal
symmetry in the asymptotic region. The symmetry is described by two
independent Virasoro algebras with different central charges.
The existence of different central char\-ges ($\alpha_3\ne 0$)
modifies the black hole entropy.
\end{itemize}

\title{Mass Matrices of Quarks and Leptons in the Approach Unifying Spins %
and Charges}
\author{A. Bor\v stnik Bra\v ci\v c${}^{1,3}$ and %
S.N. Manko\v c Bor\v stnik${}^{2,3}$}
\institute{%
${}^1$ Educational Faculty, University of Ljubljana,
 Kardeljeva plo\v s\v cad 17, 1000 Ljubljana, Slovenia\\
${}^2$ Primorska Institute for Natural Sciences and Technology, 
C. Mare\v zganskega upora 2, 6000 Koper, Slovenia\\
${}^3$ Department of Physics, University of
Ljubljana, Jadranska 19, 1000 Ljubljana, Slovenia}

\titlerunning{Mass Matrices of Quarks and Leptons in the Approach \ldots}
\authorrunning{A. Bor\v stnik Bra\v ci\v c and S.N. Manko\v c Bor\v stnik}
\maketitle

\begin{abstract} 
In the approach unifying all the internal degrees of freedom - that is the spin and 
all the charges into only the spin - proposed by one of 
us\cite{ABBnorma92,ABBnorma93,ABBnormasuper94,ABBnorma95,ABBnorma97,ABBpikanormaproceedings1,ABBholgernorma00,ABBnorma01,%
ABBpikanormaproceedings2,ABBPortoroz03}, spinors, living in $d \;(=1+13)-$dimensional
space, carry only the spin and interact with only the gravity through spin connections
and vielbeins. 
After a  break of symmetries  a spin can manifest in $d\;=(1+3)$ ''physical'' 
space as the spin  and all the known charges.
In this talk we discuss the mass matrices of quarks and leptons, predicted by the approach. 
Mass matrices follow from the starting Lagrangean, if assuming that there is a kind of 
breaking symmetry from $SO(1,13)$ to $SO(1,7)\times SU(3)\times U(1)$,
which does end up with massless spinors in $d\;(=1+7)$-dimensional space, while a further break leads 
to mass matrices. 
\end{abstract}

\section{ Introduction:}
\label{ABBintroduction}

The Standard model of the electroweak and strong interactions (extended by the inclusion of the massive 
neutrinos) fits well the existing experimental data. It assumes around 25 parameters and requests, 
the origins of which is not yet understood.

The advantage of the approach, proposed by one of us (N.S.M.B.), unifying spins and 
charges\cite{ABBnorma92,ABBnorma93,ABBnormasuper94,ABBnorma95,ABBnorma97,%
ABBpikanormaproceedings1,ABBholgernorma00,ABBnorma01,ABBpikanormaproceedings2,ABBPortoroz03} is, that it might offer possible 
answers to the open questions of the Standard electroweak model. We demonstrated in 
references\cite{ABBpikanormaproceedings1,%
ABBnorma01,ABBpikanormaproceedings2,ABBPortoroz03} that a left handed $SO(1,13)$ 
Weyl spinor multiplet includes, if the representation is interpreted
in terms of the subgroups $SO(1,3)$, $SU(2)$, $SU(3)$ and the sum of the two $U(1)$'s,  all the spinors of
the Standard model - that is the left handed $SU(2)$ doublets and the right handed  $SU(2)$ 
singlets of (with the group  $SU(3)$ charged) quarks and  (chargeless) leptons.
Right handed neutrinos -  weak and hyper chargeless - are also included.
In the gauge theory of gravity (in our case in $d=(1+13)$-dimensional space), the Poincar\' e group is gauged, 
leading to spin connections and vielbeins, which  determine the gravitational field\cite{ABBmil,ABBnorma93,ABBnorma01}.
By introducing two kinds of the Dirac operators\cite{ABBpikanormaproceedings2,ABBPortoroz03,
ABBastridragannorma,ABBbmnBled04} $\gamma^a$, there follow two kinds of the spin connection fields
and the spin connection  and vielbein fields manifest - after the appropriate compactification 
(or some other kind of making the rest of d-4 space unobservable at low energies) - in the four dimensional 
space as all the known gauge fields, as well as the Yukawa couplings.

In the present talk we demonstrate, how do the Yukawa like terms, leading to masses of quarks and leptons, appear in the 
approach unifying spins and charges. No Higgs doublets, needed in the Standard model to ''dress''
weak chargeless spinors, aught to be assumed.  
We show, how can families of quarks and leptons be generated and how consequently
the Yukawa couplings among the families of spinors appear.


\section{Spinor representation and families in terms of two kinds of  Clifford algebra objects}
 \label{technique}

 We  define two kinds of the Clifford algebra objects\cite{ABBnorma93,ABBholgernorma02,ABBtechnique03}, 
 $\gamma^a$ and $\tilde{\gamma^a}$, 
 with the properties
 \begin{eqnarray}
 \{\gamma^a,\gamma^b\}_{+} = 2\eta^{ab} =  \{\tilde{\gamma}^a,\tilde{\gamma}^b\}_{+},
 \quad \{\gamma^a,\tilde{\gamma}^b\}_{+} = 0.
 \label{clifford}
 \end{eqnarray}
 The operators $\tilde{\gamma}^a$  are introduced formally as operating on any Clifford algebra object $B$ 
 from the left hand side, but they also can be expressed in terms of the  ordinary $\gamma^a$ as  
 operating from the right hand side as follows:
 $ \tilde{\gamma}^a B : = i(-)^{n_B} B \gamma^b,$
 with $(-)^{n_B} =  +1$ or $-1$, when the object $B$ has a Clifford even or odd character, respectively.
 
 Accordingly two kinds of generators of the Lorentz transformations follow, namely 
 \begin{eqnarray}
         S^{ab} &=& i/4 (\gamma^a \gamma^b - \gamma^b \gamma^a), \nonumber\\ 
 \tilde{S}^{ab} &=& i/4 (\tilde{\gamma}^a \tilde{\gamma}^b - \tilde{\gamma}^b \tilde{\gamma}^a),\nonumber\\
 & & \{ S^{ab},\tilde{S}^{cd}\}_{-}=0.
 \label{sab}
 \end{eqnarray}

 We define spinor representations as  
 eigen states of the chosen Cartan sub algebra of the Lorentz algebra $SO(1,13)$,  with  the operators 
 $S^{ab}$ and $\tilde{S}^{ab}$ in the two Cartan sub algebra sets, with the same indices in both cases.
 By introducing the notation
 \begin{eqnarray}
 \stackrel{ab}{(\pm i)}: &=& \frac{1}{2}(\gamma^a \mp  \gamma^b),  \quad 
 \stackrel{ab}{[\pm i]}: = \frac{1}{2}(1 \pm \gamma^a \gamma^b), \;{for} \; \eta^{aa} \eta^{bb} =-1, \nonumber\\
 \stackrel{ab}{(\pm )}: &= &\frac{1}{2}(\gamma^a \pm i \gamma^b),  \quad 
 \stackrel{ab}{[\pm ]}: = \frac{1}{2}(1 \pm i\gamma^a \gamma^b), \;{for} \; \eta^{aa} \eta^{bb} =1,
 \label{eigensab}
 \end{eqnarray}
 it can be shown that  
 \begin{eqnarray}
 S^{ab} \stackrel{ab}{(k)}: &=&  \frac{k}{2} \stackrel{ab}{(k)}, \quad 
 S^{ab} \stackrel{ab}{[k]}:  =  \frac{k}{2} \stackrel{ab}{[k]}, \nonumber\\
 \tilde{S}^{ab} \stackrel{ab}{(k)}:  &= & \frac{k}{2} \stackrel{ab}{(k)},  \quad 
 \tilde{S}^{ab} \stackrel{ab}{[k]}:   =   - \frac{k}{2} \stackrel{ab}{[k]}.
 \label{eigensabev}
 \end{eqnarray}
 The above binomials are all ''eigenvectors''  of  the generators $S^{ab}$, as well as of  $\tilde{S}^{ab}$.
 We further find 
 \begin{eqnarray}
 \gamma^a \stackrel{ab}{(k)}&=&\eta^{aa}\stackrel{ab}{[-k]},\quad 
 \gamma^b \stackrel{ab}{(k)}= -ik \stackrel{ab}{[-k]}, \nonumber\\
 \gamma^a \stackrel{ab}{[k]}&=& \stackrel{ab}{(-k)},\quad \quad \quad
 \gamma^b \stackrel{ab}{[k]}= -ik \eta^{aa} \stackrel{ab}{(-k)}
 \label{graphgammaaction}
 \end{eqnarray}
 and similarly
 \begin{eqnarray}
 \tilde{\gamma^a} \stackrel{ab}{(k)}: &=& - i\eta^{aa}\stackrel{ab}{[k]},\quad
 \tilde{\gamma^b} \stackrel{ab}{(k)}: =  - k \stackrel{ab}{[k]}, \nonumber\\
 \tilde{\gamma^a} \stackrel{ab}{[k]}: &=&  \;\;i\stackrel{ab}{(k)},\quad \quad \quad
 \tilde{\gamma^b} \stackrel{ab}{[k]}: =  -k \eta^{aa} \stackrel{ab}{(k)}.
 \label{gammatilde}
 \end{eqnarray}
 We shall later make use  of the relations
 \begin{eqnarray}
 \stackrel{ab}{(k)}\stackrel{ab}{(k)}& =& 0, \quad \quad \stackrel{ab}{(k)}\stackrel{ab}{(-k)}
 = \eta^{aa}  \stackrel{ab}{[k]}, \quad 
 \stackrel{ab}{[k]}\stackrel{ab}{[k]} =  \stackrel{ab}{[k]}, \quad \quad
 \stackrel{ab}{[k]}\stackrel{ab}{[-k]}= 0, 
  \nonumber\\
 \stackrel{ab}{(k)}\stackrel{ab}{[k]}& =& 0,\quad \quad \quad \stackrel{ab}{[k]}\stackrel{ab}{(k)}
 =  \stackrel{ab}{(k)}, \quad \quad 
 \stackrel{ab}{(k)}\stackrel{ab}{[-k]} =  \stackrel{ab}{(k)},
 \quad \quad \stackrel{ab}{[k]}\stackrel{ab}{(-k)} =0,  \nonumber\\ 
 \stackrel{ab}{\tilde{(k)}} \stackrel{ab}{(k)}& =& 0, \quad  \stackrel{ab}{\tilde{(-k)}} \stackrel{ab}{(k)}
 = -i \eta^{aa}  \stackrel{ab}{[k]}, \;\; \stackrel{ab}{\tilde{(-k)}}\stackrel{ab}{[-k]}= i \stackrel{ab}{(-k)},\;\;
 \stackrel{ab}{(k)} \stackrel{ab}{[-k]} = 0, \nonumber\\
 \stackrel{ab}{\tilde{(k)}} \stackrel{ab}{[k]}& =& i \stackrel{ab}{(k)}, \;
 \stackrel{ab}{\tilde{(-k)}}\stackrel{ab}{[+k]}= 0, \; \stackrel{ab}{\tilde{(-k)}}\stackrel{ab}{(-k)}=0,
  \;\stackrel{ab}{\tilde{(k)}}\stackrel{ab}{(-k)} = -i \eta^{aa} \stackrel{ab}{[-k]}.
 \label{graphbinoms}
 \end{eqnarray}
 Here 
 %
 $\stackrel{ab}{\tilde{(\pm i)}} = \frac{1}{2} (\tilde{\gamma}^a \mp \tilde{\gamma}^b), \;
 \stackrel{ab}{\tilde{(\pm 1)}} = \frac{1}{2} (\tilde{\gamma}^a \pm i\tilde{\gamma}^b), \;
 \stackrel{ab}{\tilde{[\pm i]}} = \frac{1}{2} (1 \pm \tilde{\gamma}^a \tilde{\gamma}^b), \;
 \stackrel{ab}{\tilde{[\pm 1]}} = \frac{1}{2} (1 \pm i \tilde{\gamma}^a \tilde{\gamma}^b).$ 
 %

 We make a choice of {\em presenting  spinors as products of  binomials $\stackrel{ab}{(k)}$ or 
 $\stackrel{ab}{[k]}$, never of $\stackrel{ab}{\tilde{(k)}}$ or  $\stackrel{ab}{\tilde{[k]}}$.} 
 
 The reader should also notice that $\gamma^a$'s transform the binomial  
 $\stackrel{ab}{(k)}$ into the binomial $\stackrel{ab}{[-k]}$,
 whose eigen value with respect to $S^{ab}$ change sign, while
 $\tilde{\gamma}^a$'s transform the binomial $\stackrel{ab}{(k)}$ 
 into $\stackrel{ab}{[k]}$ with unchanged ''eigen value''
 with respect to $S^{ab}$. 
 
 We define the operators of  handedness of the group $SO(1,13)$ and of the subgroups $SO(1,3), SO(1,7), 
 SO(6)$ and $SO(4)$ so that $(\Gamma^{(d)})^2 =I$ and $(\Gamma^{(d)})^{\dagger}= \Gamma^{(d)}$ 
\begin{eqnarray}
 \Gamma^{(1,13)} &=& i 2^{7} \; S^{03} S^{12} S^{56} \cdots S^{13 \; 14}, \nonumber\\
 \Gamma^{(1,3)}  &=&  - i 2^2 S^{03} S^{12}, \nonumber\\
 \Gamma^{(1,7)} &=&  - i2^{4}  S^{03} S^{12} S^{56} S^{78},\nonumber\\
 \Gamma^{(6)} &=& - 2^3 S^{9 \;10} S^{11\;12} S^{13 \; 14},\nonumber\\
 \Gamma^{(4)} &=& 2^2 S^{56} S^{78}.\nonumber
\end{eqnarray}

 To represent one Weyl spinor in $d=1+13$, one must make a choice of the
 operators belonging to the Cartan sub algebra of $7$ elements of the group $SO(1,13)$ for both kinds of 
 the Clifford algebra objects. We make the following choice 
 \begin{eqnarray}
 S^{03}, S^{12}, S^{56}, S^{78}, S^{9 10}, S^{11\; 12},  S^{13\; 14}, \nonumber\\
 \tilde{S}^{03}, \tilde{S}^{12}, \tilde{S}^{56}, \tilde{S}^{78}, \tilde{S}^{9 10}, \tilde{S}^{11\; 12},  
 \tilde{S}^{13\; 14},
 \label{cartan}
 \end{eqnarray}
 with the same indices for both kinds of the generators.

 
 \subsection{Subgroups of SO(1,13)}
 \label{subgroups}
 
 The group $SO(1,13)$ of the rank $7$ has as possible subgroups the groups $SO(1,3)$,  
  $SU(2), SU(3)$ and the two $U(1)$'s, with the sum of the ranks of all 
 these subgroups  equal to $7$. These subgroups 
  are  candidates for describing the spin, the weak charge, the colour charge and the two hyper charges, 
  respectively (only one is needed in the Standard model). 
  The generators of these groups  can be written in terms of the generators $S^{ab}$ as follows 
 \begin{eqnarray}
 \tau^{Ai} = \sum_{a,b} \;c^{ai}{ }_{ab} \; S^{ab},
 \nonumber\\
 \{\tau^{Ai}, \tau^{Bj}\}_- = i \delta^{AB} f^{Aijk} \tau^{Ak}.
 \label{tau}
 \end{eqnarray}
 We use $A=1,2,3,4,$ to represent the subgroups describing charges  and $f^{Aijk}$ to describe 
 the corresponding structure 
 constants. Coefficients $c^{Ai}{ }_{ab}$, with $a,b \in \{5,6,...,14\}$, have to be determined so that the 
 commutation relations of Eq.(\ref{tau}) 
 hold.
 
 The weak charge  ($SU(2)$ with the generators $\tau^{1i}$) and  one $ U(1)$ charge (with the generator
 $\tau^{21}$)  content of the compact group $SO(4)$
  (a subgroup of $SO(1,13)$) can  be demonstrated when expressing:
 %
 $\tau^{11}: = \frac{1}{2} ( {\mathcal S}^{58} - {\mathcal S}^{67} ),\;
 \tau^{12}: = \frac{1}{2} ( {\mathcal S}^{57} + {\mathcal S}^{68} ),\;
 \tau^{13}: = \frac{1}{2} ( {\mathcal S}^{56} - {\mathcal S}^{78} ); \;
 \tau^{21}: = \frac{1}{2} ( {\mathcal S}^{56} + {\mathcal S}^{78} ).$
 To see the colour charge  and one additional $U(1)$ charge content in the group $SO(1,13)$ 
 we write $\tau^{3i}$ and 
 $\tau^{41}$, respectively, in terms of  the generators ${\mathcal S}^{ab}$: 
 %
 $\tau^{31}: = \frac{1}{2} ( {\mathcal S}^{9\;12} - {\mathcal S}^{10\;11} ),\;
 \tau^{32}: = \frac{1}{2} ( {\mathcal S}^{9\;11} + {\mathcal S}^{10\;12} ),\;
 \tau^{33}: = \frac{1}{2} ( {\mathcal S}^{9\;10} - {\mathcal S}^{11\;12} ),\;
 \tau^{34}: = \frac{1}{2} ( {\mathcal S}^{9\;14} - {\mathcal S}^{10\;13} ),\;
 \tau^{35}: = \frac{1}{2} ( {\mathcal S}^{9\;13} + {\mathcal S}^{10\;14} ),\;
 \tau^{36}: = \frac{1}{2} ( {\mathcal S}^{11\;14} - {\mathcal S}^{12\;13}),\; 
 \tau^{37}: = \frac{1}{2} ( {\mathcal S}^{11\;13} + {\mathcal S}^{12\;14} ),\;
 \tau^{38}: = \frac{1}{2\sqrt{3}} ( {\mathcal S}^{9\;10} + {\mathcal
 S}^{11\;12} - 2{\mathcal S}^{13\;14}); \;
 \tau^{41}: = -\frac{1}{3}( {\mathcal S}^{9\;10} + {\mathcal S}^{11\;12}
 + {\mathcal S}^{13\;14} ).$

 To reproduce the Standard model groups one must introduce the two superposition of the two $U(1)$'s 
 generators as follows:
 $Y = \tau^{41} + \tau^{21}, \quad  Y' = \tau^{41} - \tau^{21}. $

 \subsection{One Weyl representation}
 \label{oneWeyl}

 Let us choose the starting state  of one Weyl representation of the group 
 $SO(1,13)$ to be the
 eigen state of all the members of the Cartan sub algebra (Eq.(\ref{cartan})) and is left handed
 ($\Gamma^{(1,13)} =-1$) 
 \begin{eqnarray}
 &&\stackrel{03}{(+i)}\stackrel{12}{(+)}|\stackrel{56}{(+)}\stackrel{78}{(+)}
 ||\stackrel{9 \;10}{(+)}\stackrel{11\;12}{(-)}\stackrel{13\;14}{(-)} =
 \nonumber\\
 &&(\gamma^0 -\gamma^3)(\gamma^1 +i \gamma^2)| (\gamma^5 + i\gamma^6)(\gamma^7 +i \gamma^8)||
 (\gamma^9 +i\gamma^{10})(\gamma^{11} -i \gamma^{12})(\gamma^{13}-i\gamma^{14}).
 \nonumber\\
 \label{start}
 \end{eqnarray}
 The signs "$|$" and "$||$" are to point out the  $SO(1,3)$ (up to $|$), $SO(1,7)$ (up to $||$)
 and $SO(6)$ (following $||$) substructure of the starting state of the left handed multiplet of
 $SO(1,13)$. $|\psi\rangle$ can be any state, which the Clifford objects generating the state do not 
 transform into zero. From now on  we shall therefore skip it. 
 One easily finds that the eigen values of the chosen
 Cartan sub algebra elements $S^{ab}$ and $\tilde{S}^{ab}$ (Eq.(\ref{cartan})) 
are 
 ($+i/2$, $ 1/2$, $ 1/2$, $1/2$, $1/2$, $-1/2$, $-1/2$) and ($+i/2$, $ 1/2$, $ 1/2$, $1/2$, $1/2$, $-1/2$, $-1/2$),  
 respectively. This particular state is a right handed spinor with respect to $SO(1,3)$ ($\Gamma^{(1,3)} =1$), 
 with spin up 
 ($S^{12} =1/2$), it is $SU(2)$  
 singlet ($\tau^{13} = 0$ and all $\tau^{1i}$ give zero, 
 and it is the member 
 of  the $SU(3)$ triplet (subsect.\ref{subgroups}) with ($\tau^{33} =1/2, \tau^{38} = 1/(2 \sqrt{3})$),
 it has $\tau^{21} = 1/2$ and $\tau^{41}= 1/6$, while $Y=2/3$ and $Y'= -1/3$. We further find 
 that $\Gamma^{(4)} =1$ (the handedness of the group $SO(4)$, whose 
 subgroups are $SU(2)$ and $U(1)$), 
 $\Gamma^{(1,7)}= 1$ and $ \Gamma^{(6)} = -1$. The starting state (Eq.(\ref{start})) can be recognized 
 in terms of the Standard model subgroups as the right handed weak chargeless $u$-quark carrying 
 one of the three colours (See also Table~\ref{abb-snmb-tableI}).

 To obtain all the states of one Weyl spinor, one only has to apply on the starting state of Eq.(\ref{start})
 the generators $S^{ab}$. All the quarks and the leptons of one family of the Standard model appear in this multiplet 
 (together with the corresponding anti quarks and anti leptons). 
 We present in Table~\ref{abb-snmb-tableI} all the quarks of one particular colour: the right handed weak 
 chargeless $u_R,d_R$ and left handed weak charged $u_L, d_L$, 
 with the colour $\tau^{33}=1/2, \tau^{38}=1/(2\sqrt{3}),$ in 
 the  Standard model notation.  
 They all are  members of one $SO(1,7)$ multiplet. The colourless  $SO(1,7)$ multiplet is the 
 neutrino and electron multiplet (See Table~\ref{abb-snmb-tableII}).

\begin{table}
 \begin{center}
 \begin{tabular}{|r|c||c||c|c||c|c|c||c|c|c||r|r|}
 \hline
 i&$$&$|^a\psi_i>$&$\Gamma^{(1,3)}$&$ S^{12}$&$\Gamma^{(4)}$&
 $\tau^{13}$&$\tau^{21}$&$\tau^{33}$&$\tau^{38}$&$\tau^{41}$&$Y$&$Y'$\\
 \hline\hline
 && ${\rm Octet},\;\Gamma^{(1,7)} =1,\;\Gamma^{(6)} = -1,$&&&&&&&&&& \\
 && ${\rm of \; quarks}$&&&&&&&&&&\\
 \hline\hline
 1&$u_{R}^{c1}$&$\stackrel{03}{(+i)}\stackrel{12}{(+)}|\stackrel{56}{(+)}\stackrel{78}{(+)}
 ||\stackrel{9 \;10}{(+)}\stackrel{11\;12}{(-)}\stackrel{13\;14}{(-)}$
 &1&$\frac{1}{2}$&1&0&$\frac{1}{2}$&$\frac{1}{2}$&$\frac{1}{2\sqrt{3}}$&$\frac{1}{6}$&$\frac{2}{3}$&$-\frac{1}{3}$\\
 \hline 
 2&$u_{R}^{c1}$&$\stackrel{03}{[-i]}\stackrel{12}{[-]}|\stackrel{56}{(+)}\stackrel{78}{(+)}
 ||\stackrel{9 \;10}{(+)}\stackrel{11\;12}{(-)}\stackrel{13\;14}{(-)}$
 &1&$-\frac{1}{2}$&1&0&$\frac{1}{2}$&$\frac{1}{2}$&$\frac{1}{2\sqrt{3}}$&$\frac{1}{6}$&$\frac{2}{3}$&$-\frac{1}{3}$\\
 \hline
 3&$d_{R}^{c1}$&$\stackrel{03}{(+i)}\stackrel{12}{(+)}|\stackrel{56}{[-]}\stackrel{78}{[-]}
 ||\stackrel{9 \;10}{(+)}\stackrel{11\;12}{(-)}\stackrel{13\;14}{(-)}$
 &1&$\frac{1}{2}$&1&0&$-\frac{1}{2}$&$\frac{1}{2}$&$\frac{1}{2\sqrt{3}}$&$\frac{1}{6}$&$-\frac{1}{3}$&$\frac{2}{3}$\\
 \hline 
 4&$d_{R}^{c1}$&$\stackrel{03}{[-i]}\stackrel{12}{[-]}|\stackrel{56}{[-]}\stackrel{78}{[-]}
 ||\stackrel{9 \;10}{(+)}\stackrel{11\;12}{(-)}\stackrel{13\;14}{(-)}$
 &1&$-\frac{1}{2}$&1&0&$-\frac{1}{2}$&$\frac{1}{2}$&$\frac{1}{2\sqrt{3}}$&$\frac{1}{6}$&$-\frac{1}{3}$&$\frac{2}{3}$\\
 \hline
 5&$d_{L}^{c1}$&$\stackrel{03}{[-i]}\stackrel{12}{(+)}|\stackrel{56}{[-]}\stackrel{78}{(+)}
 ||\stackrel{9 \;10}{(+)}\stackrel{11\;12}{(-)}\stackrel{13\;14}{(-)}$
 &-1&$\frac{1}{2}$&-1&$-\frac{1}{2}$&0&$\frac{1}{2}$&$\frac{1}{2\sqrt{3}}$&$\frac{1}{6}$&$\frac{1}{6}$&$\frac{1}{6}$\\
 \hline
 6&$d_{L}^{c1}$&$\stackrel{03}{(+i)}\stackrel{12}{[-]}|\stackrel{56}{[-]}\stackrel{78}{(+)}
 ||\stackrel{9 \;10}{(+)}\stackrel{11\;12}{(-)}\stackrel{13\;14}{(-)}$
 &-1&$-\frac{1}{2}$&-1&$-\frac{1}{2}$&0&$\frac{1}{2}$&$\frac{1}{2\sqrt{3}}$&$\frac{1}{6}$&$\frac{1}{6}$&$\frac{1}{6}$\\
 \hline
 7&$u_{L}^{c1}$&$\stackrel{03}{[-i]}\stackrel{12}{(+)}|\stackrel{56}{(+)}\stackrel{78}{[-]}
 ||\stackrel{9 \;10}{(+)}\stackrel{11\;12}{(-)}\stackrel{13\;14}{(-)}$
 &-1&$\frac{1}{2}$&-1&$\frac{1}{2}$&0&$\frac{1}{2}$&$\frac{1}{2\sqrt{3}}$&$\frac{1}{6}$&$\frac{1}{6}$&$\frac{1}{6}$\\
 \hline
 8&$u_{L}^{c1}$&$\stackrel{03}{(+i)}\stackrel{12}{[-]}|\stackrel{56}{(+)}\stackrel{78}{[-]}
 ||\stackrel{9 \;10}{(+)}\stackrel{11\;12}{(-)}\stackrel{13\;14}{(-)}$
 &-1&$-\frac{1}{2}$&-1&$\frac{1}{2}$&0&$\frac{1}{2}$&$\frac{1}{2\sqrt{3}}$&$\frac{1}{6}$&$\frac{1}{6}$&$\frac{1}{6}$\\
 \hline\hline
 \end{tabular}
 \end{center}
\caption{\label{abb-snmb-tableI}%
 The 8-plet of quarks - the members of $SO(1,7)$ subgroup, belonging to 
one Weyl left handed ($\Gamma^{(1,13)} = -1 = \Gamma^{(1,7)} \times \Gamma^{(6)}$) spinor representation of 
 $SO(1,13)$. 
 It contains  left handed weak charged quarks and right handed weak chargeless quarks of a particular 
 colour ($(1/2,1/(2\sqrt{3}))$). Here  $\Gamma^{(1,3)}$ defines the handedness in $(1+3)$ space, 
 $ S^{12}$ defines the ordinary spin (which can also be red directly from the wave function), 
 $\tau^{13}$ defines the weak charge, $\tau^{21}$ defines the $U(1)$ charge, $\tau^{33}$ and 
 $\tau^{38}$ define the colour charge and $\tau^{41}$ another $U(1)$ charge, which together with the
 first one defines $Y$ and $Y'$. 
 The reader can find the whole Weyl representation in 
the ref.\cite{ABBPortoroz03}.}
\end{table}

 In Table~\ref{abb-snmb-tableII} we present the leptons of one family of the Standard model. All belong 
 to the same multiplet with respect to the group $SO(1,7)$ (and are the members of the same Weyl 
 representation as the quarks 
 of Table~\ref{abb-snmb-tableI}). They are colour chargeless and differ accordingly from the quarks in Table~\ref{abb-snmb-tableI} in the 
 second $U(1)$ charge.  The quarks and the leptons are 
 equivalent with respect to the group $SO(1,7)$.  

\begin{table}
 \begin{center}
 \begin{tabular}{|r|c||c||c|c||c|c|c||c|c|c||r|r|}
 \hline
 i&$$&$|^a\psi_i>$&$\Gamma^{(1,3)}$&$ S^{12}$&$\Gamma^{(4)}$&
 $\tau^{13}$&$\tau^{21}$&$\tau^{33}$&$\tau^{38}$&$\tau^{41}$&$Y$&$Y'$\\
 \hline\hline
 && ${\rm Octet},\;\Gamma^{(1,7)} =1,\;\Gamma^{(6)} = -1,$&&&&&&&&&& \\
 && ${\rm of \; leptons}$&&&&&&&&&&\\
 \hline\hline
 1&$\nu_{R}$&$\stackrel{03}{(+i)}\stackrel{12}{(+)}|\stackrel{56}{(+)}\stackrel{78}{(+)}
 ||\stackrel{9 \;10}{(+)}\stackrel{11\;12}{[+]}\stackrel{13\;14}{[+]}$
 &1&$\frac{1}{2}$&1&0&$\frac{1}{2}$&0&$0$&$-\frac{1}{2}$&0&-1\\
 \hline 
 2&$\nu_{R}$&$\stackrel{03}{[-i]}\stackrel{12}{[-]}|\stackrel{56}{(+)}\stackrel{78}{(+)}
 ||\stackrel{9 \;10}{(+)}\stackrel{11\;12}{[+]}\stackrel{13\;14}{[+]}$
 &1&$-\frac{1}{2}$&1&0&$\frac{1}{2}$&0&$0$&$-\frac{1}{2}$&0&-1\\
 \hline
 3&$e_{R}$&$\stackrel{03}{(+i)}\stackrel{12}{(+)}|\stackrel{56}{[-]}\stackrel{78}{[-]}
 ||\stackrel{9 \;10}{(+)}\stackrel{11\;12}{[+]}\stackrel{13\;14}{[+]}$
 &1&$\frac{1}{2}$&1&0&$-\frac{1}{2}$&0&$0$&$-\frac{1}{2}$&-1&0\\
 \hline 
 4&$e_{R}$&$\stackrel{03}{[-i]}\stackrel{12}{[-]}|\stackrel{56}{[-]}\stackrel{78}{[-]}
 ||\stackrel{9 \;10}{(+)}\stackrel{11\;12}{[+]}\stackrel{13\;14}{[+]}$
 &1&$-\frac{1}{2}$&1&0&$-\frac{1}{2}$&0&$0$&$-\frac{1}{2}$&-1&0\\
 \hline
 5&$e_{L}$&$\stackrel{03}{[-i]}\stackrel{12}{(+)}|\stackrel{56}{[-]}\stackrel{78}{(+)}
 ||\stackrel{9 \;10}{(+)}\stackrel{11\;12}{[+]}\stackrel{13\;14}{[+]}$
 &-1&$\frac{1}{2}$&-1&$-\frac{1}{2}$&0&0&$0$&$-\frac{1}{2}$&$-\frac{1}{2}$&$-\frac{1}{2}$\\
 \hline
 6&$e_{L}$&$\stackrel{03}{(+i)}\stackrel{12}{[-]}|\stackrel{56}{[-]}\stackrel{78}{(+)}
 ||\stackrel{9 \;10}{(+)}\stackrel{11\;12}{[+]}\stackrel{13\;14}{[+]}$
 &-1&$-\frac{1}{2}$&-1&$-\frac{1}{2}$&0&0&$0$&$-\frac{1}{2}$&$-\frac{1}{2}$&$-\frac{1}{2}$\\
 \hline
 7&$\nu_{L}$&$\stackrel{03}{[-i]}\stackrel{12}{(+)}|\stackrel{56}{(+)}\stackrel{78}{[-]}
 ||\stackrel{9 \;10}{(+)}\stackrel{11\;12}{[+]}\stackrel{13\;14}{[+]}$
 &-1&$\frac{1}{2}$&-1&$\frac{1}{2}$&0&0&$0$&$-\frac{1}{2}$&$-\frac{1}{2}$&$-\frac{1}{2}$\\
 \hline
 8&$\nu_{L}$&$\stackrel{03}{(+i)}\stackrel{12}{[-]}|\stackrel{56}{(+)}\stackrel{78}{[-]}
 ||\stackrel{9 \;10}{(+)}\stackrel{11\;12}{[+]}\stackrel{13\;14}{[+]}$
 &-1&$-\frac{1}{2}$&-1&$\frac{1}{2}$&0&0&$0$&$-\frac{1}{2}$&$-\frac{1}{2}$&$-\frac{1}{2}$\\
 \hline\hline
 \end{tabular}
 \end{center}
\caption{\label{abb-snmb-tableII}%
The 8-plet of leptons - the members of $SO(1,7)$ subgroup, 
belonging to one Weyl left 
 handed ($\Gamma^{(1,13)} = -1 = \Gamma^{(1,7)} \times \Gamma^{(6)}$) spinor representation of 
 $SO(1,13)$, is presented. 
 It contains  left handed weak charged leptons and right handed weak chargeless leptons, all 
  colour chargeless. The two $8$-plets in Table~\ref{abb-snmb-tableI} 
and~\ref{abb-snmb-tableII} are equivalent with respect to the group 
  $SO(1,7)$. They only distinguish in properties with respect to the group $SU(3)$ and $U(1)$ and consequently 
  in $Y$ and $Y'$.}
\end{table} 

 \subsection{Appearance of families}
 \label{families}

 While the generators of the Lorentz group $S^{ab}$, with  a pair of $(ab)$, 
 which does not belong
 to the Cartan sub algebra (Eq.(\ref{cartan})), transform one vector of one Weyl representation into another vector 
 of the same Weyl representation, 
 transform the generators $\tilde{S}^{ab}$ (again if the pair $(ab)$ does not belong to the Cartan set) 
 a member of one family into the same member of another family, leaving all the other quantum numbers 
 (determined by $S^{ab}$)
 unchanged\cite{ABBnorma92,ABBnorma93,ABBnorma95,ABBnorma01,ABBholgernorma00,ABBPortoroz03}. 
 This is happening since  the application of  $\gamma^a$  changes the operator 
 $\stackrel{ab}{(+)}$ (or the operator $\stackrel{ab}{(+i)}$)
 into the operator $\stackrel{ab}{[-]}$ (or the operator $\stackrel{ab}{[-i]}$, respectively), while the operator
 $\tilde{\gamma}^a$ 
 changes $\stackrel{ab}{(+)}$ (or $\stackrel{ab}{(+i)}$)  into $\stackrel{ab}{[+]}$
 (or into $\stackrel{ab}{[+i]}$, respectively), without changing the ''eigen values'' of the Cartan 
 sub algebra set of the operators $S^{ab}$.  According to what we have discussed above, the operator 
 $S^{07} S^{cd}$, with $(cd)$ belonging to the Cartan set, changes, for example, a right handed 
 weak chargeless $u_R$-quark to a left handed weak charged $u_L$-quark, contributing to only diagonal elements  
 of the $u$-quarks mass matrix, while $S^{78}\tilde{S}^{ab}$ either contribute to the diagonal elements (if $(ab)$) 
 belong to the Cartan set, or to non diagonal elements otherwise. 
 Bellow, as an example, the application of $\tilde{S}^{01}$ 
 (or $\tilde{S}^{02}, \tilde{S}^{31}, \tilde{S}^{32}$) and $\tilde{S}^{07}$ (or $\tilde{S}^{08}, \tilde{S}^{37}, 
 \tilde{S}^{38}$) or of two of these operators on the 
 state of Eq.({\ref{start}}) is presented.
 \begin{eqnarray}
 \stackrel{03}{(+i)} \stackrel{12}{(+)}| \stackrel{56}{(+)} \stackrel{78}{(+)}||
 \stackrel{9 10}{(+i)} \stackrel{11 12}{(-)} \stackrel{13 14}{(-)} \nonumber\\
 \stackrel{03}{[+i]} \stackrel{12}{[+]}| \stackrel{56}{(+)} \stackrel{78}{(+)}||  
 \stackrel{9 10}{(+i)} \stackrel{11 12}{(-)} \stackrel{13 14}{(-)}\nonumber\\
 \stackrel{03}{[+i]} \stackrel{12}{(+)}| \stackrel{56}{(+)} \stackrel{78}{[+]}||
 \stackrel{9 10}{(+i)} \stackrel{11 12}{(-)} \stackrel{13 14}{(-)} \nonumber\\
 \stackrel{03}{(+i)} \stackrel{12}{[+]}| \stackrel{56}{(+)} \stackrel{78}{[+]}||
 \stackrel{9 10}{(+i)} \stackrel{11 12}{(-)} \stackrel{13 14}{(-)}.
 \label{twofamilies}
 \end{eqnarray}
 %
 One can easily see that all the states of (\ref{twofamilies})  describe a right handed $u$-quark of 
 the same colour. They are equivalent with respect to the operators $S^{ab}$. They only differ
 in properties, determined by the operators $\tilde{S}^{ab}$. Since quarks and leptons of the 
 three measured families are equivalent representations with respect to the spin and the charge generators, 
 the proposed way of generating families seems very promising.

\section{ Weyl spinors in $d= (1+13)$ manifesting families of quarks and leptons in $d= (1+3)$} 
\label{lagrangesec}

We start with a  left handed Weyl spinor  in $(1+13)$-dimensional space. A spinor carries 
only the spin (no charges) and interacts accordingly with only the gauge gravitational fields - with 
their spin connections and vielbeins. We make use of two kinds of the Clifford algebra objects and allow accordingly 
two kinds of the gauge fields\cite{ABBnorma92,ABBnorma93,ABBnormasuper94,ABBnorma95,ABBnorma97,%
ABBpikanormaproceedings1,ABBholgernorma00,ABBnorma01,ABBpikanormaproceedings2,ABBPortoroz03}. 

One kind is the ordinary
gauge field (gauging the Poincar\' e symmetry in $d=1+13$). The corresponding spin connection field
appears for spinors as a gauge field of $S^{ab}= \frac{1}{4} (\gamma^a \gamma^b - \gamma^b \gamma^a)$, 
where $\gamma^a$ are the ordinary Dirac operators. 
The contribution of these fields to the mass matrices manifests in only the diagonal terms  
(connecting right handed weak chargeless quarks or leptons with left handed weak charged partners within 
one family of spinors). 

The second kind of gauge fields is in our approach responsible for the Yukawa couplings
among families of spinors and  might explain the origin of the families of quarks and leptons. 
The corresponding spin connection field appears for spinors as a gauge field 
of $\tilde{S}^{ab}$ ($\tilde{S}^{ab} = \frac{1}{2} (\tilde{\gamma}^a \tilde{\gamma}^b-
\tilde{\gamma}^b \tilde{\gamma}^a)$).

Accordingly we write the action for a Weyl (massless) spinor  
in $d(=1+13)$ - dimensional space as follows
\begin{eqnarray}
S &=& \int \; d^dx \; {\mathcal L}
\nonumber\\
{\mathcal L} &=& \frac{1}{2} (E\bar{\psi}\gamma^a p_{0a} \psi) + h.c. = \frac{1}{2} 
(E\bar{\psi} \gamma^a f^{\alpha}{}_a p_{0\alpha}\psi) + h.c. 
\nonumber\\
p_{0\alpha} &=& p_{\alpha} - \frac{1}{2}S^{ab} \omega_{ab\alpha} - \frac{1}{2}\tilde{S}^{ab} \tilde{\omega}_{ab\alpha}.
\label{lagrange}
\end{eqnarray}
Latin indices  
$a,b,..,m,n,..,s,t,..$ denote a tangent space (a flat index),
while Greek indices $\alpha, \beta,..,\mu, \nu,.. \sigma,\tau ..$ denote an Einstein 
index (a curved index). Letters  from the beginning of both the alphabets
indicate a general index ($a,b,c,..$   and $\alpha, \beta, \gamma,.. $ ), 
 from the middle of both the alphabets   
 the observed dimensions $0,1,2,3$ ($m,n,..$ and $\mu,\nu,..$), indices from the bottom of the alphabets
 indicate the compactified dimensions ($s,t,..$ and $\sigma,\tau,..$). We assume the signature $\eta^{ab} =
diag\{1,-1,-1,\cdots,-1\}$. 
Here $f^{\alpha}{}_a$ are  vielbeins (inverted to the gauge field of the generators of translations  
$e^{a}{}_{\alpha}$, $e^{a}{}_{\alpha} f^{\alpha}{}_{b} = \delta^{a}_{b}$,
$e^{a}{}_{\alpha} f^{\beta}{}_{a} = \delta_{\alpha}{}^{\beta}$),
with $E = det(e^{a}{}_{\alpha})$, while  
$\omega_{ab\alpha}$ and $\tilde{\omega}_{ab\alpha} $ are the two kinds of the spin connection fields, the gauge 
fields of $S^{ab}$ and $\tilde{S}^{ab}$, respectively, corresponding to the two kinds of the Clifford algebra 
objects\cite{ABBholgernorma02,ABBPortoroz03}, namely $\gamma^a$ and $\tilde{\gamma}^{a}$, both types fulfilling the same
Clifford algebra relations, but anti commuting with each other ($\{\gamma^a,\tilde{\gamma}^b\}_+ =0$),
while the two corresponding types of the generators of the Lorentz transformations commute 
($\{ S^{ab}, \tilde{S}^{cd}\}_-=0$).  We introduced these two kinds of $\gamma^a$'s in Subsect.\ref{technique}.

We saw in subsect.\ref{oneWeyl} that  one Weyl spinor in $d=(1+13)$ with the spin as the only internal degree of 
freedom, can manifest  in
four-dimensional ''physical'' space  as the ordinary ($SO(1,3)$) spinor with all the known charges 
of one family of  quarks and leptons of the Standard model.
(The reader can see this analyses also in several references, like the one\cite{ABBPortoroz03}.)

To see that the Yukawa couplings are the part of the starting Lagrangean of Eq.(\ref{lagrange}), we must rewrite
the Lagrangean in  Eq.(\ref{lagrange}) in an appropriate way\cite{ABBPortoroz03}.

First we recognize
 that $-\frac{1}{2} S^{ab}\omega_{abm} = -\sum_{A,i}\tau ^{Ai} A^{Ai}{}_{m} +{\rm 
 \; additional\; terms},$  with $m=0,1,2,3$ 
 and $\tau^{Ai}$ defined in (Eq.\ref{tau}). We expect that under appropriate break of the symmetry $SO(1,13)$
 to the symmetry of the groups of the Standard model, the additional terms would be negligible at 
 ''physical'' energies. Next we recognize from Table~\ref{abb-snmb-tableI} that only the terms $\sum_{s=7,8} \;
 \psi^{\dagger} \gamma^0 \gamma^s p_{0s} \psi $, with $p_{0s} = p_s - \frac{1}{2} S^{ab}\omega_{abs} 
 -\frac{1}{2} \tilde{S}^{ab} \tilde{\omega}_{abs}$, can appear as terms, transforming right 
 handed weak chargeless 
 $u_R$ into weak charged $u_L$ of the same spin and colour (as seen on 
Table~\ref{abb-snmb-tableI}, if we apply $\gamma^0 \gamma^7$ 
 or $\gamma^0 \gamma^8$ on $u_R$)  in the action. For the rest of terms we assume that 
 they are at low energies negligible. 
 Under these discussion and assumptions the action (\ref{lagrange}) can be rewritten  
 as follows
 \begin{eqnarray}
 S = \int  d^dx \;(\bar{\psi} \gamma^{m} (p_{m}- \sum_{A,i} \;  
 \tau^{Ai} A^{Ai}_{m}) \; \psi + {\cal L}_Y), \; {\cal L}_Y = \sum_{s=7,8}\; 
 \psi^{\dagger} \gamma^0 \gamma^{s} p_{0s} \; \psi.
 \label{lagrange1}
 \end{eqnarray}
 We neglected the additional terms, expecting that they are negligible at ''physical'' energies. 
 It is the term ${\cal L}_Y$ in the action (\ref{lagrange1}), which can be interpreted as Yukawa couplings,
 determining the mass matrices in the Standard electroweak and colour model. Let us point out 
 that the terms $-\frac{1}{2} S^{ab} \omega_{abs}$ in $p_{0s}$ contribute only to diagonal mass matrix elements 
 (they stay within a family), while the terms $-\frac{1}{2} \tilde{S}^{ab} \tilde{\omega}_{abs}$ contribute to
 diagonal as well as off diagonal matrix elements.

 We can further rewrite the mass term ${\cal L}_Y$ if recognizing that $
 (\gamma^7 \pm i \gamma^8 ) = 2 \stackrel{78}{(\pm)}.$ 
 We then find 
 \begin{eqnarray}
  {\cal L}_Y &=& \psi^{\dagger} \gamma^0 \{\stackrel{78}{(+)} p_{0+} + \stackrel{78}{(-)}p_{0-} \} \; \psi,
  \quad {\rm with} \nonumber\\
  p_{0\pm} &=& (p_{7} \mp i \; p_{8}) - \frac{1}{2} S^{ab} \omega_{ab\pm}  - 
   \frac{1}{2} \tilde{S}^{ab} \tilde{\omega}_{ab\pm}, \nonumber\\
   \omega_{ab\pm} &=& \omega_{ab7} \mp i \; \omega_{ab8},\quad 
 \tilde{\omega}_{ab\pm} = \tilde{\omega}_{ab 7} \mp i \; \tilde{\omega}_{ab8}. 
 \label{Yukawa}
 \end{eqnarray}
 We also can rewrite $\tilde{S}^{ab}$, which do not contribute to Cartan sub algebra,  
 in terms of nilpotents 
 \begin{eqnarray}
 \stackrel{ab}{\tilde{(k)}}\stackrel{cd}{\tilde{(l)}},  
 \label{lowertilde}
 \end{eqnarray}
 with indices $(ab)$ and $(cd)$ which belong to the Cartan sub algebra indices (Eq.(\ref{cartan})).
 We may write accordingly
 \begin{eqnarray}
  - \sum_{(a,b) } \frac{1}{2} \stackrel{78}{(\pm)}\tilde{S}^{ab} \tilde{\omega}_{ab\pm} =
 \sum_{(ac),(bd), \;  k,l}\stackrel{78}{(\pm)}\stackrel{ac}{\tilde{(k)}}\stackrel{bd}{\tilde{(l)}} 
 \; \tilde{A}^{kl}{}_{\pm} ((ac),(bd)),
 \label{partofY}
 \end{eqnarray}
 where the pair $(a,b)$ on the left hand side of equality sign runs over all the  indices, 
 which do not characterise  the Cartan 
 sub algebra set, with $ a,b = 0,\dots, 8$,  while the two pairs $((ac),(bd))$ 
are used to denote only the 
 Cartan sub algebra pairs
  ($((03), (12))$; $((03), (56))$ ;$((03), (78))$;
 $((12),(56))$; $((12), (78))$; $((56),(78)))$ ; $k$ and $l$ run over four 
 possible values so that $k=\pm i$, if $(ac) = (03)$ 
 and $k=\pm 1$ in all other cases, while $l=\pm 1$.
 The fields  $\tilde{A}^{kl}_{\pm} ((ac),(bd))$ can then 
 be expressed by $\tilde{\omega}_{ab \pm}$ as follows
 \begin{eqnarray}
 \tilde{A}^{++}_{\pm} ((ac),(bd)) &=& -\frac{i}{2} (\tilde{\omega}_{ac\pm} -\frac{i}{r} \tilde{\omega}_{bc\pm} 
 -i \tilde{\omega}_{ad\pm} -\frac{1}{r} \tilde{\omega}_{bd\pm} ), \nonumber\\
 \tilde{A}^{--}_{\pm} ((ac),(bd)) &=& -\frac{i}{2} (\tilde{\omega}_{ac\pm} +\frac{i}{r} \tilde{\omega}_{bc\pm} 
 +i \tilde{\omega}_{ad\pm} -\frac{1}{r} \tilde{\omega}_{bd\pm} ),\nonumber\\
 \tilde{A}^{+-}_{\pm} ((ac),(bd)) &=& -\frac{i}{2} (\tilde{\omega}_{ac\pm} + i r \tilde{\omega}_{bc\pm} 
 - i r^2 \tilde{\omega}_{ad\pm} +\frac{1}{r} \tilde{\omega}_{bd\pm} ), \nonumber\\
 \tilde{A}^{-+}_{\pm} ((ac),(bd)) &=& -\frac{i}{2} (\tilde{\omega}_{ac\pm} - i r \tilde{\omega}_{bc\pm} 
 + i r^2 \tilde{\omega}_{ad\pm} +\frac{1}{r} \tilde{\omega}_{bd\pm} ),
 \label{Awithomega}
 \end{eqnarray}
 with $r =i$, if  $ (ac)= (03)$, otherwise $r=1$. We simplify the indices $k$ and $l$ in the exponent 
 of fields $\tilde{A}^{kl}_{\pm} ((ac),(bd))$ to $\pm $, omitting $i$. 
 
 We then end up with the Lagrange density, determining in the ''physical'' space the masses 
 of quarks and leptons as follows
 \begin{eqnarray}
 {\mathcal L}_{Y} &=& \psi^+ \gamma^0 \;  
 \{  \stackrel{78}{(+)} ( p_+ - \sum_{(ab)} \frac{1}{2} ( S^{ab}\omega_{ab+} + 
 \tilde{S}^{ab}\tilde{\omega}_{ab+})) + \nonumber\\
  && \stackrel{78}{(-)} ( p_- - \sum_{(ab)} \frac{1}{2} ( S^{ab}\omega_{ab-} + 
 \tilde{S}^{ab}\tilde{\omega}_{ab-})) + \nonumber\\
  && \stackrel{78}{(+)} \sum_{\{(ac)(bd) \},k,l} \; \stackrel{ac}{\tilde{(k)}} \stackrel{bd}{\tilde{(l)}}
 \tilde{{A}}^{kl}_{+}((ab),(cd)) \;\;+  \nonumber\\
  && \stackrel{78}{(-)} \sum_{\{(ac)(bd) \},k,l} \; \stackrel{ac}{\tilde{(k)}}\stackrel{bd}{\tilde{(l)}}
 \tilde{{A}}^{kl}_{-}((ab),(cd))\}\psi,
 \label{yukawa4tilde}
 \end{eqnarray}
 with  pairs $(ab), ((ac),(bd))$, which run over all the members of the Cartan sub algebra, while  $k=\pm i,$ if 
 $ (ac) =(03)$, otherwise $k=\pm 1$ and $l= \pm 1$.
 
 Since according to Eq.(\ref{graphbinoms}) $\stackrel{78}{(k)} \stackrel{78}{(k)} = 0, 
 \stackrel{78}{(-k)} \stackrel{78}{(k)} = -  \stackrel{78}{[-k]}, \stackrel{78}{(k)}\stackrel{ab}{[k]} = 0,$
 $\stackrel{78}{(k)}\stackrel{78}{[-k]} =  \stackrel{78}{(k)}$, 
 we see, looking at Table~\ref{abb-snmb-tableI}, that only terms with $\stackrel{78}{(-)}$ contribute
 to the mass matrix of $u$-quarks (or $\nu$ in the lepton case) and only terms with $\stackrel{78}{(+)}$ 
 contribute to  the mass matrix of $d$-quarks (or electrons in the lepton case).

Let us now repeat all the assumptions we have made up to now. They are either the starting assumptions 
of our approach unifying spins and charges, or were  needed in order that we guarantee that at low energies 
we agree with the Standard model assumptions.

The assumptions:

a.i. We use the approach, unifying spins and charges, which assumes, that in $d=1+13$ massless 
spinors carry  two types of 
spins: the ordinary (in $d=1 + 13$) one, which we describe by $S^{ab} = 
\frac{1}{4}(\gamma^a \gamma^b - \gamma^b \gamma^a)$ and the additional one, described by $\tilde{S}^{ab} =
\frac{1}{4}(\tilde{\gamma}^a \tilde{\gamma}^b - \tilde{\gamma}^b \tilde{\gamma}^a)$. The two types of the 
Clifford algebra objects anti commute ($\{\gamma^a, \tilde{\gamma}^b \}_+ =0$). Spinors carry no charges in $d=1+13$. 
The appropriate break of symmetries assure that the superpositions of operators $S^{ab}$ determine at low energies, 
the ordinary spin in $d= 1+3$ and all the known charges, while $\tilde{S}^{ab}$ generate  families of spinors. 
Accordingly spinors 
interact with only the gravitational fields, the gauge fields of the Poincar\' e group ($p_{\alpha}, $
$S^{ab}$) and the gauge fields of the operators $\tilde{S}^{ab}$:
$p_{0a} = f^{\alpha}{}_{a}p_{\alpha} - \frac{1}{2} (S^{cd} \omega_{cda} + \tilde{S}^{cd} \tilde{\omega}_{cda}).$

a.ii. The break of symmetry $SO(1,13)$ into $SO(1,7)\times SU(3)\times U(1)$ occurs in a way that only massless spinors 
in $d=1+7$ with the charge $ SU(3)\times U(1)$ survive. Further breaks cause that the charge 
$Q= \tau^{41} + S^{56}$ is conserved.

\section{An example of mass matrices for  quarks of four families}
\label{example}

Let us make, for simplicity, two further assumptions besides the two (a.i-a.ii.), presented at the end of 
Sect.\ref{lagrangesec}:

b.i. The break of symmetries influences the Poincar\' e symmetry and the symmetries described by 
$\tilde{S^{ab}}$ in a similar way. We also assume that the terms which include $p_{\sigma}, \sigma=(5),..,(14)$ do not 
contribute at low energies.

b.ii. There are no terms, which would in Eq.(\ref{yukawa4tilde}) transform  
$\stackrel{\tilde{56}}{(+)}$ into $\stackrel{\tilde{56}}{[+]}$. 

b.iii. The estimation will be done on  ''a tree level''. 

Since we do not know either how does the break of symmetries occur or how does the break influence the 
strength of the fields $\tilde{\omega}_{abc}$ (we have not yet studied in details the break of symmetries 
in the proposed approach),
we can not really say, whether or not these assumptions are justified. Yet we hope that 
these simplifications allows us 
to estimate predictions of the proposed approach. 

Our approach predicts an even number of families.  The assumption b.ii. leads 
to four (instead of eight) families of quarks and leptons, presented 
in the Eq.(\ref{twofamilies}).

Integrating the Lagrange density $L_Y$ over the coordinates and the internal (spin) degrees of freedom, 
we end up 
with the mass matrices for four families of quarks and leptons (Eq.(\ref{yukawa4tilde})),  
presented in Table~\ref{abb-snmb-tableIII}. 
\begin{table}
\begin{center}
\begin{tabular}{|r||c|c|c|c|}
\hline
$$&$ I_{R} $&$ II_{R} $&$ III_{R} $&$ IV_{R}$\\
\hline\hline
&&&& \\
$I_{L}$   & $ A^I_{\mp} $ & $ \tilde{A}^{++}_{\mp} ((03),(12)) $ & $ \pm \tilde{A}^{++}_{\mp} ((03),(78))$ &
$ \mp  \tilde{A}^{++}_{\mp} ((12),(78))$ \\
&&&& \\
\hline 
&&&& \\
$II_{L}$  & $ \tilde{A}^{--}_{\mp} ((03),(12)) $ & $ A^{II}_{\mp} $ & $ \pm \tilde{A}^{-+}_{\mp} ((12),(78))$ &
$ \mp  \tilde{A}^{-+}_{\mp} ((03),(78))$ \\
&&&& \\
\hline 
&&&& \\
$III_{L}$ & $ \pm \tilde{A}^{--}_{\mp} ((03),(78)) $ & $\mp \tilde{A}^{+-}_{\mp} ((12),(78)) $ & $ A^{III}_{\mp}$ &
$  \pm \tilde{A}^{-+}_{\mp} ((03),(12))$ \\
&&&& \\
\hline 
&&&& \\
$IV_{L}$  & $\pm \tilde{A}^{--}_{\mp} ((12),(78)) $ & $\mp \tilde{A}^{+-}_{\mp} ((03),(78)) $ & 
$ \tilde{A}^{+-}_{\mp} ((03),(12))$ & $ A^{IV}_{\mp} $ \\
&&&& \\
\hline\hline
\end{tabular}
\end{center}
\caption{\label{abb-snmb-tableIII}%
 The mass matrices for four families of quarks and leptons in the approach unifying spins 
and charges, obtained under the assumptions a.i.- a.ii. and b.i.- b.iii..
The values $ A^{I'}_{-}, I'= I,II,III,IV,$ and $ \tilde{A}^{lm}_{-} ((ac),(bd)); l,m = \pm,$  determine 
matrix elements for the $u$ quarks and the neutrinos, the values $ A^{I'}_{+},  I'= I,II,III,IV,$ and $
\tilde{A}^{lm}_{+} ((ac),(bd)); l,m =\pm,$  determine  the matrix elements for the $d$ quarks and the electrons. 
Diagonal matrix elements are different for quarks then for  leptons and distinguish also between  the $u$ 
and the $d$ quarks and between the $\nu$ and the $e$ leptons (Eqs.\ref{yukawa4tilde}). 
They
also differ from family to family. Non diagonal matrix elements distinguish  among families and 
among $(u, \nu)$ and $(d,e)$.}
\end{table}

The explicit form of the diagonal matrix elements for the above choice of assumptions in terms of 
$\omega_{abc}$ and
$\tilde{\omega}_{abc}$ is  as follows
\begin{eqnarray}
A^{I}_{u} &=& -\frac{1}{2} (\omega_{56 -} + \omega_{78 -} + \frac{1}{3} A^{41}_{-}
+ i \tilde{\omega}_{03 -} + \tilde{\omega}_{12 -} +\tilde{\omega}_{56 -} + \tilde{\omega}_{78 -}
+ \frac{1}{3} \tilde{A}^{41}_{-}), \nonumber\\
A^{I}_{\nu} &=& -\frac{1}{2} (\omega_{56 -} + \omega_{78 -} -  A^{41}_{-}
+ i \tilde{\omega}_{03 -} + \tilde{\omega}_{12 -} +\tilde{\omega}_{56 -} + \tilde{\omega}_{78 -}
+ \frac{1}{3} \tilde{A}^{41}_{-}), \nonumber\\
A^{I}_{d} &=& -\frac{1}{2} (-\omega_{56 +} - \omega_{78 +} + \frac{1}{3} A^{41}_{+}
+ i \tilde{\omega}_{03 +} + \tilde{\omega}_{12 +}+ \tilde{\omega}_{56 +} + \tilde{\omega}_{78 +}
+ \frac{1}{3} \tilde{A}^{41}_{+}), \nonumber\\
A^{I}_{e} &=& -\frac{1}{2} (-\omega_{56 +} - \omega_{78 +} -  A^{41}_{+}
+ i \tilde{\omega}_{03 +} + \tilde{\omega}_{12 +}+ \tilde{\omega}_{56 +} + \tilde{\omega}_{78 +}
+ \frac{1}{3} \tilde{A}^{41}_{+}), \nonumber\\
A^{II}_{u} &= & A^{I}_{u} + (i \tilde{\omega}_{03-} + \tilde{\omega}_{12 -}), \quad
A^{II}_{\nu} =  A^{I}_{\nu} + (i \tilde{\omega}_{03-} + \tilde{\omega}_{12 -}), \nonumber\\
A^{II}_{d} &= & A^{I}_{d} + (i \tilde{\omega}_{03+} + \tilde{\omega}_{12 +}), \quad
A^{II}_{e} =  A^{I}_{e} + (i \tilde{\omega}_{03+} + \tilde{\omega}_{12 +}), \nonumber\\
A^{III}_{u} &= & A^{I}_{u} + (i \tilde{\omega}_{03-} + \tilde{\omega}_{78 -}), \quad
A^{III}_{\nu} =  A^{I}_{\nu} + (i \tilde{\omega}_{03-} + \tilde{\omega}_{78 -}), \nonumber\\
A^{III}_{d} &= & A^{I}_{d} + (i \tilde{\omega}_{03+} + \tilde{\omega}_{78 +}), \quad
A^{III}_{e} =  A^{I}_{e} + (i \tilde{\omega}_{03+} + \tilde{\omega}_{78 +}), \nonumber\\
A^{IV}_{u} &= & A^{I}_{u} + ( \tilde{\omega}_{12-} + \tilde{\omega}_{78 -}), \quad
A^{IV}_{\nu} =  A^{I}_{\nu} + ( \tilde{\omega}_{12-} + \tilde{\omega}_{78 -}), \nonumber\\
A^{IV}_{d} &= & A^{I}_{d} + ( \tilde{\omega}_{12+} + \tilde{\omega}_{78 +}), \quad
A^{IV}_{e} = A^{I}_{e} + ( \tilde{\omega}_{12+} + \tilde{\omega}_{78 +}).
\label{diagmfour}
\end{eqnarray}
The explicit form of nondiagonal matrix elements are written in Eq.(\ref{Awithomega}).

To evaluate briefly the structure of mass matrices we make one further assumption:

b.iv. Let the mass matrices be real and symmetric.\\

We then obtain for the $u$ quarks the  mass matrice as presented in 
Table~\ref{abb-snmb-tableIV}.

\begin{table}
\begin{center}
\begin{tabular}{|r||c|c|c|c|}
\hline
$u$&$ I_{R} $&$ II_{R} $&$ III_{R} $&$ IV_{R}$\\
\hline\hline
&&&& \\
$I_{L}$   & $ A^I_{u}  $ & $ \tilde{A}^{++}_{u} ((03),(12))= $ & $  \tilde{A}^{++}_{u} ((03),(78)) =$  &
$ -  \tilde{A}^{++}_{u} ((12),(78)) = $ \\
$$&$$ &$ \frac{1}{2}(\tilde{\omega}_{327} +\tilde{\omega}_{018})$&$ \frac{1}{2}(\tilde{\omega}_{387} +\tilde{\omega}_{078}) $&$
 \frac{1}{2}(\tilde{\omega}_{277} +\tilde{\omega}_{187})$ \\
&&&& \\
\hline
&&&&\\
$II_{L}$  & $ \tilde{A}^{--}_{u} ((03),(12))= $ & $ A^{II}_{u}= $ & $  \tilde{A}^{-+}_{u} ((12),(78)) = $ &
$ -  \tilde{A}^{-+}_{u} ((03),(78)) = $ \\
$$&$ \frac{1}{2}(\tilde{\omega}_{327} +\tilde{\omega}_{018})$&$A^{I}_{u} +  (\tilde{\omega}_{127} - \tilde{\omega}_{038})$
&$ \frac{1}{2}(\tilde{\omega}_{277} -\tilde{\omega}_{187}) $&$  \frac{1}{2}(\tilde{\omega}_{387} - \tilde{\omega}_{078})$ \\
\hline 
&&&& \\
$III_{L}$ & $  \tilde{A}^{--}_{u} ((03),(78)) =$ & $- \tilde{A}^{+-}_{u} ((12),(78))= $ & $ A^{III}_{u}=$ &
$   \tilde{A}^{-+}_{u} ((03),(12)) =$ \\
$$&$  \frac{1}{2}(\tilde{\omega}_{387} +\tilde{\omega}_{078})$&$ \frac{1}{2}(\tilde{\omega}_{277} -\tilde{\omega}_{187})$
&$A^{I}_{u} +  (\tilde{\omega}_{787} - \tilde{\omega}_{038})$&$ -\frac{1}{2}(\tilde{\omega}_{327} -\tilde{\omega}_{018}) $\\
&&&& \\
\hline 
&&&& \\
$IV_{L}$  & $ \tilde{A}^{--}_{u} ((12),(78)) =$ & $- \tilde{A}^{+-}_{u} ((03),(78)) = $ & 
$ \tilde{A}^{+-}_{u} ((03),(12))$ & $ A^{IV}_{u} $ \\
$$&$ \frac{1}{2}(\tilde{\omega}_{277} +\tilde{\omega}_{187})$&$ \frac{1}{2}(\tilde{\omega}_{387} -\tilde{\omega}_{078})$
&$ -\frac{1}{2}(\tilde{\omega}_{327} -\tilde{\omega}_{018}) $&$A^{I}_{u} +  (\tilde{\omega}_{127} + \tilde{\omega}_{787})$\\
&&&& \\
\hline\hline
\end{tabular}
\end{center}
\caption{\label{abb-snmb-tableIV}%
The mass matrix of four families of the $u$-quarks, obtained within 
the Approach unifying spins and charges and under 
assumptions a.i.-a.ii., b.i.-b.iv.
According to Eqs.(\ref{diagmfour}, \ref{Awithomega}) there are 12 free 
parameters, expressed in terms
of fields $A^I_{\alpha}$ and $\tilde{\omega}_{abc}$ for the  four families 
of two types of quarks and their
mixing matrix and of two types of leptons and their mixing matrix. }
\end{table}

The corresponding mass matrix for $d$ -quarks is presented 
in Table~\ref{abb-snmb-tableV}.

\begin{table}
\begin{center}
\begin{tabular}{|r||c|c|c|c|}
\hline
$d$&$ I_{R} $&$ II_{R} $&$ III_{R} $&$ IV_{R}$\\
\hline\hline
&&&& \\
$I_{L}$   & $ A^I_{d}  $ & $ \tilde{A}^{++}_{d} ((03),(12))= $ & $  - \tilde{A}^{++}_{d} ((03),(78)) =$  &
$   \tilde{A}^{++}_{d} ((12),(78)) = $ \\
$$&$$ &$ \frac{1}{2}(\tilde{\omega}_{327} - \tilde{\omega}_{018})$&$ -\frac{1}{2}(\tilde{\omega}_{387} - 
\tilde{\omega}_{078}) $&$  -\frac{1}{2}(\tilde{\omega}_{277} +\tilde{\omega}_{187})$ \\
&&&& \\
\hline
&&&&\\
$II_{L}$  & $ \tilde{A}^{--}_{d} ((03),(12))= $ & $ A^{II}_{d}= $ & $  -\tilde{A}^{-+}_{d} ((12),(78)) = $ &
$  \tilde{A}^{-+}_{d} ((03),(78)) = $ \\
$$&$ \frac{1}{2}(\tilde{\omega}_{327} -\tilde{\omega}_{018})$&$A^{I}_{d} +  (\tilde{\omega}_{127} + \tilde{\omega}_{038})$
&$ -\frac{1}{2}(\tilde{\omega}_{277} -\tilde{\omega}_{187}) $&$  -\frac{1}{2}(\tilde{\omega}_{387} +\tilde{\omega}_{078})$ \\
\hline 
&&&& \\
$III_{L}$ & $  - \tilde{A}^{--}_{d} ((03),(78)) =$ & $ \tilde{A}^{+-}_{d} ((12),(78))= $ & $ A^{III}_{d}=$ &
$   \tilde{A}^{-+}_{d} ((03),(12)) =$ \\
$$&$ -\frac{1}{2}(\tilde{\omega}_{387} -\tilde{\omega}_{078})$&$ -\frac{1}{2}(\tilde{\omega}_{277} -\tilde{\omega}_{187})$
&$A^{I}_{d} +  (\tilde{\omega}_{787} + \tilde{\omega}_{038})$&$ -\frac{1}{2}(\tilde{\omega}_{018} +\tilde{\omega}_{327}) $\\
&&&& \\
\hline 
&&&& \\
$IV_{L}$  & $ -\tilde{A}^{--}_{d} ((12),(78)) =$ & $ \tilde{A}^{+-}_{d} ((03),(78)) = $ & 
$ \tilde{A}^{+-}_{d} ((03),(12))$ & $ A^{IV}_{d} $ \\
$$&$ -\frac{1}{2}(\tilde{\omega}_{277} +\tilde{\omega}_{187})$&$ -\frac{1}{2}(\tilde{\omega}_{387} +\tilde{\omega}_{078})$
&$ -\frac{1}{2}(\tilde{\omega}_{018} +\tilde{\omega}_{327}) $&$A^{I}_{d} +  (\tilde{\omega}_{127} + \tilde{\omega}_{787})$\\
&&&& \\
\hline\hline
\end{tabular}
\end{center}
\caption{\label{abb-snmb-tableV}%
The mass matrix of four families of the $d$-quarks. Comments are the same 
as in Table~\ref{abb-snmb-tableIV}.}
\end{table}

\section{Discussions and conclusions}
\label{ABBconclusions}

In this talk we presented how from our Approach unifying spins and charges the mass matrices for the quarks 
and the leptons can follow. The approach assumes that a Weyl spinor of a chosen handedness carries in 
$d (=1+13)-$ dimensional space nothing but two kinds of spin degrees of freedom. One kind belongs to 
the Poincar\' e group in $d=1+13$, 
another kind generates families. Spinors interact with only the gravitational fields, 
manifested by the vielbeins and  spin connections, the gauge fields of the momentum 
$p_{\alpha}$ and the two kinds of the generators of the Lorentz group $S{ab}$ and $\tilde{S}^{ab}$, 
respectively. To derive mass matrices (that is to calculate the Yukawa couplings, which are 
postulated by the Standard model) in a simple and transparent way, we made several further assumptions,
some of them needed only to simplify the estimations:

i.  The break of symmetry $SO(1,13)$ into $SO(1,7)\times SU(3)\times U(1)$ occurs in a way that only 
massless spinors in $d=1+7$ with the charge $ SU(3)\times U(1)$ survive. (Our work on the compactification 
of a massless spinor in $d=1+5$ into the  $d=1+3$ and a finite disk gives us some hope that this assumption 
might be fulfilled\cite{ABBholgernorma05}.) Further breaks cause that the charge 
$Q= \tau^{41} + S^{56}$ is conserved.

ii. The break of symmetries influences the Poincar\' e symmetry and the symmetries described by 
$\tilde{S^{ab}}$ in a similar way. We also assume that the terms which include $p_s, s=5,5,..,14$ do not 
contribute at low energies.

iii. There are no terms, which would transform  
$\stackrel{\tilde{56}}{(+)}$ into $\stackrel{\tilde{56}}{[+]}$.

iv. We  made estimations on a ''tree level''.  

v. We assume the mass matrices to  be real and symmetric.

Our starting Weyl spinor representation of a chosen handedness manifests, if analysed in terms of the 
subgroups $SO(1,3), SU(3), SU(2)$ and two $U(1)'$s (the sum of the ranks of the subgroups is the rank of 
the group) of the group $SO(1,13)$, the spin and all the charges of one family of quarks and leptons. 

We use our technique\cite{ABBholgernorma02,ABBtechnique03} to present spinor representations in 
a transparent way so that one easily sees how the part of the covariant 
derivative of a spinor in $d=1+13$ manifests in $d=1+3$ as Yukawa couplings of the Standard model.

\section*{Acknowledgments} 
It is a pleasure to thank all the participants of the   workshops entitled "What comes beyond the Standard model", 
taking place  at Bled annually in  July, starting at 1998,  for fruitful discussions, although most of them 
still do not believe either in more then three families of quarks and leptons, or that  our approach 
will at the end show the way beyond the Standards models,

\title{Dark Matter From Encapsulated Atoms}
\author{C.D.~Froggatt${}^{1,2}$ and H.B.~Nielsen${}^{2}$}
\institute{%
${}^{1}$ Department of Physics and Astronomy, Glasgow University, 
Glasgow, Scotland\\
${}^{2}$ The Niels Bohr Institute, Copenhagen, Denmark}

\titlerunning{Dark Matter From Encapsulated Atoms}
\authorrunning{C.D.~Froggatt and H.B.~Nielsen}
\maketitle

\begin{abstract}
We propose that dark matter consists of collections of atoms
encapsulated inside pieces of an alternative vacuum, in which the
Higgs field vacuum expectation value is appreciably smaller than
in the usual vacuum. The alternative vacuum is supposed to have
the same energy density as our own. Apart from this degeneracy of
vacuum phases, we do not introduce any new physics beyond the
Standard Model. The dark matter balls are estimated to have a
radius of order 20 cm and a mass of order $10^{11}$ kg. However
they are very difficult to observe directly, but inside dense 
stars may expand eating up the star and cause huge explosions 
(gamma ray bursts). The ratio of dark matter to ordinary 
baryonic matter is estimated to be of the order of the ratio 
of the binding energy per nucleon in helium to the difference 
between the binding energies per nucleon in heavy nuclei and in 
helium. Thus we predict approximately five times as much dark 
matter as ordinary baryonic matter!
\end{abstract}

\section{Introduction}

Recent ``precision" cosmological measurements agree on a so-called
concordant model (see the Reviews of Astrophysics and Cosmology in
\cite{pdg}), according to which the Universe is flat with
$\Omega$, the ratio of its energy density to the critical density,
being very close to unity. The energy budget of the Universe is
presently dominated by three components: ordinary baryonic matter
($\Omega_{ordinary} \simeq 0.04$), dark matter ($\Omega_{dark}
\simeq 0.23$) and dark energy ($\Omega_{\Lambda} \simeq 0.73$).
The main evidence for the density of ordinary matter comes from
the abundances of the light elements formed in the first three
minutes by big bang nucleosynthesis (BBN). The evidence for the
dark matter density comes from galactic rotation curves, motions
of galaxies within clusters, gravitational lensing and analyses
(e.g.~WMAP \cite{spergel}) of the cosmic microwave background
radiation. The need for a form of dark energy, such as a tiny
cosmological constant $\Lambda$, is provided by the evidence for
an accelerating Universe from observations of type Ia supernovae,
large scale structure and the WMAP data.

In this paper we shall concentrate on the dark matter component.
It must be very stable, with a lifetime greater than $10^{10}$
years. The dark matter density is of a similar order of magnitude
as that of ordinary matter, with a ratio of \begin{equation}
\frac{\Omega_{dark}}{\Omega_{ordinary}} \simeq 6 \end{equation} Also the dark
matter was non-relativistic at the time of the onset of galaxy
formation (i.e.~cold dark matter).

According to folklore, no known elementary particle can account
for all of the dark matter. Many hypothetical particles have been
suggested as candidates for dark matter, of which the most popular
is the lightest supersymmetric particle (LSP): the neutralino. The
stability of the the LSP is imposed by the assumption of R-parity
conservation. The LSP density is predicted to be close to the
critical density for a heavy neutralino \cite{jkg} with mass
$m_{LSP} \sim 100-1000$ GeV, but {\it a priori} it is unrelated to
the density of normal matter.

However we should like to emphasize that the dark matter could in
fact be baryonic, if it were effectively separated from normal
matter at the epoch of BBN. This separation must therefore already
have been operative 1 second after the big bang, when the
temperature was of order 1 MeV. Our basic idea is that dark matter
consists of ``small balls" of an alternative Standard Model vacuum
degenerate with the usual one, containing close-packed nuclei and
electrons and surrounded by domain walls separating the two vacua
\cite{prl}. The baryons are supposed to be kept inside the balls
due to the vacuum expectation value (VEV) of the Weinberg-Salam
Higgs field $<\phi_{WS}>$ being smaller, say by a factor of 2, in
the alternative phase. The quark and lepton masses \begin{equation} m_f = g_f
<\phi_{WS}> \end{equation} are then reduced (by a factor of 2). We use an
additive quark mass dependence approximation for the nucleon mass
\cite{CFDweinberg}: \begin{equation} m_N = m_0 + \sum_{i=1}^3 m_{q_i}, \end{equation} where
the dominant contribution $m_0$ to the nucleon mass arises from
the confinement of the quarks. Then, assuming quark masses in our
phase of order $m_u \sim 5$ MeV and $m_d \sim 8$ MeV, we obtain a
reduction in the nucleon mass in the alternative phase by an
amount $\Delta m_N \sim 10$ MeV. The pion may be considered as a
pseudo-Goldstone boson with a mass squared proportional to the sum
of the masses of its constituent quarks: \begin{equation} M_{\pi}^2 \propto m_u
+ m_d. \end{equation} It follows that the pion mass is also reduced (by a
factor of $\sqrt{2}$) in the alternative phase. The range of the
pion exchange force is thereby increased and so the nuclear
binding energies are larger in the alternative phase, by an amount
comparable to the binding they already have in normal matter. We
conclude it would be energetically favourable for the dark matter
baryons to remain inside balls of the alternative vacuum for
temperatures lower than about 10 MeV. These dark matter nucleons
would be encapsulated by the domain walls, remaining relatively
inert and not disturbing the successful BBN calculations in our
vacuum. We should note that a model for dark matter using an
alternative phase in QCD has been proposed by Oaknin and
Zhitnitsky \cite{OZ}.

\section{Degenerate vacua in the Standard Model}

The existence of another vacuum could be due to some genuinely new
physics, but here we want to consider a scenario, which does not
introduce any new fundamental particles or interactions beyond the
Standard Model. Our main assumption is that the dark energy or
cosmological constant $\Lambda$ is not only fine-tuned to be tiny
for one vacuum but for several, which we have called
\cite{bnp,origin,MPP,bn,fn2} the Multiple Point Principle (MPP).
This entails a fine-tuning of the parameters (coupling constants)
of the Standard Model analogous to the fine-tuning of the
intensive variables temperature and pressure at the triple point
of water, due to the co-existence of the three degenerate phases:
ice water and vapour.

Different vacuum phases can be obtained by having different
amounts of some Bose-Einstein condensate. We are therefore led to
consider a condensate of a bound state of some SM particles.
Indeed, in this connection, we have previously proposed
\cite{itep,portoroz,coral,pascos04} the existence of a new exotic
strongly bound state made out of 6 top quarks and 6 anti-top
quarks. The reason that such a bound state had not been considered
previously is that its binding is based on the collective effect
of attraction between several quarks due to Higgs exchange. In
fact our calculations show that the binding could be so strong
that the bound state is on the verge of becoming tachyonic and
could form a condensate in an alternative vacuum degenerate with
our own. With the added assumption of a third Standard Model
phase, having a Higgs vacuum expectation value of the order of the
Planck scale, we obtained a value of 173 GeV for the top quark
mass \cite{fn2} and even a solution of the hierarchy problem, in
the sense of obtaining a post-diction of the order of magnitude of
the ratio of the weak to the Planck scale
\cite{itep,portoroz,coral,pascos04}. However this third Planck
scale vacuum is irrelevant for our dark matter scenario.

With the existence of just the 2 degenerate vacua domain walls
would have easily formed, separating the different phases of the
vacuum occurring in different regions of space, at high enough
temperature in the early Universe. Since we assume the weak scale
physics of the top quark and Higgs fields is responsible for
producing these bound state condensate walls, their energy scale
will be of order the top quark mass. We note that, unlike walls
resulting from the spontaneous breaking of a discrete symmetry,
there is an asymmetry between the two sides of the the wall. So,
in principle, a wall can readily contract to one side or the other
and disappear.

\section{Formation of dark matter balls in the early Universe}

We now describe our favoured scenario for how the dark matter
balls formed. Let us denote the order parameter field describing
the new bound state which condenses in the alternative phase by
$\phi_{NBS}$. In the early Universe it would fluctuate
statistically mechanically and, as the temperature $T$ fell below
the weak scale, would have become more and more concentrated
around the -- assumed equally deep -- minima of the effective
potential $V_{eff}(\phi_{NBS})$. There was then an effective
symmetry between the vacua, since the vacua had approximately the
same free energy densities. So the two phases would have formed
with comparable volumes separated by domain walls. Eventually the
small asymmetry between their free energy densities would have led
to the dominance of one specific phase inside each horizon region
and, finally, the walls would have contracted away. However it is
a very detailed dynamical question as to how far below the weak
scale the walls would survive. It seems quite possible that they
persisted until the temperature of the Universe fell to around 1
MeV.

We imagine that the disappearance of the walls in our phase --
except for very small balls of the fossil phase -- occurred when
the temperature $T$ was of the order of 1 MeV to 10 MeV. During
this epoch the collection of nucleons in the alternative phase was
favoured by the Boltzmann factor $\exp(-\Delta m_N/T)$. Thus the
nucleons collected more and more strongly into the alternative
phase, leaving relatively few nucleons outside in our phase. We
suppose that a rapid contraction of the alternative phase set in
around a temperature $T \sim 1$ MeV.

Due to the higher density and stronger nuclear binding,
nucleosynthesis occurred first in the alternative phase. Ignoring
Coulomb repulsion, the temperature $T_{NUC}$ at which a given
species of nucleus with nucleon number A is thermodynamically
favoured is given \cite{kolbturner} by:
\begin{equation}
T_{NUC} = \frac{B_A /(A-1)}{\ln(\eta^{-1}) + 1.5 \ln(m_N
/T_{NUC})}.
\end{equation}
Here $B_A$ is the binding energy of the nucleus -- in the phase in
question of course -- $\eta$= $\frac{n_B}{n_{\gamma}}$ is the
ratio of the baryon number density relative to the photon density,
and $m_N$ is the nucleon mass. In our phase, for example, the
temperature for ${^4}$He to be thermodynamically favoured turns
out from this formula to be 0.28 MeV. In the other phase, where
the Higgs field has a lower VEV by a factor of order unity, the
binding energy $ B_A$ is bigger and, with say $\eta \sim 10^{-3}$,
$^4$He could have been produced at $T \sim 1$ MeV.

We assume that the alternative phase continued to collect up any
nucleons from our phase and that, shortly after $^4$He production,
there were essentially no nucleons left in our phase. The rapid
contraction of the balls continued until there were more nucleons
than photons, $\eta > 1$, in the alternative phase and fusion to
heavier nuclei, such as $^{12}$C and $^{56}$Fe, took place, still
with $T \sim 1$ MeV. A chain reaction could then have been
triggered, resulting in the explosive heating of the whole ball as
the $^4$He burnt into heavier nuclei. The excess energy would have
been carried away by nucleons freed from the ball.

At this stage of internal fusion, the balls of the alternative
phase would have been so small that any nucleons in our phase
would no longer be collected into the balls. So the nucleons
released by the internal fusion would stay forever outside the
balls and make up {\it normal matter}. This normal matter then
underwent the usual BBN in our phase.

\section{Prediction of the ratio of dark matter to normal matter}

According to the above internal fusion scenario, the ratio of the
normal matter density to the total matter density is given by: \begin{equation}
\frac{\Omega_{ordinary}}{\Omega_{matter}} = \frac{\mbox{Number of
nucleons released}}{\mbox{Total number of nucleons}} \end{equation} The
fraction of nucleons released from the balls of alternative phase
during the internal fusion can be obtained from a simple energy
conservation argument.

Before the further internal fusion process took place, the main
content of the balls was in the form of ${^4}$He nuclei. Now the
nucleons in a ${^4}$He nucleus have a binding energy of 7.1 MeV in
normal matter in our phase, while a typical ``heavy'' nucleus has
a binding energy of 8.5 MeV for each nucleon \cite{ring}. Let us,
for simplicity, assume that the ratio of these two binding
energies per nucleon is the same in the alternative phase and use
the normal binding energies in our estimate below. Thus we take
the energy released by the fusion of the helium into heavier
nuclei to be 8.5 MeV - 7.1 MeV = 1.4 MeV per nucleon. Now we can
calculate what fraction of the nucleons, counted as {\it a priori}
initially sitting in the heavy nuclei, can be released by this 1.4
MeV per nucleon. Since they were bound inside the nuclei by 8.5
MeV relative to the energy they would have outside, the fraction
released should be (1.4 MeV)/(8.5 MeV) = $0.16_5$ = 1/6. So we
predict that the normal baryonic matter should make up 1/6 of the
total amount of matter, dark as well as normal baryonic. According
to astrophysical fits \cite{spergel}, giving 23\% dark matter and
4\% normal baryonic matter relative to the critical density, the
amount of normal baryonic matter relative to the total matter is
$\frac{4\%}{23\% + 4\%} = 4/27 = 0.15$. This is in remarkable
agreement with our prediction.

\section{Properties of dark matter balls}

The size of the balls depends sensitively on the order of
magnitude assumed for the wall energy density, which we take to be
of the weak scale or about 100 GeV. Let us first consider the
stability condition for these balls. For a ball of radius R, the
wall tension $s$ is given by \begin{equation} s \approx (100\ \mathrm{GeV})^3
\end{equation} which provides a pressure $\frac{s}{R}$ that must be balanced
by the electron pressure. The energy needed to release a nucleon
from the alternative vacuum into our vacuum is approximately 10
MeV. So the maximum value for the electron Fermi level inside the
balls is $\sim 10$ MeV, since otherwise it would pay for electrons
and associated protons to leave the alternative vacuum. Thus the
maximum electron pressure is of order (10 MeV)$^4$.

In order that the pressure from the wall should not quench this
maximal electron pressure, we need to satisfy the stability
condition: \begin{equation} s/R = \frac{(100\ \mathrm{GeV})^3}{R} < (10\
\mathrm{MeV})^4 = 10^{-8}\ \mathrm{GeV}^4. \end{equation} This means the ball
radius must be larger than a critical radius given by: \begin{equation} R
> R_{crit} = 10^{14}\ \mathrm{GeV}^{-1} = 2\ \mathrm{cm}. \end{equation}
If the balls have a radius smaller than $R_{crit}$, they will
implode. These critical size balls have a nucleon number density
of \begin{equation} n_e = (10\ \mathrm{MeV})^3 = \frac{1}{(20\ \mathrm{fm})^3}
\simeq 10^{35}\ \mathrm{cm}^{-3}. \end{equation}  So, with $R_{crit} = 2$ cm,
it contains of order $N_e \simeq 10^{36}$ electrons and
correspondingly of order $N_B \simeq 10^{36}$ baryons, with a mass
of order $M_B \simeq 10^9$ kg.

We estimate the typical radius of a dark matter ball in our
scenario to be of order 20 cm. It contains of order $N_B = 3\times
10^{37}$ baryons and has a mass of order $M_B = 10^{11}$ kg =
$10^{-19}M_{\odot} = 10^{-14}M_{\oplus}$. Therefore dark matter
balls can not be revealed by microlensing searches, which are only
sensitive to massive astrophysical compact objects with masses
greater than $10^{-7}M_{\odot}$ \cite{afonso}. Since the dark
matter density is 23\% of the critical density
$\rho_{\mathrm{crit}} = 10^{-26}$ kg/m$^3$, a volume of about
$10^{37}$ m$^3$ = (20 astronomical units)$^3$ will contain on the
average just one dark matter ball.

Assuming the sun moves with a velocity of 100 km/s relative to the
dark matter and an enhanced density of dark matter in the galaxy
of order $10^5$ higher than the average, the sun would hit of
order $10^8$ dark matter balls of total mass $10^{19}$ kg in the
lifetime of the Universe. A dark matter ball passing through the
sun would plough through a mass of sun material similar to its own
mass. It could therefore easily become bound into an orbit say or
possibly captured inside the sun, but be undetectable from the
earth. On the other hand, heavy stars may capture some dark matter
balls impinging on them.

In the lifetime of the Universe, the earth would hit $10^4$ or so
dark matter balls. However they would have gone through the earth
without getting stopped appreciably. It follows that DAMA
\cite{dama} would not have any chance of seeing our dark matter
balls, despite their claim to have detected a signal for dark
matter in the galactic halo. However EDELWEISS \cite{edelweiss},
CRESST \cite{cresst} and CDMS \cite{cdms} do not confirm the
effect seen by DAMA. It is also possible that DAMA saw something
other than dark matter. Geophysical evidence for the dark matter
balls having passed through the earth would also be extremely
difficult to find.

We conclude that the dark matter balls are very hard to see
directly. On the other hand, we could imagine that dark matter
balls had collected into the interior of a collapsing star. Then,
when the density in the interior of the star gets sufficiently
big, the balls could be so much disturbed that they would explode.
The walls may then start expanding into the dense material in the
star, converting part of the star to dark matter. As the wall
expands the pressure from the surface tension diminishes and lower
and lower stellar density will be sufficient for the wall to be
driven further out through the star material. This could lead to
releasing energy of the order of 10 MeV per nucleon in the star,
which corresponds to of the order of one percent of the Einstein
energy of the star! Such events would give rise to really huge
energy releases, perhaps causing supernovae to explode and
producing the canonballs suggested by Dar and De Rujula
\cite{deRujula} to be responsible for the cosmic gamma ray bursts.
We should note that a different (SUSY) phase transition inside the
star has already been suggested \cite{clavelli} as an explanation
for gamma ray bursts.

A dark matter ball can also explode due to the implosion of its
wall. Such an implosive instability might provide a mechanism for
producing ultra high energy cosmic rays from seemingly empty
places in the Universe. This could help to resolve the
Greisen-Zatsepin-Kuzmin \cite{G,ZK} cut-off problem.

\section{Conclusion}

Under the assumption that there be at least two different phases
of the vacuum with very closely the same tiny energy density or
cosmological constant, we have put forward an idea for what dark
matter could be. Indeed we suggest that dark matter consists of
baryons hidden inside pieces of an alternative vacuum with a
smaller Higgs field VEV. The SM might provide such a second vacuum
degenerate with our own, due to the condensation of an exotic $6t
+ 6\overline{t}$ strongly bound state. The ratio of dark matter to
ordinary matter is expressed as a ratio of nuclear binding
energies and predicted to be about 5. Big bang nucleosynthesis is
supposed to proceed as usual in our vacuum relatively undisturbed
by the crypto-baryonic dark matter encapsulated in a few balls of
the alternative vacuum.

We estimate that a typical dark matter ball has a radius of about
20 cm and a mass of order $10^{11}$ kg. The dark matter balls are
very difficult to detect directly, but they might be responsible
for gamma ray bursts or ultra high energy cosmic rays.

\section*{Acknowledgements}

We acknowledge discussions with D.~Bennett and Y.~Nagatani in the
early stages of this work. CDF would like to acknowledge the hospitality
of the Niels Bohr Institute and support from the Niels Bohr Institute
Fund and PPARC.

\title{Dirac Sea for Bosons Also and SUSY for Single Particles\thanks{%
The authors have sent this paper too late to be included in the
Proceedings to the 7th Workshop 'What Comes Beyond the Standard Models', Bled,
19. -- 31. July 2004. The editors therefore desided to include the
improved version in the Proceedings for 2005 (this volume).
This contribution is also available as preprints
YITP-05-23; OIQP-05-06; hep-th/0505238.}}
\author{Y. Habara${}^1$\footnote{%
Adress after 1 April 2005, Department of %
Physics, Kyoto University, Kyoto 606-8502, Japan.}, H.B. Nielsen${}^2$ and 
M. Ninomiya${}^3$\footnote{Also working at Okayama Institute for Quantum Physics.}}
\institute{%
${}^1$ Department of Physics, Osaka University, Osaka 560-0043, Japan\\
${}^2$ Niels Bohr Institute, University Copenhagen, 17 Blegdamsvej Copenhagen \o, Denmark\\
${}^3$ Yukawa Institute for Theoretical Physics, Kyoto University,Kyoto 606-8502, Japan}

\titlerunning{Dirac Sea for Bosons Also and SUSY for Single Particles}
\authorrunning{Y. Habara, H.B. Nielsen and M. Ninomiya}
\maketitle

\begin{abstract}
  We consider the long standing problem in field theories of bosons that the
  boson vacuum does not consist of a `sea', unlike the fermion vacuum. We show
  with the help of supersymmetry considerations that the boson vacuum indeed
  does also consist of a sea in which the negative energy states are all
  ``filled", analogous to the Dirac sea of the fermion vacuum, and that a hole
  produced by the annihilation of one negative energy boson is an
  anti-particle. This might be formally coped with by introducing the notion
  of a double harmonic oscillator, which is obtained by extending the
  condition imposed on the wave function. Next, we present an attempt to
  formulate the supersymmetric and relativistic quantum mechanics utilizing
  the equations of motion.
\end{abstract}

\section{Introduction}

In the Quantum Field Theory there is a long-standing historical mystery why
there is no negative energy sea for bosons, contrary to fermions when we pass
from 1st quantized theory into 2nd quantized one. Needless to say nowadays one
uses a well functioning method that in the negative energy states creation
operator and destruction operators should be formally exchanged. This
rewriting can be used both bosons and fermions. In this formal procedure, as
for fermions, the true vacuum is the one that the negative energy states are
completely filled for one particle in each state due to Pauli principle. By
filling all empty negative states to form Dirac sea \cite{dirac} we define
creation operators $\tilde{d}^+(\vec{p},s,w)$ with positive energy $w$ for
holes which is equivalent to destruction operators $d(-\vec{p},-s,-w)$. In
particular for boson case the associated filling of the negative energy state
to form a sea as the true vacuum has never been heard.\footnote{The
  interesting historical description can be found in ref.\cite{weinberg}}

In this report one of the main new contents present a new method of 2nd
quantization of boson field theories in the analogous manner to filling of
empty Dirac sea for fermions.

At the very end when the true vacuum in the 2nd quantized boson theory is
formed according to our method\cite{nn,hnn} we will come exactly the same
theory as the usual one. Thus our approach cannot be incorrect as far of
quantum field theories are concerned. However application to the string
theories may be very interesting open question. Although we have to introduce
such an unfamiliar notion to form negative energy sea (Dirac) sea for boson
that we have to subtract ``minus one boson'' or create ``a boson hole" in
boson sea.

Validity and importance of our new method to 2nd quantize boson theories may
become more clear by considering supersymmetric field theories. Due to
supersymmetry one should expect the boson vacuum structure similar to the
fermion one. We in fact utilize the N=2 matter multiplet called a
hypermultiplet\cite{sohnius} and construct the Noether current from the
supersymmetric action. By requiring that the entire system be supersymmetric,
we derive the properties of the boson vacuum, while the fermion vacuum is
taken to be the Dirac sea.

In the supersymmetric theories there seems to exist no truly quantum
mechanical relativistic, i.e. 1 particle model so that one starts from the
field theories. We propose in the present article a method to construct
sypersymmetric quantum mechanics, although we have not yet completed the
procedure\footnote{In ref.~\cite{hnn2} we proposed the complete formulation of
the supersymmetric quantum mechanics by utilizing the 8-component matrix
notation.}.

The present article is organized as follows: In the next section 2 we utilize
the N=2 supersymmetry to further clarify the vacuum structures of boson as
well as fermions; both vacuua should be expected to be very similar by the
supersymmetry. We utilize N=2 matter multiplet called hypermultiplet and
explicitly show that the boson true vacuum can be formulated so as to form the
negative energy sea, i.e. Dirac sea. In section 3 we start with describing the
harmonic oscillator with the requirement of analyticity of the wave function
instead of the usual square integrability condition. This naturally leads to
the boson negative energy (Dirac sea) states as well as the usual positive
ones. We then argue how to treat the Dirac sea for bosons. In the next section
4 we turn our attention to the 1 particle supersymmetric theory which may be
considered 1st excited states in the boson vacuum, i.e. filled Dirac seas for
bosons as well as fermions in our formalism. This theory is viewed as
supersymmetric relativistic quantum mechanics. We propose a prescription how
to realize the supersymmetry in a single particle level. The final section 5
is devoted to the conclusions.

\section{Boson and Dirac Seas in a Hypermultiplet Model with $N=2$ Supersymmetry}

In the present section, we consider an ideal world in which supersymmetry
holds exactly. Then, it is natural to believe that, in analogy to the true
fermion vacuum, the true boson vacuum is a state in which all negative energy
states are occupied. To investigate the details of the vacuum structure of
bosons, we utilize the $N=2$ matter multiplet called a
hypermultiplet~\cite{sohnius,west}. In fact, we construct the Noether current
from the supersymmetric action, and by requiring that the entire system be
supersymmetric, we derive the properties of the boson vacuum, while the
fermion vacuum is taken to be the Dirac sea.

Hereafter, the Greek indices $\mu ,\nu ,\cdots$ are understood to run from 0
to 3, corresponding to the Minkowski space, and the metric is given by
$$\eta^{\mu \nu}=diag(+1,-1,-1,-1).$$

\subsection{$N=2$ matter multiplet: Hypermultiplet}

Let us summarize the necessary part of $N=2$ supersymmetric field theory in
the free case, in order to reveal the difficulty of making supersymmetric
description. The hypermultiplet is the simplest multiplet that is
supersymmetric and involves a Dirac fermion. It is written

\begin{align*}
        \phi =(A_1, A_2;\> \psi ;\> F_1, F_2), \tag{2.1}
\end{align*}

\noindent where $A_i$ and $F_i\> (i=1,2)$ denote complex scalar fields, and the Dirac field is given by $\psi$. The multiplet (2.1) transforms under a supersymmetric transformation as 

\begin{align*}
  & \delta_{\xi}A_i=2\bar{\xi}_i\psi , \\
  & \delta_{\xi}\psi =-i\xi_iF_i-i\gamma^{\mu}\partial_{\mu}\xi_iA_i, \\
  & \delta_{\xi}F_i=2\bar{\xi_i}\gamma^{\mu}\partial_{\mu}\psi ,\tag{2.2}
\end{align*}

\noindent where $\gamma^{\mu}$ denotes the four-dimensional gamma matrices, with $\{ \gamma^{\mu},\gamma^{\nu}\} =2\eta^{\mu \nu}$. Then, we obtain the Lagrangian density of the hypermultiplet, 

\begin{align}
        \mathcal{L} & =\frac{1}{2}\partial_{\mu}A_i^{\dagger}\partial^{\mu}
        A_i+\frac{1}{2}F_i^{\dagger}F_i+\frac{i}{2}\bar{\psi}\gamma^{\mu}
        \partial_{\mu}\psi -\frac{i}{2}\partial_{\mu}\bar{\psi}\gamma^{\mu}
        \psi +m\left[\frac{i}{2}A_i^{\dagger}
        F_i-\frac{i}{2}F_i^{\dagger}A_i+\bar{\psi}\psi \right]. \tag{2.4}
\end{align}

To derive the Noether currents whose charges generate the supersymmetry transformation, we consider a variation under the supersymmetry transformation (2.2), 

\begin{align*}
        \delta_{\xi}\mathcal{L}=\bar{\xi}_i\partial_{\mu}K_i^{\mu}
        +\partial_{\mu}\bar{K}_i^{\mu}\xi_i, \tag{2.5}
\end{align*}

\noindent where $K_i^{\mu}$ is given by 

\begin{align*}
        K_i^{\mu}\equiv \frac{1}{2}(\gamma^{\mu}\gamma^{\nu}\psi 
        \partial_{\nu}A_i^{\dagger}+im\gamma^{\mu}\psi A_i^{\dagger}). 
        \tag{2.6}
\end{align*}

\noindent Thus, the Noether current $J_i^{\mu}$ is written as 

\begin{align*}
        & \bar{\xi}_iJ_i^{\mu}+\bar{J}_i^{\mu}\xi_i=
        \frac{\delta L}{\delta_{\xi}(\partial_{\mu}\phi)}\delta_{\xi}\phi 
        -\left(\bar{\xi}_iK_i^{\mu}+\bar{K}_i^{\mu}\xi_i\right) \\
        & \quad \Longrightarrow J_i^{\mu}=\gamma^{\nu}\gamma^{\mu}\psi 
        \partial_{\nu}A_i^{\dagger}-im\gamma^{\mu}\psi A_i^{\dagger}. \tag{2.7}
\end{align*}

We could attempt to think of treating the bosons analogous to the fermions by imagining that the creation operators $a^{\dagger}(\vec{k})$ of the anti-bosons were really annihilation operators in some other formulation, but, as we shall see, such an attempt leads to some difficulties. If we could indeed do so, we would write also a boson field in terms of only annihilation operators formally as follows 

\begin{align*}
        & A_i(x)=\int \frac{d^3\vec{k}}{\sqrt{(2\pi )^32k_0}}\left\{
        a_{i+}(\vec{k})e^{-ikx}+a_{i-}(\vec{k})e^{ikx}\right\}, \tag{2.8} \\
        & \psi(x)=\int \frac{d^3\vec{k}}{\sqrt{(2\pi )^32k_0}}\sum_{s=\pm}
        \left\{b(\vec{k},s)u(\vec{k},s)e^{-ikx}+d(\vec{k},s)v(\vec{k},s)e^{ikx}
        \right\}. \tag{2.9}
\end{align*}

\noindent Here, $k_0\equiv \sqrt{\vec{k}^2+m^2}$ is the energy of the particle, and $s\equiv \frac{\vec{\sigma}\cdot \vec{k}}{|\vec{k}|}$ denotes the helicity. Particles with positive and negative energy are described by $a_{i+}(\vec{k}), b(\vec{k},s)$ and $a_{i-}(\vec{k}), d(\vec{k},s)$, respectively. The commutation relations between these field modes are derived as 

\begin{align*}
        & \left[a_{i+}(\vec{k}),a_{j+}^{\dagger}(\vec{k}^{\prime})\right]=
        +\delta_{ij}\delta^3(\vec{k}-\vec{k}^{\prime}), 
        \quad \left[a_{i-}(\vec{k}),a_{j-}^{\dagger}(\vec{k}^{\prime})\right]=
        -\delta_{ij}\delta^3(\vec{k}-\vec{k}^{\prime}), \tag{2.10} \\
        & \left\{b(\vec{k},s),b^{\dagger}(\vec{k}^{\prime},s^{\prime})\right\}
        =+\delta_{ss^{\prime}}\delta^3(\vec{k}-\vec{k}^{\prime}), 
        \quad \left\{d(\vec{k},s),d^{\dagger}(\vec{k}^{\prime},s^{\prime})
        \right\}=+\delta_{ss^{\prime}}\delta^3(\vec{k}-\vec{k}^{\prime}), 
        \tag{2.11}
\end{align*}

\noindent with all other pairs commuting or anti-commuting. Note that the right-hand side of the commutation relation (2.10) for negative energy bosons, has the opposite sign of that for positive energy bosons. In the ordinary method, recalling that the Dirac sea is the true fermion vacuum, we can use $d^{\dagger}$ as the creation operator and $d$ as the annihilation operator for negative energy fermions. Then, the operators $d^{\dagger}$ and $d$ are re-interpreted as the annihilation operator and creation operator for positive energy holes. In this manner, we obtain the particle picture in the real world. In this procedure, negative energy fermions are regarded as actually existing entities. 

For bosons, in contrast to the fermions, we rewrite the second equation of (2.10) as 

\begin{align*}
        & \> a_{i-}\equiv \tilde{a}_{i-}^{\dagger}, \quad a_{i-}^{\dagger}
        \equiv \tilde{a}_{i-}, \\
        & \left[\tilde{a}_{i-}(\vec{k}),\tilde{a}_{j-}^{\dagger}
        (\vec{k}^{\prime})\right]=+\delta_{ij}\delta^3
        (\vec{k}-\vec{k}^{\prime}). \tag{2.12}
\end{align*}

\noindent This implies that we can treat negative energy bosons in the same manner as positive energy bosons. Consequently, the true vacua for positive and negative energy bosons, which are denoted $||0_+\rangle$ and $||0_-\rangle$, respectively\footnotemark, are given by 

\footnotetext{In the following, we denote the vacua by, for example in the boson case, $|0_{\pm}\rangle$ in the system of single particle, and $||0_{\pm}\rangle$ in the system with many particles.}

\begin{align*}
        & a_{i+}||0_+\rangle =0, \\
        & \tilde{a}_{i-}||0_-\rangle =0. \tag{2.13}
\end{align*}

\noindent Thus, in the true vacuum, meaning the one on which our experimental world is built, both the negative and positive energy vacua are empty when using the particle $a_{i+}$ and anti-particle $\tilde{a}_{i-}$ annihilation operators respectively. However, in order to have a supersymmetry relation to the analogous negative energy states for the fermions, we would have liked to consider, instead of $||0_-\rangle$, a vacuum so that it were empty with respect to the negative energy bosons described by $a_{i-}$ and $a_{i-}^{\dagger}$. That is to say we would have liked a empty vacuum obeying $a_{i-}||0_{\text{wanted}}\rangle =0$. Because of (2.13) it is, however, immediately seen that this $||0_{\text{wanted}}\rangle$ cannot exist. In true nature, we should rather be in a situation or a ``sector" in which we have a state with $a_{i-}^{\dagger}||0_-\rangle =\tilde{a}_{i-}||0_-\rangle =0$. It could be called a ``sector with a top" $||0_-\rangle$.

Perhaps the nicest way of describing this extension is by means of the double harmonic oscillator to be presented in Section 3 below, but let us stress that all we need is a formal extrapolation to also include the possibility of negative numbers of bosons.

\subsection{Supersymmetry invariant vacuum}

As described in Subsection 2.1, when considered in terms of supersymmetry, there is a difference between the boson and fermion pictures. In the present subsection, we give preliminary considerations to the problem determining the nature of a boson sea that would correspond to the Dirac sea for the fermion case. To this end, we impose the natural condition within the supersymmetric theory that the vacuum be supersymmetry invariant. 

We first rewrite the supersymmetry charges $\mathcal{Q}_i$ derived from the supersymmetry currents described by Eq.(2.7) in terms of the creation and annihilation operators as 

\begin{align*}
        \mathcal{Q}_i & =\int d^3\vec{x} J_i^0(x)=i\int d^3\vec{k} 
        \sum_{s=\pm} \left\{b(\vec{k},s)u(\vec{k},s)
        a_{i+}^{\dagger}(\vec{k})-d(\vec{k},s)v(\vec{k},s)a_{i-}^{\dagger}
        (\vec{k})\right\}, \\
        \bar{\mathcal{Q}}_i & =\int d^3\vec{x} \bar{J}_i^0(x)=-i\int 
        d^3\vec{k} \sum_{s=\pm} \left\{b^{\dagger}(\vec{k},s)
        \bar{u}(\vec{k},s)a_{i+}(\vec{k})-d^{\dagger}(\vec{k},s)\bar{v}
        (\vec{k},s)a_{i-}(\vec{k})\right\}. \tag{2.14}
\end{align*}

\noindent By applying these charges, the condition for the vacuum to be supersymmetric can be written 

\begin{align*}
        \mathcal{Q}_i||0\rangle=\bar{\mathcal{Q}}_i||0\rangle=0. \tag{2.15}
\end{align*}

\noindent We then decompose the total vacuum into the boson and fermion vacua, $||0_{\pm}\rangle$ and $||\tilde{0}_{\pm}\rangle$, writing 

\begin{align*}
        ||0\rangle \equiv ||0_+\rangle \otimes ||0_-\rangle \otimes 
        ||\tilde{0}_+\rangle \otimes ||\tilde{0}_-\rangle , \tag{2.16}
\end{align*}

\noindent where $\otimes$ denotes the direct product, and $||\tilde{0}_-\rangle$ is the Dirac sea, given by 

\begin{align*}
        ||\tilde{0}_-\rangle =\bigg\{\prod_{\vec{p},s}d^{\dagger}
        (\vec{p},s)\bigg\}||\tilde{0}\rangle .
\end{align*}

\noindent Here, $||\tilde{0}_+\rangle$ represents an empty vacuum, annihilated by the ordinary $b$ operator, while $||\tilde{0}_-\rangle$, given by Eq.(23), represents the Dirac sea, which is obtained through application of all $d^{\dagger}$. The condition for the bosonic vacuum reads 

\begin{align*}
        & a_{i+}(\vec{k})||0_+\rangle =0,\quad a_{i-}^{\dagger}(\vec{k})
        ||0_-\rangle =0. \tag{2.17}
\end{align*}

\noindent It is evident that the vacuum of the positive energy boson $||0_+\rangle$ is the empty one, vanishing under the annihilation operator $a_{i+}$. On the other hand, the vacuum of the negative energy boson $||0_-\rangle$ is defined such that it vanishes under the operator $a_{i-}^{\dagger}$ that creates the negative energy quantum. This may seem very strange. One could call the strange ``algebra" looked for a ``sector with top", contrary to the more usual creation and annihilation systems which could rather be called ``sectors with a bottom".

In the next section, using the fact that the algebras (2.10) constitute that is essentially a harmonic oscillator system with infinitely many degrees of freedom, we investigate in detail the vacuum structure by considering the simplest one-dimensional harmonic oscillator system. In fact, we will find the explicit form of the vacuum $||0_-\rangle$ that is given by a coherent state of the excited states of all the negative energy bosons.

\section{Negative Energy (Dirac-like) Sea for Bosons}

When looking for solutions to the Klein-Gordon equation for energy (and momentum) it is well-known that, we must consider not only the positive energy particles but also the negative energy ones. In the previous section, we found that in order to implement the analogy to the Dirac sea for fermions suggested by supersymmetry, we would have liked to have at our disposal the possibility to organize an analogon of the Dirac sea (for fermions). In the present section we introduce the concept of a ``sector with top" as an extension of the harmonic oscillator spectrum to a negative energy sector. Thereby we have to extend the ordinary meaning of the wave function (in this case for the harmonic oscillator). Performing this we find that the vacuum of the negative energy sector leads to a ``boson sea", corresponding to the Dirac sea of fermions.

\subsection{Analytic wave function and double harmonic oscillator}

As is well known, the eigenfunction $\phi (x)$ of a one-dimensional 
Schr\"{o}\-dinger equation in the usual treatment should satisfy the square integrability condition, 

\begin{align*}
        \int_{-\infty}^{+\infty}dx \> |\phi (x)|^2<+\infty . \tag{3.1}
\end{align*}

\noindent If we apply this condition to a one-dimensional harmonic oscillator, we obtain as the vacuum solution only the empty one satisfying 

\begin{align*}
        a_+|0\rangle =0.
\end{align*}

\noindent Thus, we are forced to extend the condition for physically allowed wave functions in order to obtain ``boson sea" analogous to the Dirac sea. In fact we extend the condition (3.1), replacing it by the condition under which, when we analytically continuate $x$ to the entire complex plane, the wave function $\phi (x)$ is analytic and only an essential singularity is allowed as $|x|\! \to \! \infty$. In fact, for the harmonic oscillator, we can prove the following theorem: 

\vspace{0.5cm}

\noindent i) The eigenvalue spectrum $E$ for the harmonic oscillator 

\begin{align*}
        \left( -\frac{1}{2}\frac{d^2}{dx^2}+\frac{1}{2}x^2 \right) 
        \phi (x)=E\phi (x). \tag{3.2}
\end{align*}

\noindent is given by 

\begin{align*}
        E=\pm \left(n+\frac{1}{2}\right), \quad n\in \mathcal{Z}_+\cup \{0\}. 
        \tag{3.3}
\end{align*}

\noindent ii) The wave functions for positive energy states are the usual ones 

\begin{align*}
        & \phi_n(x)=A_nH_n(x)e^{-\frac{1}{2}x^2}, \nonumber \\
        & E=n+\frac{1}{2},\qquad n=0_+,1,2,\cdots . \tag{3.4}
\end{align*}

\noindent Here $H_n(x)$ is the Hermite polynomial while $A_n=\left( \sqrt{\pi}2^nn!\right)^{-\frac{1}{2}}$. For negative energy states, the eigenfunctions are given by 

\begin{align*}
        & \phi_{-n}(x)=A_nH_n(ix)e^{+\frac{1}{2}x^2}, \\
        & E=-\left( n+\frac{1}{2}\right) ,\qquad n=0_-,1,2,\cdots . \tag{3.5}
\end{align*}

\noindent iii) The inner product is defined as 

\begin{align*}
        \langle n|m\rangle =\int_{\Gamma}dx \> \phi_n(x^{\ast})^{\ast}
        \phi_m(x), \tag{3.6}
\end{align*}

\noindent where the contour is denoted by $\Gamma$. The $\Gamma$ should be chosen so that the integrand should go down to zero at $x=\infty$, but there remains some ambiguity in the choice of $\Gamma$. However if one chooses the same $\Gamma$ for all negative $n$ states, the norm of these states have an alternating sign.

\vspace{0.5cm}

The above i)-iii) constitute the theorem. Proof of this theorem is rather trivial, and we skip it by referring the refs.~\cite{nn,hnn}, but some comments are given in the following.

Going from (3.4) to (3.5) corresponds to the replacement $x\! \to \! ix$, so the creation and annihilation operators are transformed as 

\begin{align*}
        [a_+,a_+^{\dagger}]=+1 & \to [a_-,a_-^{\dagger}]=-1.
\end{align*}

\noindent Here, $a_+$ and $a_+^{\dagger}$ are ordinary operators in the positive energy sector, and $a_-$ and $a_-^{\dagger}$ are operators in the negative energy sector which create and annihilate negative energy quanta respectively.

It is useful to summarize the various results obtained to this point in operator form. We write each vacuum and excited state in the positive and negative energy sectors, respectively, as 

\begin{align*}
        & \phi_{+0}(x)=e^{-\frac{1}{2}x^2}\simeq |0_+\rangle , \tag{3.7} \\
        & \phi_{-0}(x)=e^{+\frac{1}{2}x^2}\simeq |0_-\rangle , \tag{3.8} \\
        & \phi_n(x)\simeq |n\rangle , \qquad n\in \mathcal{Z}-\{0\}. \tag{3.9}
\end{align*}

The important point here is that there exists a gap between the positive and negative sectors. Suppose that we write the states in order of their energies as 

\begin{figure}[ht]
        \begin{center}
        \begin{picture}(360,40)
        \put(40,20){$\cdots$}
        \put(65,23){$\xrightarrow[]{a^{\dagger}}$}
        \put(65,14){$\xleftarrow[\> a \>]{}$}
        \put(85,18){$|\! -\! 1\rangle$}
        \put(115,23){$\xrightarrow[]{a^{\dagger}}$}
        \put(115,14){$\xleftarrow[\> a \>]{}$}
        \put(135,18){$|0_-\rangle$}
        \put(160,23){$\xrightarrow[]{a^{\dagger}}$}
        \put(180,18){$0$}
        \put(190,14){$\xleftarrow[\> a \>]{}$}
        \put(210,18){$|0_+\rangle$}
        \put(235,23){$\xrightarrow[]{a^{\dagger}}$}
        \put(235,14){$\xleftarrow[\> a \>]{}$}
        \put(255,18){$|\! +\! 1\rangle$}
        \put(285,23){$\xrightarrow[]{a^{\dagger}}$}
        \put(285,14){$\xleftarrow[\> a \>]{}$}
        \put(310,20){$\cdots$.}
        \end{picture}
        \end{center}
        \label{sequence}
\end{figure}%

\noindent As usual, the operators causing transitions in the right and left directions are $a_{\pm}^{\dagger}$ and $a_{\pm}$, respectively. However, between the two vacua $|0_-\rangle$ and $|0_+\rangle$ there is a ``wall" of the classical number $0$, and due to its presence, these two vacua cannot be transformed into each other under the operations of $a_{\pm}$ and $a_{\pm}^{\dagger}$. In going to the second quantized theory with interactions, there appears to be the possibility of such a transition. However, it turns out that the usual polynomial interactions do not induce such a transition.

Next, we comment on the definition of the inner product of states. As explained above, there exists a gap such that no transition between the two sectors can take place. Thus, we can define the inner product of only states in the same sector. The inner product that in the positive energy sector provides the normalization condition is, as usual, given by 

\begin{align}
        \langle n|m\rangle & \equiv \int_{-\infty}^{+\infty}dx\> 
        \phi_n^{\dagger}(x)\phi_m(x)=\delta_{nm}, \qquad n,m=0_+,1,2,\cdots . 
        \tag{3.10}
\end{align}

\noindent However, the eigenfunctions in the negative energy sector are obtained as Eq.(3.5), so we propose  a path of integration such that the integration is convergent, since we impose the condition that the wave functions are analytic. Then, we define the inner product in terms of a path $\Gamma$, which we make explicit subsequently: 

\begin{align}
        \langle n|m\rangle \equiv -i\int_{\Gamma}dx\> \phi_n^{\ast}(x^{\ast})
        \phi_m(x)=(-1)^n\delta_{nm}, \qquad n,m=0_-,-1,-2,\cdots \tag{3.11}
\end{align}

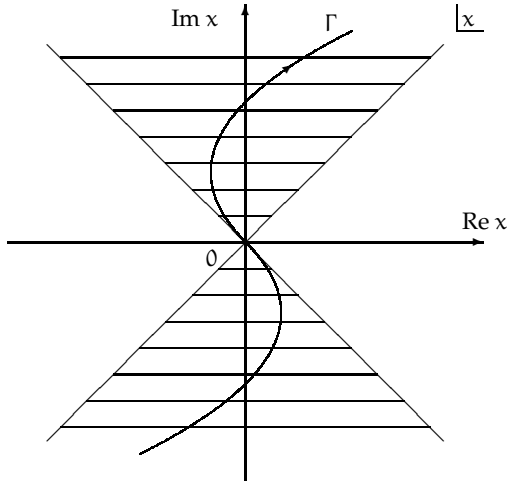
\begin{figure}[ht]
        \begin{center}
        \begin{picture}(180,180)
        \put(0,90){\vector(1,0){180}}
        \put(90,0){\vector(0,1){180}}
        \put(170,170){\line(1,0){10}}
        \put(170,170){\line(0,1){10}}
        \put(172,172){$x$}
        \put(75,81){$0$}
        \put(172,95){Re$\> x$}
        \put(62,172){Im$\> x$}
        \put(15,15){\line(1,1){150}}
        \put(15,165){\line(1,-1){150}}
        \put(80,100){\line(1,0){20}}
        \put(70,110){\line(1,0){40}}
        \put(60,120){\line(1,0){60}}
        \put(50,130){\line(1,0){80}}
        \put(40,140){\line(1,0){100}}
        \put(30,150){\line(1,0){120}}
        \put(20,160){\line(1,0){140}}
        \put(80,80){\line(1,0){20}}
        \put(70,70){\line(1,0){40}}
        \put(60,60){\line(1,0){60}}
        \put(50,50){\line(1,0){80}}
        \put(40,40){\line(1,0){100}}
        \put(30,30){\line(1,0){120}}
        \put(20,20){\line(1,0){140}}
        \qbezier(90,90)(50,130)(130,170)
        \qbezier(90,90)(130,50)(50,10)
        \put(120,170){$\Gamma$}
        \put(98,150){\vector(4,3){10}}
        \end{picture}
        \end{center}
        \caption[path $\Gamma$]{Path $\Gamma$ for which the inner product 
        (3.11) converges.}
        \label{gamma}
\end{figure}%

\noindent Here, it is understood that the complex conjugation yielding $\phi^{\ast}(x^{\ast})$ is taken so that the inner product is invariant under deformations of the path $\Gamma$ within the same topological class in the lined regions as shown in Fig.~\ref{gamma}.

Therefore, the definition of the inner product in the negative energy sector is not essentially different from that of the positive energy sector, except for the result of the alternating signature $(-1)^{n}\delta_{nm}$, which is, however, very crucial. On the other hand, if we adopt $\phi^{\ast}(x)$ instead of $\phi^{\ast}(x^{\ast})$ in (3.11), we can also obtain the positive definite inner product $\langle n|m\rangle =\delta_{nm}$ for the negative energy sector at the sacrifice of the path-independence of $\Gamma$.

\subsection{Boson vacuum in the negative energy sector}

The vacua $|0_+\rangle$ and $|0_-\rangle$ in the positive and negative energy sectors are 

\begin{align*}
        & |0_+\rangle \simeq e^{-\frac{1}{2}x^2}, \tag{3.12} \\
        & |0_-\rangle \simeq e^{+\frac{1}{2}x^2}. \tag{3.13}
\end{align*}

\noindent In order to demonstrate how $|0_-\rangle$ represents a sea, we derive a relation between the two vacua (3.12) and (3.13) analogous to that in the fermion case. In fact, by comparing the explicit functional forms of each vacuum, we easily find the relation 

\begin{align*}
        & e^{+\frac{1}{2}x^2}=e^{x^2}\cdot e^{-\frac{1}{2}x^2}, \qquad 
        e^{x^2}=e^{\frac{1}{2}(a+a^{\dagger})^2} \nonumber \\
        & \Longrightarrow 
        |0_-\rangle =e^{\frac{1}{2}(a+a^{\dagger})^2}|0_+\rangle . \tag{3.14}
\end{align*}

\noindent This relation is preferable for bosons for the following reason. In the fermion case, due to the exclusion principle, the Dirac sea is obtained by exciting only one quantum of the empty vacuum. Contrastingly, because in the boson case there is no exclusion principle, the vacuum $|0_-\rangle$ in the negative energy sector is constructed as a sea by exciting all even number of quanta, i.e. an infinite number of quanta. 

\subsection{Boson sea}

In the present subsection, we investigate the boson vacuum structure in detail, utilizing the second quantized theory for a complex scalar field. Firstly, we clarify the properties of the unfamiliar vacuum $||0_-\rangle$ in the negative energy sector, using the result of Subsection 3.1. To this end, we study the details of the infinite-dimen\-sional harmonic oscillator, which is identical to a system of a second quantized complex scalar field. The representation of the algebra (2.10) that is formed by $a_+,a_-$ and their conjugate operators is expressed as 

\begin{align*}
        & a_+(\vec{k})=\left( A(\vec{k})+\frac{\delta}{\delta A(\vec{k})}
        \right), \quad a_+^{\dagger}(\vec{k})=\left( A(\vec{k})
        -\frac{\delta}{\delta A(\vec{k})}\right), \tag{3.15} \\
        & a_-(\vec{k})=i\left( A(\vec{k})+\frac{\delta}{\delta A(\vec{k})}
        \right),\quad a_-^{\dagger}(\vec{k})=i\left( A(\vec{k})
        -\frac{\delta}{\delta A(\vec{k})}\right). \tag{3.16}
\end{align*}

\noindent The Hamiltonian and Schr\"{o}dinger equation of this system as the infinite-dimen\-sional harmonic oscillator read 

\begin{align*}
        & H=\int \frac{d^3\vec{k}}{(2\pi )^3}\left\{ -\frac{1}{2}
        \frac{\delta^2}{\delta A^2(\vec{k})}+\frac{1}{2}A^2(\vec{k})\right\} 
        , \tag{3.17} \\
        & H\Phi [A]=E\Phi [A]. \tag{3.18}
\end{align*}

\noindent Here, $\Phi [A]$ denotes a wave functional of the wave function $A(\vec{k})$. We are now able to write an explicit wave functional for the vacua of the positive and negative enegy sectors: 

\begin{align*}
        & ||0_+\rangle \simeq \Phi_{0_+}[A]=e^{-\! \frac{1}{2} \int \! 
        \frac{d^3\vec{k}}{(2\pi )^3}A^2(\vec{k})}, \tag{3.19} \\
        & ||0_-\rangle \simeq \Phi_{0_-}[A]=e^{+\! \frac{1}{2} \int \! 
        \frac{d^3\vec{k}}{(2\pi )^3}A^2(\vec{k})}. \tag{3.20}
\end{align*}

\noindent We can find a relation between these two vacua via Eq.(3.14): 

\begin{align*}
        ||0_-\rangle & =e^{\int \! \frac{d^3\vec{k}}{(2\pi )^3}A^2(\vec{k})}
        ||0_+\rangle \nonumber \\
        & =e^{-\frac{1}{2}\! \int \! \frac{d^3\vec{k}}{(2\pi )^3}
        \left\{ a_-(\vec{k})+a_-^{\dagger}(\vec{k})\right\}^2}||0_+\rangle . 
        \tag{3.21}
\end{align*}

\noindent From this equation, we see that the negative energy vacuum $||0_-\rangle$ is a coherent state constructed from the empty vacuum $||0_+\rangle$ of the positive energy sector by creating all the even number negative energy bosons through the action of $a_-^{\dagger}(\vec{k})$. In this sense, $||0_-\rangle$ is the sea in which all the negative energy boson states are filled.

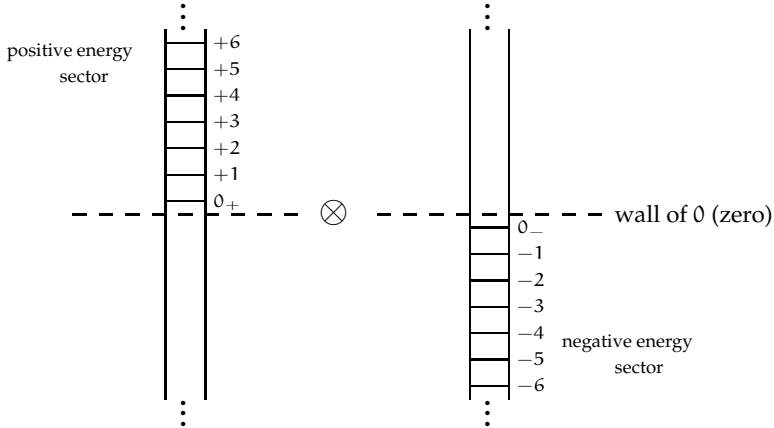
\begin{figure}[ht]
        \begin{center}
        \begin{picture}(200,160)
        \put(35,0){\line(0,1){140}}
        \put(50,0){\line(0,1){140}}
        \put(150,0){\line(0,1){140}}
        \put(165,0){\line(0,1){140}}
        \put(93,67){\Large $\otimes$}
        \put(40,-10){\Large $\vdots$}
        \put(40,140){\Large $\vdots$}
        \put(155,-10){\Large $\vdots$}
        \put(155,140){\Large $\vdots$}
        \put(-25,130){\scriptsize positive energy}
        \put(-5,120){\scriptsize sector}
        \put(185,20){\scriptsize negative energy}
        \put(205,10){\scriptsize sector}
        \multiput(35,75)(0,10){7}{\line(1,0){15}}
        \multiput(150,5)(0,10){7}{\line(1,0){15}}
        \multiput(0,70)(10,0){9}{\line(1,0){5}}
        \multiput(115,70)(10,0){9}{\line(1,0){5}}
        \put(53,73){\scriptsize $0_+$}
        \put(53,83){\scriptsize $+1$}
        \put(53,93){\scriptsize $+2$}
        \put(53,103){\scriptsize $+3$}
        \put(53,113){\scriptsize $+4$}
        \put(53,123){\scriptsize $+5$}
        \put(53,133){\scriptsize $+6$}
        \put(168,63){\scriptsize $0_-$}
        \put(168,53){\scriptsize $-1$}
        \put(168,43){\scriptsize $-2$}
        \put(168,33){\scriptsize $-3$}
        \put(168,23){\scriptsize $-4$}
        \put(168,13){\scriptsize $-5$}
        \put(168,3){\scriptsize $-6$}
        \put(205,67){wall of $0$ (zero)}
        \end{picture}
        \end{center}
        \caption[tower of states]{Physical states in the two sectors.}
        \label{tower}
\end{figure}

To avoid the misconceptions that the positive and negative energy sectors may simultaneously coexist and that there is no distinction between them, we depict them in Fig.~\ref{tower}.

To end the present section, a comment about the inner product of the states in the second quantized theory is in order. If we write the $n$-th excited state as $||n\rangle \simeq \Phi_n[A]$, the inner product is defined by 

\begin{align}
        \langle n||m\rangle =\int \mathfrak{D}A\> \Phi_n^{\ast}[A^{\ast}]
        \Phi_m[A], \tag{3.22}
\end{align}

\noindent where on the right-hand side there appears a functional integration over the scalar field $A(\vec{k})$. Recalling the definition of a convergent inner product in the first quantization, (3.11), it might be thought that the integration over $A$ should be properly taken for $n,m=0_-,-1,-2,\cdots$ in order to make the integration convergent.

\section{Towards Supersymmetric Relativistic Quantum \\Mechanics}

In the proceeding section 3 we investigated the structure of the vacuua in the multiparticle system, i.e. field theory in terms of the supersymmetry. The present section we attempt to formulate the realization of the supersymmetric 1 particle system, i.e. relativistic quantum mechanics. This formulation in our 2nd qunantization method is really starting point: the 1 particle system appears as the excitation of the unfilled Dirac seas for bosons and fermions.

We first express the supersymmetry transformation for the chiral multiplet in the N=1 supersymmetric theory, 

\begin{align*}
        & \delta_{\xi}A=\sqrt{2}\xi \psi , \\
        & \delta_{\xi}\psi =i\sqrt{2}\sigma^{\mu}\bar{\xi}\vec{\partial}_{\mu}A
        +\sqrt{2}\xi F, \tag{4.1}\\
        & \delta_{\xi}F=i\sqrt{2}\bar{\xi}\bar{\sigma}^{\mu}
        \vec{\partial}_{\mu}\psi ,
\end{align*}

\noindent in terms of the matrices. For this purpose we write the chiral multiplet as the vector notation: 

\begin{align*}
        \vec{\Phi}=
        \left( \begin{array}{c}
        A \\ \psi_{\beta} \\ F
        \end{array} \right). \tag{4.2}
\end{align*}

Then the supersymmetric generators at the 1st quantized level in an off-shell representation which induces the SUSY transformation (4.1) is given by 

\begin{align*}
        Q_{\alpha}=
        \left( \begin{array}{ccc}
        0 & \sqrt{2}\delta_{\alpha}^{\> \beta} & 0 \\
        0 & 0 & \sqrt{2}\epsilon_{\beta \alpha} \\
        0 & 0 & 0
        \end{array} \right), \quad 
        \bar{Q}_{\dot{\alpha}}=
        \left( \begin{array}{ccc}
        0 & 0 & 0 \\
        -i\sqrt{2}\sigma^{\mu}_{\beta \dot{\alpha}}\vec{\partial}_{\mu} 
        & 0 & 0 \\
        0 & i\sqrt{2}\bar{\sigma}^{\mu \dot{\gamma}\beta}
        \epsilon_{\dot{\gamma}\dot{\alpha}}\vec{\partial}_{\mu} & 0
        \end{array} \right). \tag{4.3}
\end{align*}

The Lorentz-invariant inner product is defined as natural one, 

\begin{align*}
        & \langle \Phi |\Phi \rangle \equiv \int d^3\vec{x} \> 
        \vec{\Phi}^{\dagger}I\vec{\Phi}, \tag{4.4}
\end{align*}

\noindent with a matrix I,

\begin{align*}
        & I\equiv \left( \begin{array}{ccc}
        i\overleftrightarrow{\partial_0} & 0 & 0 \\
        0 & -\bar{\sigma}^{0\dot{\alpha}\alpha} & 0 \\
        0 & 0 & 0 \\
        \end{array} \right). \tag{4.5}
\end{align*}

It is easy to check that the SUSY algebra $\left\{Q_{\alpha},\bar{Q}_{\dot{\alpha}}\right\}=-2i\sigma^{\mu}_{\alpha\dot{\alpha}}\vec{\partial}_{\mu}$ as well as hermiticity condition $\left<Q\Phi|\Phi\right>_{\dot{\alpha}}=\left<\Phi|\bar{Q}\Phi\right>_{\dot{\alpha}}$ hold by means of the equations of motion 

\begin{align}
        \vec{\partial}^2 A=0, \quad \sigma^{\mu}\vec{\partial}_{\mu}\bar{\psi}
        =0, \quad F=0. \tag{4.6}
\end{align}

\noindent For instance, by utilizing the equations of motion 

\begin{align*}
        (Q_{\alpha})^{\dagger}I=I\bar{Q}_{\dot{\alpha}}, \tag{4.7}
\end{align*}

\noindent where the dagger $\dagger$ on the left hand side denotes the usual hermitian conjuate of the matrix.

In passing to the 2nd quantized theory, the SUSY generator $\mathcal{Q}$ is expressed as 

\begin{align}
        \xi^{\alpha}\mathcal{Q}_{\alpha}+\bar{\mathcal{Q}}_{\dot{\alpha}}
        \bar{\xi}^{\dot{\alpha}}=\int d^3\vec{x} \> \Phi^{\dagger}I
        (\xi^{\alpha}Q_{\alpha}+\bar{Q}_{\dot{\alpha}}\bar{\xi}^{\dot{\alpha}})
        \Phi \tag{4.8}
\end{align}

If we wish, we may use the Dirac representation by doubling the number of fields and through the equations of motion we rederive the $\mathcal{Q}_{\alpha}$ and $\bar{\mathcal{Q}}_{\dot{\alpha}}$ in terms of the creation and annihilation operators which coincide with the equations (2.14).

Under the action of this SUSY generator (4.8) the vacuum

\begin{align}
        ||0_+\rangle =||0\rangle \otimes ||0_-\rangle 
        \otimes ||\tilde{0}_+\rangle \otimes ||\tilde{0}_-\rangle \tag{4.9}
\end{align}

\noindent is variant. It is worthwhile to notice that 1 particle states are related each other as 

\begin{align*}
        & ||+\! 1,\vec{k}\rangle =a_+^{\dagger}(\vec{k})||0_+\rangle 
        \Longleftrightarrow
        ||+\! \tilde{1},\vec{k}\rangle =b^{\dagger}(\vec{k})||\tilde{0}_+
        \rangle , \\
        &||-\! 1,\vec{k}\rangle =a_-(\vec{k})||0_-\rangle \Longleftrightarrow
        ||-\! \tilde{1},\vec{k}\rangle =d(\vec{k})||\tilde{0}_-\rangle . 
        \tag{4.10}
\end{align*}

\noindent Thus it is clear that the supersymmetry at the level of 1 particle states is perfectly realized.

The unsolved problem to the authors is to derive the classical action which should lead to the supersymmetric quantum mechanics~\cite{hnn2}. It is still unknown how to obtain this classical action: in the usual theories, as we know, the Schr\"odin\-ger equation (4.6) that is derived from the classical action describing world line, is obtained by replacing the coordinates of the space-time by operators. However, we have not succeeded so far to obtain the classical action in SUSY case.

To close this section, we note that the above argument can easily be done in a superspace formalism, because each entry of the above matrix representation is nothing but the degree of superspace coordinate $\theta$. As is well known, instead of (4.3), we may use the SUSY generators in a superspace formalism in terms of chiral superfield in the following form 

\begin{align*}
        & Q_{\alpha}=\frac{\partial}{\partial \theta^{\alpha}}
        -i\sigma_{\alpha \dot{\alpha}}^{\mu}\bar{\theta}^{\dot{\alpha}}
        \partial_{\mu}, \quad \bar{Q}_{\dot{\alpha}}=   
        -\frac{\partial}{\partial \bar{\theta}^{\dot{\alpha}}}
        +i\theta^{\alpha}\sigma_{\alpha \dot{\alpha}}^{\mu}\partial_{\mu}.
\end{align*}

\noindent By making use of the Lorentz invariant inner product

\begin{align*}
        \langle \Phi |\Phi \rangle =\int d^3\vec{x} d^2\theta d^2\bar{\theta}
        \> \Phi^{\dagger}\theta \sigma^0 \bar{\theta}\Phi
\end{align*}

\noindent
where we inserted $\theta\sigma^0\bar{\theta}$ instead of the matrix I in equations (4.4) and (4.5). In this way we go on the argument in the superspace formalism exactly parallel to our previous formalism where the matrix representation is used.

\section{Conclusions}

In this paper, we have proposed the idea that the boson vacuum forms a sea, like the Dirac sea for fermions, in which all the negative energy states are filled. This was done by introducing a double harmonic oscillator, which stems from an extension of the concept of the wave function. Furthermore, analogous to the Dirac sea where due to the exclusion principle each negative energy state is filled with one fermion, in the boson case we also discussed a modification of the vacuum state so that one could imagine two types of different vacuum fillings for all the momenta. The usual interpretation of an anti-particle, as a hole in the negative energy sea, turns out to be applicable not only for the case of fermions but also for that of bosons. Thus, we have proposed a way of resolving the long-standing problem in field theory that the bosons cannot be treated analogously to the Dirac sea treatment of the fermions. Our presentation relies on the introduction of the double harmonic oscillator, but that is really just to make it concrete. What is really needed is that we formally extrapolate to have negative numbers of particles, precisely what is described by our ``double harmonic oscillator", which were extended to have negative numbers of excitation quanta. Supersymmetry also plays a substantial role in the sense that it provides us with a guideline for how to develop the method. In fact, our method is physically very natural when we consider supersymmetry, which, in some sense, treats bosons and fermions on an equal footing. 

Our picture of analogy between fermion and boson sea description is summarized by Table 2.

\begin{center}
        \begin{tabular}{|c|c|c|} \hline
        & \multicolumn{2}{|c|}{\textbf{\large Fermions}} \\
        & \small{positive single particle energy} & 
        \small{negative single particle energy} \\
        & $E>0$ & $E<0$ \\ \hline
        & & \\
        empty $||\tilde{0}_+\rangle$ & true & not realized in nature \\
        & & \\ \hline
        & & \\
        filled $||\tilde{0}_-\rangle$ & not realized in nature & true \\
        & & \\ \hline
        \multicolumn{3}{c}{} \\ \hline
        & \multicolumn{2}{|c|}{\textbf{\large Bosons}} \\
        & \small{positive single particle energy} & 
        \small{negative single particle energy} \\
        & $E>0$ & $E<0$ \\ \hline
        ``sector with bottom" & & \\
        analogous to ``empty" & true & not realized in nature \\
        $||0_+\rangle$ & & \\ \hline
        ``sector with top" & & \\
        analogous to ``filled" & not realized in nature & true \\
        $||0_-\rangle$ & & \\ \hline
        \multicolumn{3}{c}{} \\
        \multicolumn{3}{c}{Table 2: Analogy between fermion and boson sea 
        description} \\
        \end{tabular}
\end{center}

Next, we presented an attempt to formulate the supersymmetric quantum mechanics. We found that our representation of SUSY generators in the matrix form is properly constructed only when we utilize the equations of motion (Schr\"{o}\-dinger equations) and can be interpreted as the SUSY generators in the second quantization. Using these generators, the 1 particle states of boson and fermion are related within the positive and negative energy sectors respectively. Therefore, we could conclude that our SUSY generators in the matrix form are useful to formulate the supersymmetric and relativistic quantum mechanics.

\section*{Acknowledgements}
Y.H. is supported by The 21st century COE Program ``Towards a New Basic Science; Depth and Synthesis", of the Department of Physics, Graduate School of Science, Osaka University. This work is supported by Grants-in-Aid for Scientific Research on Priority Areas, Number of Areas 763, ``Dynamics of strings and Fields", from the Ministry of Education of Culture, Sports, Science and Technology, Japan.

%

\title{Searching for Boundary Conditions in Kaluza-Klein-like Theories}
\author{D. Lukman${}^1$, N. S. Manko\v c Bor\v stnik${}^{1,2}$ and 
H. B. Nielsen${}^3$}
\institute{%
${}^1$ Faculty of Mathematics and Physics, University of
 Ljubljana, Jadranska 19, 1000 Ljubljana, Slovenia  \\
${}^2$ Primorska Institute for Natural Sciences and Technology,\\ 
6000 Koper, Slovenia\\
${}^3$ Department of Physics, Niels Bohr Institute,
Blegdamsvej 17, Copenhagen, DK-2100}

\titlerunning{Searching for Boundary Conditions in Kaluza-Klein-like Theories}
\authorrunning{D. Lukman, N. S. Manko\v c Bor\v stnik and H. B. Nielsen}
\maketitle

\begin{abstract}
In the approach by one of us (N.S.M.B.)\cite{norma,pikanorma} unifying spins and charges, 
the gauge fields origin only 
in the gravity and the spins and the charges in only the spin. This approach is also a kind of the genuine 
Kaluza-Klein theory, suffering the problem of getting chiral fermions in the ''physical space''.
In ref.\cite{hnhep03} we  discussed a possible way for solving this problem by an 
appropriate choice of  boundary conditions. In this contribution we discuss further possible choices of
boundary conditions.
\end{abstract}

\section{Introduction}
\label{orbif:introduction}

 Genuine Kaluza-Klein-like theories, assuming nothing but  a gravitational field 
in $d$-dimensional space (no additional gauge or scalar fields), 
which after the spontaneous compactification of a $(d-4)$-dimensional part of space manifest in 
four dimensions as all the known gauge fields including gravity, have difficulties\cite{witten} with 
masslessness of fermionic fields at low energies. It looks  
namely very difficult to avoid after the compactification of a part of space  the appearance of  
representations of both handedness in this part of space and consequently also in the 
(1+3)-dimensional space. Accordingly, the gauge fields can hardly couple chirally in the (1+3) - dimensional 
space.

In an approach by one of us\cite{pikanorma,norma} 
it has long been the wish to obtain the gauge fields from only gravity, so that ''everything'' would become gravity.
This approach has taken the inspiration from looking for unifying all the internal degrees of freedom, that is 
the spin and all the charges, into only the spin. This approach is also a kind of the genuine Kaluza-Klein theory, suffering
the same problems, with the problem of getting chiral fermions included, unless we can solve them.

There are several attempts in the literature, which use boundary conditions to select massless fields of
a particular handedness\cite{Horawa,Kawamura,Hebecker,Buchmueller}. Boundary conditions are chosen by choosing 
discrete orbifold symmetries.

In this contribution we study  a toy model with a Weyl spinor, which caries in $d(=1+5)$ -dimensional 
space with the symmetry $M^{(1+3)}\times $ a flat finite disk only the spin as the internal degree of freedom. 
The only back ground field is the gravitational gauge field with vielbeins and spin connections, which
manifest the rotational symmetry on the flat disk. 

We require that the Kaluza-Klein charge of spinors is proportional to the total angular momentum
on the disk and search for spinors, which manifest in $M^{(1+3)}$ masslessness (have no partners of opposite
handedness) and are chirally coupled by the Kaluza-Klein charge to the corresponding Kaluza-Klein field. 

This contribution mainly comments the ref.\cite{hnhep03}.

\section{ Weyl spinors in gravitational fields with spin connections and vielbeins}
\label{Weyl}

We let\footnote{Latin indices  
$a,b,..,m,n,..,s,t,..$ denote a tangent space (a flat index),
while Greek indices $\alpha, \beta,..,\mu, \nu,.. \sigma,\tau ..$ denote an Einstein 
index (a curved index). Letters  from the beginning of both the alphabets
indicate a general index ($a,b,c,..$   and $\alpha, \beta, \gamma,.. $ ), 
from the middle of both the alphabets   
the observed dimensions $0,1,2,3$ ($m,n,..$ and $\mu,\nu,..$), indices from the bottom of the alphabets
indicate the compactified dimensions ($s,t,..$ and $\sigma,\tau,..$). We assume the signature $\eta^{ab} =
diag\{1,-1,-1,\cdots,-1\}$.
} a spinor interact with a gravitational field\cite{hnhep03} through vielbeins
$f^{\alpha}{}_{a}$ (inverted vielbeins to 
$e^{a}{}_{\alpha}$ with the properties $e^a{}_{\alpha} f^{\alpha}{}_b = \delta^a{}_b,\; 
e^a{}_{\alpha} f^{\beta}{}_a = \delta^{\beta}_{\alpha} $ ) and   
spin connections, namely 
$\omega_{ab\alpha}$, which is the gauge field of $S^{ab}= \frac{i}{4}(\gamma^a \gamma^b - \gamma^b \gamma^a)$.
We choose the basic states in the  space of spin degrees of freedom to be 
eigen states of the Cartan sub algebra of the operators: $S^{03}, S^{12}, S^{56}$.

The covariant momentum of a spinor is taken to be
\begin{eqnarray}
p_{0 a} &=& f^{\alpha}{}_{a}p_{0 \alpha}, \quad p_{0 \alpha} \psi = p_{ \alpha} - \frac{1}{2} S^{cd} 
\omega_{cd \alpha}, 
\label{covp}
\end{eqnarray}
when applied to a spinor function $\psi$. 

The corresponding Lagrange density ${\cal L}$  for   a Weyl has the form
$${\cal L} = E \frac{1}{2} [(\psi^{\dagger}\gamma^0 \gamma^a p_{0a} \psi) + (\psi^{\dagger} \gamma^0\gamma^a p_{0 a}
\psi)^{\dagger}]$$ 
and leads to
\begin{eqnarray}
{\cal L} &=& E\psi^{\dagger}\gamma^0 \gamma^a   ( p_{a} - \frac{1}{2} S^{cd}  \Omega_{cda})\psi,
\label{weylL}
\end{eqnarray}
with $ E = \det(e^a{}_{\alpha}) $, 
$ \Omega_{cda} =\frac{1}{2}( \omega_{cda} + (-)^{cda}  \omega^*{}_{cda})$, and with
$(-)^{cda}$, which is $-1$, if two indices are equal, and is $1$ otherwise (if all three indices are different). 
(In $d=2$ case
$\Omega_{abc}$ is always pure imaginary.)

The Lagrange density (\ref{weylL}) leads to the Weyl equation
\begin{eqnarray}
\gamma^0 \gamma^a  P_{0a}\psi =0, \quad P_{0a}=   (f^{\alpha}{}_{a} p_{\alpha} -\frac{1}{2} S^{cd} \Omega_{cda}).
\label{Weylgen}
\end{eqnarray}
We assume that a two dimensional space is  a flat  disk 
\begin{eqnarray}
f^{\sigma}{}_{s} = \delta^{\sigma}{}_{s},\; \omega_{56 s} =0,
\label{disk}
\end{eqnarray}
with the rotational symmetry and with the radius $\rho_0$. 

\section{Solutions on the disk}
\label{solutions}

Spinors manifest masslessness in $d=(1+3)$-dimensional space, if they solve the Weyl equation (\ref{Weylgen})
with $a=5,6$, so that the term $E\psi^{\dagger}\gamma^0 \gamma^s  p_{0s}\psi,\; s=5,6$ (the only term, which would 
manifest as a mass term in $d=1+3$) is equal to zero. Otherwise they manifest a mass in  $d=(1+3)$-dimensional space.

We are accordingly looking for massless ($m=0$) and massive ($ m \ne 0$) solutions  of the  equation on a flat disk
 \begin{eqnarray}
   \gamma^5  e^{2i \phi S^{56} }(p_{0(5)}  + 
   2i S^{56} p_{0 (6)})\psi = -m \psi
 \label{weyld=2m0}
 \end{eqnarray}
and  require that the solutions are eigenvectors of 
the total angular momentum operator on the disk- $M^{56}= L^{56} + S^{56}$. 

We find as the eigenfunctions of $M^{56}$ 
\begin{eqnarray}
\psi^{m}{}_{+(n+1/2)}(\rho, \varphi) &=& \alpha^{+}{}_{n}(\rho) e^{i n \varphi}(+) + \beta^{+}{}_{n+1}(\rho) e^{i (n+1) 
\varphi} [-],\nonumber\\
\psi^{m}{}_{-(n+1/2)}(\rho, \varphi) &=& \alpha^{-}{}_{n+1}(\rho) e^{-i (n+1) \varphi}(+) + \beta^{-}{}_{n}(\rho) e^{-i n 
\varphi} [-].  
 \label{solutionweylm}
 \end{eqnarray}
Index $m$ denotes that spinors carry a mass $m$. The functions $\alpha^{\pm}{}_{k}(\rho), \; \beta^{\pm}{}_{k}(\rho),\; 
k=n, n+1, $ $n=0,1,2,3...,$ must solve Eq.(\ref{weyld=2m0}) (for either 
the massive - $\psi^{m}{}_{\pm(n+\frac{1}{2})}$ -  or the  massless - $\psi^{m=0}{}_{\pm(n+\frac{1}{2})}$ - case), 
which in the polar coordinates read
 \begin{eqnarray}
i (\frac{\partial}{\partial \rho} - \frac{n}{\rho}) \alpha^{+}{}_{n}(\rho) &=& m \;\beta^{+}{}_{n+1} (\rho),\nonumber\\ 
i (\frac{\partial}{\partial \rho} + \frac{n+1}{\rho}) \beta^{+}{}_{n+1} (\rho) &=& m \;\alpha^{+}{}_{n}(\rho).
 \label{weyld=2mpolar0}
 \end{eqnarray}
One immediately finds that for $m=0$ the functions  $\alpha^{+}{}_{n}(\rho) = \rho^n$, 
$\beta^{+}{}_{n+1} = 0$ and $\alpha^{-}{}_{n+1}=0,$ $\beta^{-}{}_{n} =\rho^n$, $n=0,1,2...,$ solve 
Eq.(\ref{weyld=2mpolar0}), 
so that the right handed solutions and the left handed solutions in $d=2$ are as follows
\begin{eqnarray}
\psi^{m=0}{}_{+(n+1/2)}(\rho, \varphi) = A_{+(n+\frac{1}{2})} \rho^{n} e^{i n \varphi}(+), \nonumber\\
\psi^{m=0}{}_{-(n+1/2)}(\rho, \varphi) = A_{-(n+\frac{1}{2})} \rho^{n} e^{-i n \varphi} [-], 
 \label{solutionweylm=0}
 \end{eqnarray}
 with $A_{\pm(n+\frac{1}{2})}$  constants with respect to $\rho$ and $\varphi$.
In the massive case, for the two types of functions $\alpha^{\pm}{}_{m}$ and $\beta^{\pm}{}_{m}$  
the Bessel functions of the first order can be taken as follows:
$\alpha^{+}{}_{n}(\rho)=\beta^{-}{}_{n}(\rho) = J_n $ and $\beta^{+}{}_{n+1} (\rho)= \alpha^{-}_{n+1}(\rho) = -iJ_{n+1}$.

\subsection{Discussions on the solutions}
\label{discussionssolutions}

We started with a Hermitean operator (Eq.(\ref{weyld=2m0})), whose eigenvalues must be real ($m$ is a real number) and 
solutions form accordingly a complete basis for either massless or massive cases. Eigenfunctions have in the massive case 
the following orthogonality relation
\begin{eqnarray}
\int^{\rho_0}_{0} \rho d\rho \int^{2 \pi}_{0} d \varphi \; \psi^{m \dagger}{}_{\pm (n+1/2)}(\rho, \varphi)
\psi^{m}{}_{\pm(n'+1/2)}(\rho, \varphi) = \delta^{n n'},\nonumber\\
\int^{\rho_0}_{0} \rho d\rho \int^{2 \pi}_{0} d \varphi\; \psi^{m \dagger}{}_{\pm (n+1/2)}(\rho, \varphi)
\psi^{m}{}_{\mp(n'+1/2)}(\rho, \varphi)  =0,
\label{orth}
\end{eqnarray}
for any finite $\rho_0$. If $\rho_0$ is infinite, the integral in the first of the above equations is infinite 
for $n=n'$. 

When we transform Eq.(\ref{weyld=2mpolar0}) into a second order equation and if we write for the massive case 
$\alpha^{+}{}_{n}(\rho)=\beta^{-}{}_{n}(\rho) = J_n $ and $\beta^{+}{}_{n+1} (\rho)= \alpha^{-}_{n+1}(\rho) = -iJ_{n+1}$,
we end up with the Bessel equation
\begin{eqnarray}
(x^2 \frac{d^2}{d x^2} + x \frac{d}{dx} +(x^2 -n^2)) J_n = \hat{O} J_n=0, \quad {\rm for \; } m \ne 0, \quad m\rho =x.
\label{bessel}
\end{eqnarray}
For the massless case we obtain the equation
\begin{eqnarray}
(\rho^2 \frac{d^2}{d \rho^2} + \rho \frac{d}{d\rho} +(-n^2)) A^{\pm}_{n} \rho^n  =0, \quad {\rm for \; } m = 0.
\label{esecmassless}
\end{eqnarray}
Solutions of both equations form a complete set, and any solution of the second equation can be expressed 
with the solutions of the first equation and opposite. 

The Bessel functions of different n are not orthogonal. This can easily be understood, since  
$(\frac{\hat{Q}}{x})^{\dagger} = \frac{\hat{Q}}{x}$ but $\hat{Q}^{\dagger} \ne \hat{Q}$, so that we can not 
expect that $(x^2 \frac{d^2}{d x^2} + x \frac{d}{dx} +x^2)J_n= n^2 J_n$ would guarantee the orthogonality of solutions, 
since indeed $(x \frac{d^2}{d x^2} +  \frac{d}{dx} +x -\frac{n^2}{x})J_n= 0. J_n$ is the right relation 
(Any $ f(x)^{\dagger}= f(x)$, but the part with derivatives requires $\frac{\hat{Q}}{x}$). This means that 
all $J_n$ belong to the same eigen value 0, and are accordingly not orthogonal in general.
But $\psi^{m}_{\pm(n+1/2)}$ are orthogonal and they can be orthonormalized in any finite interval $\rho_0 $ smaller
than infinite.

Also the solutions of Eq.(\ref{esecmassless}, for the massless case) can only be normalized  in a finite 
interval of $\{0, \rho_0\}$ and for nonnegative integer $n$ only.

\subsection{Our boundary condition on a finite flat disk\cite{hnhep03}}

The boundary condition for a flat finite disk, used in the ref.\cite{hnhep03}, is as follows
\begin{eqnarray}
(1-i n^{(\rho)}{}_{\alpha} n^{(\varphi)}{}_{\beta} 
f^{\alpha}{}_{a}f^{\beta}{}_{b} \gamma^a \gamma^b )\psi|_{\rho= \rho_0}=0,
\label{orbif:diskboundary}
\end{eqnarray}
where $\psi$ is the solution of the Weyl equation in $d=1+5$ and
\begin{eqnarray} 
n^{(\rho)}&=& (0,0,0,0,\cos \varphi, \sin \varphi),\nonumber\\
n^{(\varphi)} &=& (0,0,0,0,-\sin \varphi, \cos \varphi)\nonumber 
\end{eqnarray}
are the two unit vectors 
perpendicular and tangential to the boundary (at $\rho_0$), respectively.
(The operator $(1-i n^{(\rho)}{}_{\alpha} n^{(\varphi)}{}_{\beta} 
f^{\alpha}{}_{a}f^{\beta}{}_{b} \gamma^a \gamma^b )$ does not commute with the operator 
$\gamma^5  e^{2i \phi S^{56} }(p_{0(5)}  + 
   2i S^{56} p_{0 (6)})$ of Eq.(\ref{weyld=2m0})). 

 According to this boundary condition (Eq.(\ref{orbif:diskboundary}))
only the masses, for which $\beta^{\pm}{}_{k} (\rho_0)=0$, are allowed, since the term with 
$(+)$ is at $\rho_0$ multiplied by zero, while the term with $[-]$ is multiplied by (1+1).
In the massless case, the boundary condition requires that $A_{-(n+\frac{1}{2})} =0$, so that only right handed
spinors with the spin part $(+)$ survive. 
There are accordingly infinite number of massive and  of massless solutions. To different solutions different
total angular momenta correspond and in the massive case also different masses. 

Masses are, due to the boundary condition, discretized: Only masses, for which $m^{\pm}{}_{n,k} \rho_0 = \alpha_{n,k}$, 
where $n$ determines the Bessel function $J_n$ and $k$ determines the $k$'th zero of $J_n,$ are allowed.

One finds:
\begin{eqnarray}
m^{+}{}_{0,1} \rho_0 = 3.83.. = m^{-}{}_{1,1} \rho_0, \nonumber\\
m^{-}{}_{0,1} \rho_0 = 2.40..\nonumber\\
m^{+}{}_{1,1} \rho_0 =5.13..= m^{-}{}_{2,1} \rho_0, \nonumber\\
\dots
\label{masses}
\end{eqnarray}
For a small enough $\rho_0$ a gap between massless and massive states can be very high in comparison with
observable energies in the ''physical'' world.

We easily see that a current through the boundary 
\begin{eqnarray}
n^{(\rho)}{}_{\alpha} j^{\alpha}|_{\rho= \rho_0} = \psi^+ \gamma^0 \gamma^a f^{\alpha}{}_{a} n^{(\rho)}{}_{\alpha}
\psi|_{\rho=\rho_0}, 
\label{diskcurrentthroughwall}
\end{eqnarray}
is in all the cases (massless and massive) equal to zero. In the massive case, the current  is proportional to the terms
$\alpha^{\pm}{}_{k} (\rho_0) \beta^{\pm}{}_{k \pm 1}(\rho_0)$, which are zero, since always
either $\alpha^{\pm}{}_{k} (\rho_0)$ or $\beta^{\pm}{}_{k \pm 1}(\rho_0)$ is zero on the boundary.
In the massless case $\beta^{+}{}_{m}$ is zero all over. (The same fact makes  the requirement that the 
term of the type $\frac{1}{2}\int d^2x (\frac{\partial}{\partial x^{s}} (\delta \psi^{\dagger}\gamma^0 \gamma^{s} \psi + 
\psi^{\dagger}\gamma^0 \gamma^{s} \delta \psi)=0$ is trivially fulfilled.)

\subsection{Discussions on the extended version of boundary conditions}

Let us try to extend the above boundary conditions as follows
\begin{eqnarray}
(1-i n^{(\rho)}{}_{\alpha} n^{(\varphi)}{}_{\beta} 
f^{\alpha}{}_{a}f^{\beta}{}_{b} \gamma^a \gamma^b -2  F(n^{(\rho)}{}_{\alpha} p^{\alpha}.
n^{(\rho)}{}_{\alpha} x^{\alpha} ))
\psi|_{\rho= \rho_0}=0,
\label{diskboundary}
\end{eqnarray}
where we are searching for  a function $F(n^{(\rho)}{}_{\alpha} p^{\alpha}.
n^{(\rho)}{}_{\alpha} x^{\alpha} )$, which would  allow only one massless solution of a chosen handedness 
and leave massive solutions untouched. For our flat disk this condition simplifies to $F(i\rho\frac{ d}{d\rho})$.
In the massless case this boundary condition requires if, for example, $F(\rho \frac{d}{d\rho})= 
2(i\rho \frac{d}{d\rho})^2$, that
\begin{eqnarray}
n^2 \psi^{m=0}_{+(n+\frac{1}{2})}|_{\rho= \rho_0}=0,\nonumber\\
(1 + n^2) \psi^{m=0}_{-(n+\frac{1}{2})}|_{\rho=\rho_0} =0.
\label{diskboundarykm0}
\end{eqnarray}
It then follows that only one left handed solution is nonzero, namely $\psi^{m=0}_{+(\frac{1}{2})}$. 

For massive cases the same requirement
\begin{eqnarray}
(1-\Gamma^{(2)} +2 (x \frac{d}{dx})^2) \psi^{m}{}_{\pm(n+\frac{1}{2})}|_{\rho=\rho_0} =0,
\label{diskboundarykm}
\end{eqnarray}
leads to $(x^2 \frac{d^2}{dx^2}+ x\frac{d}{dx}) J_{n}|_{\rho=\rho_0} =0 = 
((x^2-n^2) ) J_{n}|_{\rho=\rho_0})$  and at the same time to 
$(1+x^2 \frac{d^2}{dx^2}+ x\frac{d}{dx}) J_{n+1}|_{\rho=\rho_0} =0 = 
(1+ (x^2-(n+1)^2) ) J_{n+1}|_{\rho=\rho_0}),$ which only can be true, if $J_n|_{\rho=\rho_0} =0 =
J_{n+1}|_{\rho=\rho_0}$. But this is not the case, since  two Bessel functions  never 
have a zero at the same value of $x$. 

One can hardly expect that any other function of the derivatives 
$F(\rho \frac{d}{d\rho})$ would lead to requirements, which would allow all the above massive solutions to 
survive. It looks like therefore, that if we insist to have infinite many (discrete) massive solutions, 
 then also all possible massless solutions 
of a chosen handedness will be allowed. 

\subsection{Orbifolds on the boundary}

Let us now assume that the boundary on the disk is not a manifold $S^1$ but instead an orbifold $S^1/Z_2$.

 We then require:
\begin{eqnarray}
\psi^{m}{}_{\pm(n+1/2)}(\rho_0, -\varphi) = P\psi^{m}{}_{\pm(n+1/2)}(\rho_0, \varphi), 
 \label{orbifoldcon}
 \end{eqnarray}
with $P=\pm1$. This way of identification of two points on a circle means for $0\le \varphi < \pi $ 
that $x^{(5)} $ goes to $-x^{(5)}$ and $x^{(6)} $ goes to $ x^{(6)}$ at $\rho = \rho_0$, 
if $\varphi$ goes to $-\varphi$. Since this kind of 
orbifold conditions break the rotational symmetry on the disk, the total angular momentum $M^{56}$ is not a symmetry
of the disk and 
for massless spinors this boundary condition allows only $\psi^{m=0}{}_{\pm 1/2}$ to survive,
while no massive solution exists.

Since the two massless solutions carry two different charges ($M^{56} =\pm1/2$ for 
$\psi^{m=0}{}_{\pm 1/2}$, respectively), this two solutions are  
mass protected and chirally coupled to the Kaluza-Klein charge, and accordingly acceptable.
The fact that there are no massive solution makes this orbifold condition unacceptable.

\section{Conclusions}
We discuss in this contribution the influence of a choice of boundary conditions on 
possible solutions of Weyl equations in a space $M^{1+5}$. We start with a spinor of a chosen 
handedness in $d=1+5$-dimensional space  and assume that
the space factorizes into $M^{(1+3)} \times$ a flat finite disk with the radius $\rho_0$ and 
with boundaries, which should  
allow massless spinors of only a chosen handedness. 
The spinor, whose the only internal degree of freedom  
is the spin,  interacts  with the gauge gravitational field represented by  
spin connections ($\omega_{ab \alpha}$) and vielbeins ($f^{\alpha}{}_{a}$). 
The disk (manifesting the rotational symmetry) is flat ($f^{\sigma}{}_{s}=\delta^{\sigma}{}_{s}$, $\omega_{st\sigma}=0$). 
We look for massless  spinors  in $(1+3)$ ''physical'' space, which are mass protected and chirally coupled
to a Kaluza-Klein gauge field through a quantized (proportional to an integer) Kaluza-Klein charge. 

To be massless in $(1+3)$ space, spinors must  obey the Weyl equation on a disk: $\gamma^0\gamma^s 
f^{\sigma}{}_{s} p_{\sigma}\psi=0, s=\{5,6\}, \sigma =\{(5),(6) \}$. 
The boundary condition on the disk makes the current of (massless and massive) spinors in the perpendicular 
direction to the boundary to be zero and guarantees that  massless spinors are mass protected. We have found one  
boundary condition:  
$$(1-i n^{(\rho)}{}_{\alpha} n^{(\varphi)}{}_{\beta} 
f^{\alpha}{}_{a}f^{(\beta)}{}_{b} \gamma^a \gamma^b )\psi|_{\rho= \rho_0}=0,$$
which allows massless solutions
of only one handedness and makes massive solutions to have discrete values of masses. There are infinite many 
massive solutions and there are  more then
one massless solutions in the ''physical'' space, manifesting different charges. Other discussed boundary conditions  
do not allow any massive solutions.
We demonstrated in ref.\cite{hnhep03}, that the background gauge field, chosen to obey isometry relations and 
respecting accordingly the rotational symmetry on the disk, fulfills the general equations of motion, which follow 
in $(1+5)$ from the action, linear in the Riemann curvature. The effective Lagrangean in $d=(1+3)$ is for the flat space
the ordinary Lagrangean for the $U(1)$ field. The current (for massless or massive spinors) is in the
$(1+3)$-dimensional space proportional to the total angular momentum on the disk $(M^{56})$, which 
is accordingly determining the charge of spinors (proportional to $n+1/2, n=0,1,2,..$). 
 Consequently massless spinors are mass protected and chirally coupled to the Kaluza-Klein gauge fields.

\title{Second Quantization of Spinors and Clifford Algebra Objects}
\author{N.S. Manko\v c Bor\v stnik${}^{1,2}$ and H. B. Nielsen${}^3$}
\institute{%
${}^1$ Department of Physics, University of
Ljubljana, Jadranska 19, 1000 Ljubljana, Slovenia \\
${}^2$ Primorska Institute for Natural Sciences and Technology,
C. Mare\v zganskega upora 2, 6000 Koper, Slovenia\\
${}^3$ Department of Physics, Niels Bohr Institute,
Blegdamsvej 17,
Copenhagen, DK-2100}

\titlerunning{Second Quantization of Spinors and Clifford Algebra Objects}
\authorrunning{N.S. Manko\v c Bor\v stnik and H. B. Nielsen}
\maketitle

\begin{abstract} 
Using a technique \cite{SQCholgernorma2002,SQCholgernorma2003} to construct a basis for spinors 
 in terms of the Clifford algebra objects, we define    
the creation and annihilation operators for spinors and families of spinors as an odd Clifford algebra objects.
The proposed ''second quantization'' procedure works for all dimensions and any signature and 
might also help to understand the origin of charges of quarks and 
leptons\cite{SQCnorma92,SQCnorma93,SQCnormaixtapa2001,SQCpikanorma2003} as well as of their families. 
\end{abstract}

\section{Introduction}
\label{snmb-sqc1:introduction}

We presented in a paper\cite{SQCholgernorma2002} the technique to construct a spinor 
basis as products of nilpotents and projectors  formed from the objects $\gamma^a$ for which we only 
need to know that they obey the Clifford algebra. Nilpotents and projectors are odd and even objects 
of $\gamma^a$'s, respectively, and are chosen to be eigenstates of a Cartan sub algebra of the 
Lorentz group in the sense that the left multiplication of nilpotents and projectors by the Cartan 
sub algebra elements multiplies these objects by a number.  The technique enables to 
construct a spinor basis for any dimension $d$
and any signature in a simple and transparent way. Equipped with graphic representation of basic states,  
the technique offers an elegant means of seeing all the quantum numbers of states with respect to the Lorentz 
group, as well as the transformation properties of states under Clifford algebra objects. 

Multiplying products of nilpotents and projectors from the left hand side  
by any of the Clifford algebra objects, we get a linear combination of these ``basic '' 
elements back: our basis spans a left ideal, and has $2^{d/2}$ elements for $d$ even and $2^{(d-1)/2}$
elements for $d$ odd. 

The proposed technique was initiated and developed by one of the authors of this paper, when  proposing an  
approach\cite{SQCnorma92,SQCnorma93,SQCnormaixtapa2001} in which all the internal degrees of freedom of either 
spinors or vectors can be described in the space of $d$-anti commuting (Grassmann) coordinates, if the 
dimension of ordinary space is also $d$.

{\em We show in this paper how can products of nilpotents and projectors be used to define the creation and 
annihilation operators for a spinor representation.} We discuss  even dimensional spaces.

We  assume an arbitrary
signature of space time so that our metric tensor $\eta^{ab}$, with $a,b \in \{0,1,2,3,5,\cdots d \}$ 
is diagonal with values $\eta^{aa} = \pm 1$, depending on the chosen signature ($+1$ for time-like coordinates
and $-1$ for space-like coordinates).

\section{Technique to generate spinor representations in terms of Clifford algebra objects}
\label{SQCtechnique}

We shall briefly repeat the main points of the technique for generating spinor representations from 
Clifford algebra objects, following the
reference\cite{SQCholgernorma2002}. We ask the reader to look for details and proofs in this reference.

We assume the objects $\gamma^a$, which fulfil the Clifford algebra
\begin{equation}
\{\gamma^a, \gamma^b \}_+ = I \;\;2\eta^{ab}, \quad {\rm for} \quad a,b \quad \in \{0,1,2,3,5,\cdots,d \},
\label{clif}
\end{equation}
for any $d$, even or odd.  $I$ is the unit element in the Clifford algebra, 
while
 $$\{\gamma^a, \gamma^b \}_{\pm} = \gamma^a \gamma^b \pm \gamma^b \gamma^a .$$

We assume  the ``Hermiticity'' property for $\gamma^a$'s 
\begin{eqnarray}
\gamma^{a\dagger} = \eta^{aa} \gamma^a,
\label{cliffher}
\end{eqnarray}
in order that 
$\gamma^a$ are compatible with (\ref{clif}) and formally unitary, i.e. ${\gamma^a}^{\dagger}\gamma^a=I$.

We also define the Clifford algebra objects 
\begin{equation}
S^{ab} = \frac{i}{4} [\gamma^a, \gamma^b ] := \frac{i}{4} (\gamma^a \gamma^b - \gamma^b \gamma^a)
\label{SQCsab}
\end{equation}
which close the Lie algebra of the Lorentz group  
$$ \{S^{ab},S^{cd}\}_- = i (\eta^{ad} S^{bc} + \eta^{bc} S^{ad} - \eta^{ac} S^{bd} - \eta^{bd} S^{ac}).$$
One finds from Eq.(\ref{cliffher}) that $(S^{ab})^{\dagger} = \eta^{aa} \eta^{bb}S^{ab}$ and that $\{S^{ab}, S^{ac} \}_+
= \frac{1}{2} \eta^{aa} \eta^{bc}$.

Recognizing from Eq.(\ref{SQCsab}) and the Lorentz algebra relation that two Clifford algebra objects 
$S^{ab}, S^{cd}$ with all indices different 
commute, we  select (out of infinitely many possibilities) the Cartan sub algebra of the algebra of the 
Lorentz group as follows 
\begin{eqnarray}
S^{0d}, S^{12}, S^{35}, \cdots, S^{d-2\; d-1}, \quad {\rm if } \quad d &=& 2n,
\nonumber\\
S^{12}, S^{35}, \cdots, S^{d-1 \;d}, \quad {\rm if } \quad d &=& 2n +1.
\label{choicecartan}
\end{eqnarray}
To make the technique simple, we introduce the graphic representation\cite{SQCholgernorma2002}
as follows
\begin{eqnarray}
\stackrel{ab}{(k)}:&=& 
\frac{1}{2}(\gamma^a + \frac{\eta^{aa}}{ik} \gamma^b),\nonumber\\
\stackrel{ab}{[k]}:&=&
\frac{1}{2}(1+ \frac{i}{k} \gamma^a \gamma^b),
\label{signature}
\end{eqnarray}
where $k^2 = \eta^{aa} \eta^{bb}$.
One can easily check by taking into account the Clifford algebra relation (Eq.\ref{clif}) and the
definition of $S^{ab}$ (Eq.\ref{SQCsab})
that if one multiplies from the left hand side by $S^{ab}$ the Clifford algebra objects $\stackrel{ab}{(k)}$
and $\stackrel{ab}{[k]}$,
it follows that
\begin{eqnarray}
S^{ab}\stackrel{ab}{(k)}=\frac{1}{2}k \stackrel{ab}{(k)},\nonumber\\
S^{ab}\stackrel{ab}{[k]}=\frac{1}{2}k \stackrel{ab}{[k]}.
\label{grapheigen}
\end{eqnarray}
This means that
$\stackrel{ab}{(k)}$ and $\stackrel{ab}{[k]}$ acting from the left hand side on anything (on a
vacuum state $|\psi_0\rangle$, for example ) are eigenvectors of $S^{ab}$.

We further find 
\begin{eqnarray}
\gamma^a \stackrel{ab}{(k)}&=&\eta^{aa}\stackrel{ab}{[-k]},\nonumber\\
\gamma^b \stackrel{ab}{(k)}&=& -ik \stackrel{ab}{[-k]}, \nonumber\\
\gamma^a \stackrel{ab}{[k]}&=& \stackrel{ab}{(-k)},\nonumber\\
\gamma^b \stackrel{ab}{[k]}&=& -ik \eta^{aa} \stackrel{ab}{(-k)}
\label{SQCgraphgammaaction}
\end{eqnarray}
It follows that
\begin{eqnarray}
S^{ac}\stackrel{ab}{(k)}\stackrel{cd}{(k)} &=& 
-\frac{i}{2} \eta^{aa} \eta^{cc}
\stackrel{ab}{[-k]}\stackrel{cd}{[-k]},\nonumber\\ 
S^{ac}\stackrel{ab}{[k]}\stackrel{cd}{[k]} &=& \frac{i}{2}  
\stackrel{ab}{(-k)}\stackrel{cd}{(-k)}, \nonumber\\
S^{ac}\stackrel{ab}{(k)}\stackrel{cd}{[k]} &=& -\frac{i}{2} \eta^{aa}  
\stackrel{ab}{[-k]}\stackrel{cd}{(-k)}, \nonumber\\
S^{ac}\stackrel{ab}{[k]}\stackrel{cd}{(k)} &=& \frac{i}{2} \eta^{cc}  
\stackrel{ab}{(-k)}\stackrel{cd}{[-k]}.\nonumber
\end{eqnarray}
It is useful to deduce the following relations
\begin{eqnarray}
\stackrel{ab}{(k)}^{\dagger}=\eta^{aa}\stackrel{ab}{(-k)},\quad
\stackrel{ab}{[k]}^{\dagger}= \stackrel{ab}{[k]},
\label{graphher}
\end{eqnarray}
and
\begin{eqnarray}
\stackrel{ab}{(k)}\stackrel{ab}{(k)}& =& 0, \quad \quad \stackrel{ab}{(k)}\stackrel{ab}{(-k)}
= \eta^{aa}  \stackrel{ab}{[k]}, \quad \stackrel{ab}{(-k)}\stackrel{ab}{(k)}=
\eta^{aa}   \stackrel{ab}{[-k]},\quad
\stackrel{ab}{(-k)} \stackrel{ab}{(-k)} = 0 \nonumber\\
\stackrel{ab}{[k]}\stackrel{ab}{[k]}& =& \stackrel{ab}{[k]}, \quad \quad
\stackrel{ab}{[k]}\stackrel{ab}{[-k]}= 0, \;\;\quad \quad  \quad \stackrel{ab}{[-k]}\stackrel{ab}{[k]}=0,
 \;\;\quad \quad \quad \quad \stackrel{ab}{[-k]}\stackrel{ab}{[-k]} = \stackrel{ab}{[-k]}
 \nonumber\\
\stackrel{ab}{(k)}\stackrel{ab}{[k]}& =& 0,\quad \quad \quad \stackrel{ab}{[k]}\stackrel{ab}{(k)}
=  \stackrel{ab}{(k)}, \quad \quad \quad \stackrel{ab}{(-k)}\stackrel{ab}{[k]}=
 \stackrel{ab}{(-k)},\quad \quad \quad 
\stackrel{ab}{(-k)}\stackrel{ab}{[-k]} = 0
\nonumber\\
\stackrel{ab}{(k)}\stackrel{ab}{[-k]}& =&  \stackrel{ab}{(k)},
\quad \quad \stackrel{ab}{[k]}\stackrel{ab}{(-k)} =0,  \quad \quad 
\quad \stackrel{ab}{[-k]}\stackrel{ab}{(k)}= 0, \quad \quad \quad \quad
\stackrel{ab}{[-k]}\stackrel{ab}{(-k)} = \stackrel{ab}{(-k)}.
\label{SQCgraphbinoms}
\end{eqnarray}
We recognize in  the first equation of the first row and the first equation of the second row
the demonstration of the nilpotent and the projector character of the Clifford algebra objects $\stackrel{ab}{(k)}$ and 
$\stackrel{ab}{[k]}$, respectively. 

{\em Whenever the Clifford algebra objects apply from the left hand side,
they always transform } $\stackrel{ab}{(k)}$ {\em to} $\stackrel{ab}{[-k]}$, {\em never to} $\stackrel{ab}{[k]}$,
{\em and similarly } $\stackrel{ab}{[k]}$ {\em to} $\stackrel{ab}{(-k)}$, {\em never to} $\stackrel{ab}{(k)}$.

According to ref.\cite{SQCholgernorma2002},  we define a vacuum state $|\psi_0>$ so that one finds
\begin{eqnarray}
< \;\stackrel{ab}{(k)}^{\dagger}
 \stackrel{ab}{(k)}\; > = 1, \nonumber\\
< \;\stackrel{ab}{[k]}^{\dagger}
 \stackrel{ab}{[k]}\; > = 1.
\label{graphherscal}
\end{eqnarray}

Taking the above equations into account it is easy to find a Weyl spinor irreducible representation
for $d$-dimensional space, with $d$ even or odd. (We advise the reader to see the reference\cite{SQCholgernorma2002}.) 

For $d$ even, we simply set the starting state as a product of $d/2$, let us say, only nilpotents 
$\stackrel{ab}{(k)}$, one for each $S^{ab}$ of the Cartan sub algebra  elements (Eq.(\ref{choicecartan})),  applying it 
on an (unimportant) vacuum state\cite{SQCholgernorma2002}. 
Then the generators $S^{ab}$, which do not belong 
to the Cartan sub algebra, applied to the starting state from the left hand side, 
 generate all the members of one
Weyl spinor.  
\begin{eqnarray}
\stackrel{0d}{(k_{0d})} \stackrel{12}{(k_{12})} \stackrel{35}{(k_{35})}\cdots \stackrel{d-1\;d-2}{(k_{d-1\;d-2})}
\psi_0 \nonumber\\
\stackrel{0d}{[-k_{0d}]} \stackrel{12}{[-k_{12}]} \stackrel{35}{(k_{35})}\cdots \stackrel{d-1\;d-2}{(k_{d-1\;d-2})}
\psi_0 \nonumber\\
\stackrel{0d}{[-k_{0d}]} \stackrel{12}{(k_{12})} \stackrel{35}{[-k_{35}]}\cdots \stackrel{d-1\;d-2}{(k_{d-1\;d-2})}
\psi_0 \nonumber\\
\vdots \nonumber\\
\stackrel{0d}{[-k_{0d}]} \stackrel{12}{(k_{12})} \stackrel{35}{(k_{35})}\cdots \stackrel{d-1\;d-2}{[-k_{d-1\;d-2}]}
\psi_0 \nonumber\\
\stackrel{od}{(k_{0d})} \stackrel{12}{[-k_{12}]} \stackrel{35}{[-k_{35}]}\cdots \stackrel{d-1\;d-2}{(k_{d-1\;d-2})}
\psi_0 \nonumber\\
\vdots 
\label{graphicd}
\end{eqnarray}

\section{Creation and annihilation operators for spinors}
\label{creation}

If $\hat{b}^{\dagger}_i$ is a creation operator, which creates a spinor state, when operating on a vacuum state
$|\psi_0>$ and $\hat{b}_i=(\hat{b}^{\dagger}_i)^{\dagger}$ is the 
corresponding  annihilation operator, then for a set of creation operators 
$\hat{b}^{\dagger}_i$ 
and the corresponding annihilation operators $\hat{b}_i$ 
it must be 
\begin{eqnarray}
\{\hat{b}_i,\hat{b}^{\dagger}_{j}\}_+|\psi_0>&=&\delta_{ij},\nonumber\\
\{\hat{b}^{\dagger}_i,\hat{b}^{\dagger}_{j}\}_+|\psi_0>&=&0,\nonumber\\ 
\{\hat{b}_{i},\hat{b}_{j}\}_+|\psi_0>&=& 0, \nonumber\\
\hat{b}^{\dagger}_{i}|\psi_0>&\ne& 0,\nonumber\\
 \hat{b}_{i}|\psi_0>&=& 0.
\label{bbplus}
\end{eqnarray}

We first shall pay attention on only the internal degrees of freedom - the spin. In this case, let 
$\hat{b}^{\dagger}_i$ be a creation operator, which creates one of the ($2^{d/2-1}$) Weyl basic states, when operating 
on a vacuum state
 and $\hat{b}_i=(\hat{b}^{\dagger}_i)^{\dagger}$ is the 
corresponding  annihilation operator.

Let us   make a choice of the starting state for a 
Weyl representation in $d=2(2n+1)$ 
dimensional space as follows
\begin{eqnarray}
\stackrel{03}{(+i)} \stackrel{12}{(+)} \stackrel{35}{(+)}\cdots \stackrel{d-1\;d}{(+)},
\label{start2(2n+1)}
\end{eqnarray}
so that it is made out of products of odd number of nilpotents ($(2n+1)$) and has 
accordingly an odd Clifford character. In the case of $d=4m$ 
the starting state will be made again of an odd number of nilpotents and of one projector
\begin{eqnarray}
\stackrel{03}{(+i)} \stackrel{12}{(+)} \stackrel{35}{(+)}\cdots \stackrel{d-3\;d-2}{(+)}\stackrel{d-1\;d}{[+]}.
\label{start4n}
\end{eqnarray}
Again the Clifford character of the starting state is odd. To simplify the notation in the above two equations and
making accordingly the whole presentation more transparent, we made a choice of the signature, although the signature 
could be any. The choice of 
the signature determines whether $(+i)$ or $(+)$ have to appear as nilpotents. For one time and $d-1$ space coordinates 
the only $(+i)$ appears in the first nilpotent factor.
We define
\begin{eqnarray}
\hat{b}^{\dagger}_1:&=& \stackrel{03}{(+i)} \stackrel{12}{(+)} \stackrel{35}{(+)}\cdots
\stackrel{d-1\;d}{(+)},\nonumber\\
\hat{b}_1:&=& \stackrel{d-1\;d}{(-)} \cdots \stackrel{35}{(-)} \stackrel{12}{(-)}
\stackrel{03}{(-i)},\quad {\rm for}\; d=2(2n+1),\nonumber\\
\hat{b}^{\dagger}_1:&=& 
\stackrel{03}{(+i)} \stackrel{12}{(+)} \stackrel{35}{(+)}\cdots \stackrel{d-3\;d-2}{(+)}\stackrel{d-1\;d}{[+]},
\nonumber\\
\hat{b}_1:&=&\stackrel{d-1,d}{[+]} \stackrel{d-2\;d-3}{(-)} \cdots \stackrel{35}{(-)} \stackrel{12}{(-)}
\stackrel{03}{(-i)},
\quad {\rm for}\; d=4m.
\label{bstart}
\end{eqnarray}
We must define also the vacuum state. We make a choice
\begin{eqnarray}
|\psi_0>&=& \stackrel{03}{[-i]} \stackrel{12}{[-]} \stackrel{35}{[-]}\cdots
\stackrel{d-1\;d}{[-]}|0>, \quad {\rm for}\; d=2(2n+1),\nonumber\\
|\psi_0>&=& \stackrel{03}{[-i]} \stackrel{12}{[-]} \stackrel{35}{[-]}\cdots
\stackrel{d-3\;d-2}{[-]}\stackrel{d-1\;d}{[+]}|0>, \quad {\rm for}\; d=4m.
\label{vac}
\end{eqnarray}

{\em Statement 1:} $(\hat{b}^{\dagger}_1)^2 =0 $ and $(\hat{b}_1)^2 =0 $.

{\em Proof:} The proof is self evident since a square of any nilpotent is zero.

{\em Statement 2:} $\hat{b}^{\dagger}_1|\psi_0> \ne 0 $ and $\hat{b}_1 |\psi_0> =0 $.

{\em Proof:} Since according to Eq.(\ref{SQCgraphbinoms}) $\stackrel{ab}{(+)} \stackrel{ab}{[-]} = \stackrel{ab}{(+)}$ 
and $\stackrel{ab}{[+]} \stackrel{ab}{[+]} = \stackrel{ab}{[+]}$, it follows that  
$\hat{b}^{\dagger}_1|\psi_0> \ne 0 $ for both kinds of dimensions ($d=2(2n+1)$ and $d=4m$). Since 
according to the same equation $\stackrel{ab}{(-)} \stackrel{ab}{[-]} = 0$, it follows that 
$\hat{b}_1|\psi_0> = 0 $ for both kinds of dimensions ($d=2(2n+1)$ and $d=4m$).

{\em Statement 3:} $\{\hat{b}_1,\hat{b}^{\dagger}_1\}_+ |\psi_0> =1. $ 

{\em Proof} $\hat{b}_1 |\psi_0> =0 $ while according to Eq.(\ref{graphherscal}) the state 
$<\psi_0|\hat{b}_1 \hat{b}^{\dagger}_1|\psi_0>$ can always be normalized to one.

For this particular one creation and one annihilation operator, creating and annihilating the staring state 
of one Weyl representation, we have proved that all the requirements from Eq.(\ref{bbplus}) are fulfilled.

All the states of one Weyl representation follow from the starting state by the application of $S^{ab}$, 
with $(a,b)$ which do not characterize the Cartan sub algebra set. To reach all the states of one Weyl 
representation from the starting state, 
at most  a product of $(d-2)/4$ for $d=2(2n+1)$ and $d/4$ for $d=4m$ ($m$ and $n$ are integers)  
different $S^{ab}$ have to be applied on a starting state.   
None of this $S^{ab}$ may belong to the Cartan sub algebra set.  
We accordingly have
\begin{eqnarray}
\hat{b}^{\dagger}_i &\propto & S^{ab} ..S^{ef} \hat{b}^{\dagger}_1,\nonumber\\
\hat{b}_i&\propto & \hat{b}_1 S^{ef}..S^{ab},
\label{bi}
\end{eqnarray}
with $S^{ab\dagger} = \eta^{aa} \eta^{bb} S^{ab}$.

{\em Statement 1.a.:} $(\hat{b}^{\dagger}_i)^2 =0 $ and $(\hat{b}_i)^2 =0 $, for all $i$.

{\em Proof:} To prove this statement one must recognize that $S^{ac}$ (or $S^{bc}$, $S^{ad}, S^{bd}$) 
transforms  $\stackrel{ab}{(+)} \stackrel{cd}{(+)}$ to $\stackrel{ab}{[-]} \stackrel{cd}{[-]}$, that is 
an even number of nilpotents $(+)$ in the starting state will be transformed into projectors $[-]$ in 
the case of $d=2(2n+1)$. For $d=4m$ it happens also that $S^{ac}$ (or $S^{bc}$, $S^{ad}, S^{bd}$) 
transforms  $\stackrel{ab}{(+)} \stackrel{cd}{[+]}$ into $\stackrel{ab}{[-]} \stackrel{cd}{(-)}$. 
Therefore for either 
 $d=2(2n+1)$ or  $d=4m$ at least one of factors, defining a particular creation operator, will be a nilpotent: for 
 $ d=2(2n+1)$ a nilpotent factor is just one (or odd number) of the nilpotents of the starting state, for $d=4m$ a 
 nilpotent factor can also be $\stackrel{d-1\;d}{(-)}$. A square of at least one  nilpotent factor (we may have odd
 number of nilpotents) is enough 
 to guarantee that the square of the corresponding $(\hat{b}^{\dagger}_{i})^2$ is zero. Since $\hat{b}_{i} = 
 (\hat{b}^{\dagger}_{i})^{\dagger} $, the proof is valid also for annihilation operators.
  
 {\em Statement 2.a.:} $\hat{b}^{\dagger}_i|\psi_0> \ne 0 $ and $\hat{b}_i |\psi_0> =0 ,$ for all $i$.
 
 {\em Proof:} One must recognize  that $\hat{b}^{\dagger}_i$ distinguishes from  $\hat{b}^{\dagger}_1$ in 
 (an even number of) those nilpotents $(+)$, which have been transformed into $[-]$. When $\stackrel{ab}{[-]}$ from 
 $\hat{b}^{\dagger}_i$ meets $\stackrel{ab}{[-]}$ from $|\psi_0>$, the product gives $\stackrel{ab}{[-]}$ back, 
 and accordingly nonzero contribution. And the Statement 2.a.  
 is proven for $d=2(2n+1)$. For $d=4m$ also the factor $\stackrel{d-1\;d}{[+]}$ can be transformed. 
 It is transformed into $\stackrel{d-1\;d}{(-)}$ which, when applied to a vacuum state, gives again 
 a nonzero contribution ($\stackrel{d-1\;d}{(-)} \stackrel{d-1\;d}{[+]} = \stackrel{d-1\;d}{(-)}$,
 Eq.(\ref{SQCgraphbinoms})).
 In the case of $\hat{b}_i$ we  recognize that in $\hat{b}^{\dagger}_i$ at least one factor is nilpotent; 
that of the same type as in the starting $\hat{b}^{\dagger}_1$ - $(+)$ - or in the case of $d=4m$ it can be also 
$\stackrel{d-1\;d}{(-)}$. Performing the Hermitean conjugation $(\hat{b}^{\dagger}_{i})^{\dagger}$,
$(+)$ transforms into $(-)$, while  $\stackrel{d-1\;d}{(-)}$ transforms into 
$\stackrel{d-1\;d}{(+)}$ in $\hat{b}_i$. 
Since $(-)[-]$ gives zero and $\stackrel{d-1\;d}{(+)} \stackrel{d-1\;d}{[+]}$ also gives zero, the statement 
is proven.

{\em Statement 2.b.:} $\{\hat{b}^{\dagger}_i,\hat{b}^{\dagger}_j\}_+  =0$, for each pair  $(i,j). $ 
 
 {\em Proof:} There are several possibilities, which we have to discuss. A trivial one is, if both
 $\hat{b}^{\dagger}_{i}$ and $\hat{b}^{\dagger}_{j}$ have a nilpotent factor (or more than one) for the 
 same pair of indices, say $\stackrel{kl}{(+)}$. Then the product of such two $\stackrel{kl}{(+)}\stackrel{kl}{(+)}$ 
gives zero. It also happens, that  $\hat{b}^{\dagger}_{i}$  has a nilpotent at the place $(ij)$ 
($\stackrel{03}{[-]}\cdots \stackrel{i j}{(+)} \cdots \stackrel{lm}{[-]}\cdots $) while $\hat{b}^{\dagger}_{j}$ 
has a nilpotent at the place $(l m)$ ($\stackrel{03}{[-]}\cdots \stackrel{i j}{[-]} \cdots 
\stackrel{lm}{(+)}\cdots $). Then in the term $\hat{b}^{\dagger}_{i} \hat{b}^{\dagger}_{j}$ the product  
$\stackrel{lm}{[-]} \stackrel{lm}{(+)}$ makes the term equal to zero, while in the term 
$\hat{b}^{\dagger}_{j} \hat{b}^{\dagger}_{i}$ the product  
$\stackrel{ij}{[-]} \stackrel{ij}{(+)}$ makes the term equal to zero. In the case that $d=4m$, it appears 
that $\hat{b}^{\dagger}_{i}  = \stackrel{03}{[-]}\cdots \stackrel{i j}{(+)} 
\cdots \stackrel{d-1\;d}{[+]}$ and $\hat{b}^{\dagger}_{j} = \stackrel{03}{[-]}
\cdots \stackrel{i j}{[-]} \cdots \stackrel{d-1\;d}{(-)}$. Then in  the term $\hat{b}^{\dagger}_{i} 
\hat{b}^{\dagger}_{j}$ the factor $\stackrel{d-1\;d}{[+]}\stackrel{d-1\;d}{(-)}$ makes it zero, while 
in $\hat{b}^{\dagger}_{j} 
\hat{b}^{\dagger}_{i}$ the factor $\stackrel{ij}{[-]}\stackrel{ij}{(+)}$ makes it zero. Since there are no  
further possibilities, the proof is complete.

{\em Statement 2.c.:} $\{\hat{b}_i,\hat{b}_j\}_+ =0, $ for each pair$(i,j)$. 
  
 {\em Proof:} The proof goes similarly as in the case with creation operators. Again we treat several possibilities.
$\hat{b}_{i}$ and $\hat{b}_{j}$ have a nilpotent factor (or more than one) with the same indices,  
 say $\stackrel{kl}{(-)}$. Then the product of such two $\stackrel{kl}{(-)}\stackrel{kl}{(-)}$ 
gives zero. It also happens, that  $\hat{b}_{i}$  has a nilpotent at the place $(ij)$ 
($\cdots \stackrel{lm}{[-]} \cdots \stackrel{ij}{(-)}\cdots \stackrel{03}{[-]} $) while $\hat{b}_{j}$ 
has a nilpotent at the place $(l m)$ ($\cdots \stackrel{lm}{(-)} \cdots 
\stackrel{ij}{[-]}\cdots \stackrel{03}{[-]}$). Then in the term $\hat{b}_{i} \hat{b}_{j}$ the product  
$\stackrel{ij}{(-)} \stackrel{ij}{[-]}$ makes the term equal to zero, while in the term 
$\hat{b}_{j} \hat{b}_{i}$ the product  
$\stackrel{lm}{(-)} \stackrel{lm}{[-]}$ makes the term equal to zero. In the case that $d=4m$, it appears 
that $\hat{b}_{i}  = \stackrel{d-1\;d}{[+]}\cdots \stackrel{ij}{(-)} 
\cdots \stackrel{03}{[-]}$ and $\hat{b}_{j} = \stackrel{d-1\;d}{(+)}
\cdots \stackrel{i j}{[-]} \cdots \stackrel{03}{[-]}$. Then in  the term $\hat{b}_{i} 
\hat{b}_{j}$ the factor $\stackrel{ij}{(-)}\stackrel{ij}{[-]}$ makes it zero, while 
in $\hat{b}_{j} 
\hat{b}_{i}$ the factor $\stackrel{d-1\;d}{(+)}\stackrel{d-1\;d}{[+]}$ makes it zero. 
The proof is thus complete.

{\em Statement 3.a.:} $\{\hat{b}_i,\hat{b}^{\dagger}_j\}_+ |\psi_0> = \delta_{ij}. $ 

We must recognize that $\hat{b}_{i} = \hat{b}_1 S^{ef}..S^{ab}$ and $\hat{b}^{\dagger}_{i} = 
S^{ab}..S^{ef} \hat{b}_1$. Since any $\hat{b}_{i}|\psi_0> =0$ (Statement 2.a.), we only have to 
treat the term $\hat{b}_{i} \hat{b}^{\dagger}_j$. We find $\hat{b}_{i} \hat{b}^{\dagger}_j \propto 
\cdots \stackrel{lm}{(-)}\cdots \stackrel{03}{(-)} S^{ef} \cdots S^{ab} S^{lm}\cdots S^{pr}
 \stackrel{03}{(+)}\cdots \stackrel{lm}{(+)} \cdots $. If we treat the term  $\hat{b}_{i} \hat{b}^{\dagger}_i$, 
generators  $S^{ef}\cdots S^{ab} S^{lm}\cdots S^{pr}$ are proportional to a number and we normalize 
$<\psi_0|\hat{b}_i \hat{b}^{\dagger}_i|\psi_0>$ to one.

When $S^{ef}\cdots S^{ab} S^{lm}\cdots S^{pr}$ are proportional to several products of $S^{cd}$, these 
generators change $\hat{b}^{\dagger}_{1}$ into $\stackrel{03}{(+)}\cdots \stackrel{kl}{[-]} \cdots 
\stackrel{np}{[-]} \cdots$, making the product $\hat{b}_i,\hat{b}^{\dagger}_j$ equal to zero, due to  factors 
of the type $\stackrel{kl}{(-)} \stackrel{kl}{[-]}$.  In the case of $d=4m$ also a factor $\stackrel{d-1\;d}{[+]} 
\stackrel{d-1\;d}{(-)}$ might occur, which also gives zero. And the proof is completed.

We proved accordingly that for the definition of the creation and annihilation operators in 
Eqs.(\ref{start2(2n+1)},\ref{start4n}) all the requirements of Eq.(\ref{bbplus}) are fulfilled. 
This could only be achieved for an odd Clifford character (which  means an odd number of nilpotents) 
of creation and annihilation operators. For an even number of factors of a nilpotent type in the
starting state and accordingly in the starting $\hat{b}^{\dagger}_1$, an annihilation operator $\hat{b}_{i}$ 
would appear with all factors of the type $[-]$, which on the vacuum state (Eq.(\ref{vac})) 
of an even Clifford character would not give zero. 

We treated up to now only the internal space of spins (which also could determine the charges\cite{SQCpikanorma2003}).
Spinor states, however, have also a part, originating in the coordinate space. This means that basic vectors 
have additional index, and so must have creation and annihilation operators
\begin{eqnarray}
|{}^{\alpha}\psi_i> = \hat{b}^{\alpha \dagger}_i|\psi_0>:&=& |\varphi^{\alpha}> \hat{b}^{\dagger}_i|\psi_0>, 
\label{xp}
\end{eqnarray}
which in the coordinate representation reads
\begin{eqnarray}
<x|{}^{\alpha}\psi_i> = <x|\hat{b}^{\alpha \dagger}_i|\psi_0>:&=&<x |\varphi^{\alpha}> \hat{b}^{\dagger}_i|\psi_0>. 
\label{xpx}
\end{eqnarray}
If we orthonormalize functions $|\varphi^{\alpha}> $  ($\int d^d x \; <\varphi^{\alpha}|x><x|\varphi^{\beta}> 
= \delta^{\alpha \beta} $), we can generalize Eq.(\ref{bbplus}) to
\begin{eqnarray}
\{\hat{b}^{\alpha}_i,\hat{b}^{\beta \dagger}_{j}\}_+|\psi_0>&=&\delta_{ij} \delta^{\alpha \beta},\nonumber\\
\{\hat{b}^{\alpha \dagger}_i,\hat{b}^{\beta \dagger}_{j}\}_+|\psi_0>&=&0,\nonumber\\ 
\{\hat{b}^{\alpha}_{i},\hat{b}^{\beta}_{j}\}_+|\psi_0>&=& 0, \nonumber\\
\hat{b}^{\alpha \dagger}_{i}|\psi_0>&\ne& 0,\nonumber\\
 \hat{b}^{\alpha}_{i}|\psi_0>&=& 0.
\label{bbplusgen}
\end{eqnarray}

\section{Conclusion}
\label{snmb-sqc1:conclusion}

In this contribution we have defined creation and annihilation operators for spinors in 
$d-$dimensional space in terms of factors of the 
Clifford algebra objects: nilpotents and projectors - ''eigen vectors'' of the Cartan sub algebra 
of the group $SO(q, d-q)$ - for a particular choice of the basis for one Weyl spinor representation. 
We let the creation and annihilation operators to include the part, determining (when operating on 
a vacuum state), which the spinor properties in the coordinate space. 
We proved that the creation and annihilation operators fulfil all the requirements, which fermionic
creation and annihilation operators must. Transformations into any other basis is straightforward.

The proposed presentation of creation and annihilation operators may help to clarify the nature 
of spinor-type objects. Since the charges of a spinor might follow from the spinor part, belonging to more 
than $(1+3)-$dimensional space, our way of defining the creation and annihilation operators would accordingly 
describe all the  internal degrees of spinors - spins and charges.

\section*{Acknowledgement}

This work was supported by the Ministry of Education, 
Science and Sport of Slovenia. 

\author{R. Mirman \thanks{E-mail:sssbb@cunyvm.cuny.edu}}
\title{Are there Interesting Problems That Could Increase Understanding of
Physics and Mathematics?}
\institute{%
14U\\
155 E 34 Street\\
New York, NY  10016}
\titlerunning{Are there Interesting Problems \ldots ?}
\authorrunning{R. Mirman}
\maketitle

\begin{abstract}
Recent developments in physics and mathematics, including group theory and
fundamental physics, suggest problems which might provide leads into useful
explorations of both physics and mathematics.
\end{abstract}

\section{Can infinite dimension lead to dimension 3+1?}\label{s1}

Suppose the dimension of space is infinite and the universe evolves in some
way to bring it down to 3+1, the only one in which physics is
possible~(\cite{gf}, chap.~7, p.~122; \cite{bnb}). Consider the following
problem. Conformal groups exist in every dimension. These have inhomogeneous
generators (including at least --- or only? --- one time translation, the
Hamiltonian in 3+1 space) which produce complex solutions. Because of
transversions they contain nonlinearities, as with general relativity. This
is exactly the same as for 3+1 space. Yet, by the argument for the
dimension, transformations acting on these must give inconsistencies. Why is
there a difference for 3+1 space? Don't the same arguments go through? This
might provide hints of how to break an infinite-dimensional space down to
3+1. If so it could be quite useful~(\cite{cnfr}. One possibility is that
these extra dimensions can appear as internal symmetries. Is this
mathematically possible? Is it physically possible?

\section{Group theory, geometry, particle physics}\label{s2}

Physical objects must be representations of the Poincar\'{e}
group~(\cite{ml}). This has two invariants. For spin-1/2, and that only,
these two equations can be replaced by a single one, Dirac's equation.
Interactions, properly, should be put in the momentum operators, and thus
into the invariants. The electromagnetic interaction however can be put into
Dirac's equation. Is this true for all interactions? Would they be different
if put into the equation than into the momentum operators? Could this place
restrictions on them? In n dimensions what is the generalization of the
Poincar\'{e} group? How many invariants does it have? Can these ever be
replaced by a single equation, equivalent to Dirac's equation? Are there
restrictions on interactions in general spaces? Interactions are
nonlinearities. Thus the group generators (equivalent to the angular
momentum operators for the rotation group) are nonlinear, that is the
operators themselves are functions of the solutions of the equations that
they give, as with gravitation. But nonlinearities may be limited since the
generators must obey the commutation relations. That could pick out certain
forms of interactions as the only ones allowable. More speculatively they
might put conditions on coupling constants. How is the generalization of the
Poincar\'{e} group embedded in the generalization of the conformal group for
these spaces? Can it be or is dimension 3+1 unique in this way also? Might
such embedding, if possible, give extra generators that can be interpreted
as internal symmetry ones? Might it put conditions like those on
interactions if any, including coupling constants?

Can some property of (generalized?) geometry be found that gives internal
symmetry? Why should SU(3), or is it su(3), arise? Is it the group or the
algebra that is relevant? Has experiment shown which is correct? Long ago
attempts were made to explain elementary particles in terms of
representations of su(6). With the present knowledge of particle physics
does this make sense? Does it fit experimental values? Can it be done? How
can su(6), consisting of internal su(3) and spin su(2), be interpreted? Can
it be~(\cite{cnfr}, sec.~IV.3.b.i, p.~199)? Might it lead anywhere?

\section{Physics of elementary particles}\label{s3}

It is an old problem to unify interactions. What does this mean? In
particular there are two massless objects, electromagnetism and gravitation.
Can they be cast into forms that make them seem more similar? Might they
both be (massless?) representations of some larger group (perhaps a
generalized conformal group)? Their interactions are very different. For one
the coupling constant is a pure number, for the other it is not. What is a
reasonable mass scale that can be used to make the gravitational coupling
constant into a pure number? For what mass would the two pure numbers for
these interactions be equal? Does that mass have physical significance? Is
it related to a (known) object? These question can also be asked for the
other interactions~(\cite{cnfr}, sec.~I.7.b.ii, p.~59; sec.~IV.3.d, p.~206).

Why are there both baryons and leptons? One possible way of looking at this
is that leptons are unaffected by strong interactions --- they are neutral
under these. Why? As an example perhaps providing clues consider
electromagnetism. Particles that can be gauge transformed are charged
(giving minimal coupling~(\cite{ml}, sec.~4.2, p.~57). But not all can be.
These are neutral. For leptons half the particles can, half cannot, perhaps
interestingly. A clue might be found looking at pions and nucleons. Here in
one isospin multiplet are particles that can, and ones that cannot, be gauge
transformed (and for pions transformed oppositely). Thus the isospin
operator not only changes the isospin ``z'' value, but also turns gauge
transformability on and off. What is the form of such an operator? How can
it be included in the set of group operators? Might the argument that the
proton can't decay~(\cite{cnfr}, sec.~IV-4, p.~212;~\cite{bnc}) suggest ways
of looking at interactions of leptons that can be used to analyze these
questions?

It is interesting that while hadrons are labeled by states of group
representations this does not seem true of leptons. Yet these also come in
pairs, something like the neutron and proton but with mass differences far
greater. And they are similar in their family structure. Is it possible to
find a set of algebra labels, or an algebra whose representations give such
labels, that include both hadrons and leptons, likely in different
representations, but both having the same family structure and with leptons
in pairs? Might there be a group with that Lie algebra? What would its
additional, beyond that of the algebra, significance be? The electromagnetic
interaction breaks isospin and SU(3), but in a definite way. It ``knows''
about these groups, although is not invariant under them. Does the weak
interaction also have ``knowledge'' of these groups, although perhaps that
is better hidden? Might it be possible to choose linear combinations of
hadron and lepton states so that they also are labeled by algebra (or group)
representation states such that the weak interaction breaks the symmetry in
a definite way? Thus leptons would be brought into a generalization of
SU(3). In that sense all particles would then be unified, except massless
ones. How would they fit in?

Why is symmetry broken? One reason might be incompatibility of
decompositions. If, for example, SU(6) were physically relevant we would
have to take the noncanonical decomposition, since its states belong to
SU(3) and SU(2). Were we to assume that there is an inhomogeneous SU(6),
ISU(6) --- the ``momenta'' forming a representation (not necessarily the
adjoint or defining representations) might be decomposed differently from
the simple part, perhaps canonically. Taking one of these as the Hamiltonian
we might get it transforming as a basis vector of a nonscalar
representation, thus breaking the symmetry, as does electromagnetism. These
give interesting mathematical questions, some perhaps relevant to physics.

The mass level formula~(\cite{cnfr}, app.~B, p.~246) for the elementary
particles clearly holds, but is quite mysterious. In particular it holds for
charged particles, less so for neutral ones (just the opposite of what is
expected). And it involves the fine-structure constant $\alpha$. Can a model
be developed to give such a formula? Is it compatible with the quark model,
or does it show that the quark model is wrong? This plus the GMO formula
gives two for particle masses. Why should there be two? How can they be
reconciled? Does this place restrictions on allowable mass values? It
clearly offers clues as to the underlying theory of elementary particles.
But while this is clear it is completely unclear what these clues are. Or is
it?

Geometry, especially through its transformations, expressed by group theory,
determines much about physics: the dimension, the allowed angular momentum
values, the spin-statistics theorem~(\cite{gf}, chap.~8, p.~146), the nature
of electromagnetism and gravitation~(\cite{ml};~\cite{bna}), the meaning of
gauge transformations, the impossibility of classical physics~(\cite{gf})
and the need, and nature of, quantum mechanics~(\cite{qm}), the reason for
interactions, for example. How far can this be pushed? Can geometry give all
fundamental laws of physics? How much information does the conformal group,
the largest invariance group of geometry (of 3+1 flat space), give? Can it
be extended to furnish further information, to place further restrictions on
physics? Are there larger relevant groups? Are there reasons for the
behavior of massive objects, as there are for massless ones~(\cite{ml})? How
much of the laws of physics come from geometry? All?

\section{Gravity}\label{s5}

It is well-known that general relativity must be the theory of gravitation,
thus the quantum theory of gravity~(\cite{ml}, chap.~11,
p.~183;~\cite{bna}). It is thus determined (but not the coupling constant
--- or is it two~(\cite{ml}, sec.~9.3.4, p.~161)?) except for the function
of the mass it couples to. This is the energy-momentum tensor. But what is
this? There are strong arguments for its forms~(\cite{ml}, sec.~9.2,
p.~153). However it is not clear that it must have these forms. Or is it? In
particular powers of the massive statefunctions (which are the vectors used
in the derivation of general relativity) might be possible (as well as
powers of the energy-momentum tensor). This could require new coupling
constants. Or is there something wrong with that? Extra coupling constants
can be introduced allowing more freedom, perhaps? Are there arguments that
rule these possibilities out so uniquely determine the function of matter
that gravity couples to? Presumably the function that the electromagnetic
potential couples to is completely determined (by minimal
coupling~(\cite{ml}, sec.~5.3.1, p.~81).

The theory of gravity is found by considering its effects on tensors, as
shown in all books on relativity. These tensors are of course
statefunctions~(\cite{ml}, sec.~4.2.7, p.~61). This raises the question of
whether gravity can act on, and be acted on by, scalar particles like the
pion and kaon. This has never been experimentally tested. Can it be? Might
these move freely in a gravitational field? Might the mass that produces a
field not include the masses of these? It is clear that the standard
derivation of the equations of gravity does not apply to these. How can they
be fitted into the formalism, or can they? Must they?

Massless Poincar\'{e} transformations give arbitrary functions of space
(gauge transformations~(\cite{ml}, sec.~3.4, p.~43)) requiring that
gravitation be nonlinear and that the electromagnetic potential be coupled
to matter, and with minimal coupling. The reasons for these interactions and
self-interactions are thus clear. Can these be extended to limit the
interaction of gravity with massive matter? Can the interactions of massive
objects (like protons and pions) be similarly explained? The conformal group
also includes arbitrary space-dependent transformations (transversions). In
a sense it thus also seems to lead to interactions. But there is no obvious
connection to real interactions. Can it be used to explain the actual
physical interactions? What information does it give about them?

\section{Understanding quantum mechanics}\label{s4}

The uncertainty principle refers to distributions, giving the spread of one
in terms of that of its conjugate one~(\cite{qm}, sec.~III.2.h, p.~112).
This is how it is derived. Yet it is often applied to a single experiment.
Is this ever correct? Can it be? Can its application to some single
experiments be shown to follow from its (correct) application to
distributions?

\section{Group theory}\label{grt}

For the conformal group the number of commuting operators is greater than
for SO(6) and SU(4), the complex forms of the algebras that are isomorphic
to that of the conformal group (but realized differently). For general
spaces how do the number of commuting operators compare? Is dimension 3+1
special in this regard?

Might noncanonical decompositions be relevant? The usual way of labeling
states is with a canonical decomposition~(\cite{sm}). Thus for SU(6) the
states are labeled by their SU(5) group representations, giving SU(6)
$\supset$ SU(5) $\times$ U(1), with SU(5) states labeled by their SU(4)
representations, so SU(6) $\supset$ SU(4) $\times$ U(1) $\times$ U(1) and so
on. But we can also use SU(6) $\supset$ SU(3) $\times$ SU(2), with this
(say) the internal symmetry and spin groups, plus other labels since these
representations occur several times in an SU(6) one. These give different
spectra, some of which might be relevant. For larger groups there is even
more, say with some subgroups labeled canonically, others noncanonically,
giving many choices so a great variety of spectra. This is relevant to the
intermediate boson model of nuclei. Might it be relevant to elementary
particles? Are there other areas in which it might be relevant?

Orthogonal (rotation) groups are defined as those keeping lengths and angles
of real lines constant. These can have complex parameters, not only real
angles, giving complex orthogonal groups CO(n,m), in particular the complex
Lorentz group CO(3,1)~(\cite{gf}, sec.~7.2, p.~124). It is relevant to
proving the TCP theorem, and also finding the dimension. These have not been
characterized, especially their representations. But they might have
important applications. For example they are relevant to the analytic
continuation of representations, uniting the representations of O(n,m), m+n
= r, for all n and m, that is all the real extensions of a single complex
group. It is interesting to wonder whether finite groups, like space
groups~(\cite{pt}, chap.~III, p.~132) can be embedded in them, thus uniting
into one larger (continuous) group space groups (and of different space
dimensions). Their representations could be relevant to, for example,
crystallography. Also they could be relevant to particle physics. They are
likely to be relevant to nuclear physics, especially intermediate boson type
models. ICO(n,m) groups might also be explored, with the Poincar\'{e} group
as a subgroup, and perhaps the conformal group can also be in some way. The
question of extending the conformal group to one with complex parameters
might be looked at.

For the rotation group generators there is an internal part, giving spin.
This then must also be true for the Lorentz group. What is the (physical)
meaning of the additional generators? The largest invariance group of space
is the conformal group. Internal parts can also be added to its
generators~(\cite{cnfr}, sec.~III.1.e, p.~115). What (physical) meaning
might they have?

In considering these questions it is essential to realize the richness of
group theory, which is not at all recognized. The form of representation
states (usually the physical objects) depends in fundamental ways on how
operators are realized~(\cite{ia}, sec.~V.3.c, p.~157), that is which
variables they are functions of. We are mostly familiar with semisimple
(especially compact) groups, realized only one way. This much restricts our
thinking. The Lorentz group has several different realizations, that have
been given~(\cite{cnfr} app.~A, p.~223), and perhaps there are more. Its
representations and states have quite different forms for these
realizations. The number of commuting generators (labeling operators)
depends not only on the realization but also on which are chosen diagonal.
Thus ISO(2), the inhomogeneous rotation group on two dimensions has either
one commuting generator, $L_z$, if that is chosen diagonal, or two, the
momenta, if those are chosen. The conformal group, abstractly isomorphic to
SO(6) and to SU(4), the transformation groups on 6 real coordinates and 4
complex ones, has its algebra realized over 4 real dimensions (these are
functions of 4 real coordinates)~(\cite{cnfr}, sec.~III.5.b, p.~162). These
then give nonlinear generators (transversions) and 4 commuting operators
(momenta), although SO(6) and SU(4) have 3, when realized over their
defining spaces~(\cite{ia}, sec.~XIV.1.c, p.~404). That is why the conformal
group can contain the Poincar\'{e} group as a subgroup. Even SU(2) can be
realized this way, with a nonlinear (transversion) generator and a momentum
one~(\cite{cnfr}, sec.~III.3, p.~120). This is true in general, with greater
richness for larger groups (even mixed realizations, with the variables on
which the generators depend possibly different for different generators, so
different choices of the number of labels). Even the most elementary groups
are richer than generally known. But this should be known. Inhomogeneous
groups, like the Poincar\'{e} group, add richness. For example consider the
inhomogeneous ISO(6) group. Suppose that it was realized over a space of 4
real dimensions (an inhomogeneous conformal group). Can it be? There would
then be a large choice of diagonal generators (labeling operators). What
would the representations be like? These groups, like the conformal group
and its inhomogeneous generalizations (there are an infinite number since
the momenta can form any representation of the homogeneous part), can have
internal generators~(\cite{cnfr}, sec.~III.1.e, p.~115), allowing more types
of representations and states. We can also consider groups (and algebras?)
with nonsymmorphic transformations (glide translations and screw
rotations~(\cite{pt}, sec.~III.4, p.~141), and perhaps generalizations if
for example the transformations of the conformal group can be used to
generalize these). It might be possible to have ``internal spaces'' (like
SU(3)?) with these. Might they have different, additional, more complicated,
spectra? Could some correspond to particle states? Further we can consider
topology. For inhomogeneous groups, even for ISO(2), rotations in
2-dimensions (or equivalently inhomogeneous unitary groups), we assume that
the transformations are in a plane, giving for the states expressions of the
form $exp(ikx)$. But suppose instead of a plane we represent the group over
a cylinder, a saddle, a torus, if possible, or generalizations for higher
dimensions. What would the states be like? Suppose we cut the cylinder along
a line. That would make variables like k discrete. We can now continue to
larger dimensions, giving representations over hyperspheres, Klein bottles,
hyperbolic spaces, ones with several holes and so on. It is not clear which,
if any, could have physical significance, but there is reason to believe
extra richness is needed (as it certainly is if such groups can explain
elementary particles), and these are ways of getting it. They will almost
certainly give interesting special functions and undoubtedly much
interesting mathematics. These are among the problems worth considering, and
in addition they may lead somewhere (physically!).

\section{Conclusions}\label{cn}

Group theory and geometry suggest many problems in pure mathematics and
fundamental physics. There are many other fundamental questions, including
ones on cosmology, astrophysics and particle physics, that are raised
elsewhere~(\cite{imp}.

\section*{Acknowledgements}

This discussion could not have existed without Norma Mankoc Borstnik.


\hyphenation{Sch-warz-schild}

\title{Noncommutative Nonsingular Black Holes}
\author{P. Nicolini\thanks{%
e-mail: Piero.Nicolini@cmfd.univ.trieste.it}}
\institute{%
Department of Theoretical Physics,
Josef Stefan Institute, Ljubljana, Slovenia\thanks{%
Department  of Mathematics, Polytechnic of Turin,
Turin, Italy},\thanks{Department  of Mathematics and
Informatics, University of Trieste, Trieste, Italy},
\thanks{INFN, National Institute for Nuclear Physics,
Trieste, Italy}}

\titlerunning{Noncommutative Nonsingular Black Holes}
\authorrunning{P. Nicolini}
\maketitle

\begin{abstract}
Adopting noncommutative spacetime coordinates, we determined a new
solution of Einstein equations for a static, spherically symmetric
matter source.  The limitations of the conventional Schwarzschild
solution, due to curvature singularities, are overcome. As a
result, the line element is endowed of a regular DeSitter core at
the origin and of two horizons even in the considered case of
electrically neutral, nonrotating matter. Regarding the Hawking
evaporation process, the intriguing new feature is that the black
hole is allowed to reach only a finite \textit{maximum}
temperature, before cooling down to an \textit{absolute zero}
extremal state. As a consequence the quantum back reaction is
negligible.
\end{abstract}

\section{Introduction}
In spite of  decades of efforts, a complete and satisfactory
quantum theory of Gravity does not yet exist. Thus a great
interest has arisen   towards the class of model theories able to
reproduce quantum gravitational effects, at least in some limit.
Quantum Field Theory on a noncommutative manifold or shortly
Noncommutative Field Theory (NFT) belongs to such class of model
theories. Indeed we retained NFT the low energy limit of String
Theory, which is the most promising candidate to be the quantum
theory of Gravity.

The starting point of the NFT is the adoption of a noncommutative
geometry, namely a manifold whose coordinates may fail to commute
in analogy to the conventional noncommutativity among conjugate
variables in quantum mechanics \begin{equation} \left[
\textbf{x}^i,\textbf{x}^j\right] = i\
\theta^{ij}\,\,\,\,\,\,\,i,j=1,...,n \label{ncrules} \end{equation}
 with
$\theta^{ij}$ an antisymmetric (constant) tensor of dimension $($
length $)^2$. Eq. (\ref{ncrules}) provides an uncertainty in any
measurement of the position of a point on the noncommutative
manifold.  Indeed we cannot speak of point anymore but rather of
delocalized positions according to the noncommutative uncertainty.
The physical motivation for assuming a noncommutative geometry
relies on the bad short distance behavior of field theories,
gravitation included. In fact this is a typical feature of
theories dealing with point like objects, a problem that has not
been completely solved by String Theory too. NFT could provide the
solution, since a noncommutative manifold is endowed of a natural
cut off due to the position uncertainty. This aspect is in
agreement with the long held belief that spacetime must change its
nature at distances comparable to the Planck scale. Quantum
Gravity has an uncertainty principle which prevents one from
measuring positions to better accuracies than the Planck length,
indeed the shortest physically meaningful length. In spite of this
promising programme and the issue of a seminal paper dated in the
early times \cite{sny}, the interest towards NFT is rather recent.
Indeed a significant push forward was given only when, in the
context of String Theory, it has been shown that target spacetime
coordinates become \textit{noncommuting} operators on a $D$-brane
\cite{sw}. This feature promoted the interpretation of NFT, among
the class of nonlocal field theories \cite{Pais:1950za}, as the
low energy limit of the theory of open strings.

The inclusion of noncommutativity in field theory in flat space is
the subject of a large literature. On the contrary, the purpose of
the paper is to introduce  noncommutativity effects in the
gravitational field, with the hope that noncommutativity could
solve the long dated problems of curvature singularities in
General Relativity.  This investigation is motivated by some
mysterious feature of the physics of quantum black holes. Indeed
the interest towards their complete understanding have been
increasing since the remarkable Hawking discovery about the
possibility for them to emit radiation \cite{hawking}. The general
formalism employed is known as quantum field theory in curved
space \cite{bd}. Such a formalism provides, in terms of a quantum
stress tensor $\left<T_{\mu\nu}\right>$ \cite{swave}, a
satisfactory description of black holes evaporation until the
graviton density is small with respect to matter field quanta
density. In other words quantum geometrical effects have to be
retained negligible, a condition that is no more valid in the
terminal stage of the evaporation. The application of
noncommutativity to gravity could provide, in an effective way,
the still missing description of black holes in those extreme
regimes, where stringy effects are
 considered relevant \cite{corr}.

\section{Noncommutative field theory models}

There exist many formulations of NFT, based on different ways of
implementing non local deformations in field theories, starting
from (\ref{ncrules}). The most popular approach is founded on the
replacement of the point-wise multiplication of fields in the
Lagrangian by a non-local Weyl-Wigner-Moyal $\ast$-product
\cite{Moyal}. In spite of its mathematical exactitude, the
$\ast$-product NFT suffers non trivial limitations. The Feynman
rules, obtained directly from the classical action, lead to
unchanged propagators, while the only modifications, concerning
vertex contributions, are responsible of the non-unitarity of the
theory and of UV/IR mixing. In other words UV divergences are not
cured but accompanied by surprisingly emerging IR ones. While
unitariety can be restored, the restriction of noncommutative
corrections only to interaction terms is a  non intuitive feature,
which appears in alternative formulations too \cite{Bahns:2002vm}.

Against this background, the coordinate coherent states approach,
based on an oscillator representation of noncommutative spacetime,
leads to a UV finite, unitary and Lorentz invariant field theory
\cite{ae}. The starting point of this formulation is to promote
the relation (\ref{ncrules}) to an equation between Lorentz
tensors \begin{equation} \left[ \textbf{x}^\mu,\textbf{x}^\nu \right] = i\
\theta^{\mu\nu}\hspace{1.5cm}\mu,\ \nu = 0 ,...,n \label{ncx}
 \end{equation}
Thus $\theta^{\mu\nu}$, now an antisymmetric Lorentz tensor, can
be represented in terms of a block diagonal form
\begin{equation}
\theta^{\mu\nu}= diag \left( \hat{\theta }_1 ,\hat{ \theta } _2
,..., \hat{ \theta } _{n/2} \right)\hspace{1.5cm}\mu,\ \nu = 0
,...,n \label{diag}
\end{equation}
where $\hat{ \theta } _i =\theta_i \left(
\begin{array}{cc}
0 & 1 \\
-1 & 0%
\end{array}%
\right) $. In case of odd dimensional manifold the last term on
the diagonal is zero. In other words, for the covariance of
(\ref{ncx}) we provide a foliation of spacetime into
noncommutative planes, defined by (\ref{diag}). There is a
condition to be satisfied: field theory Lorentz invariance and
unitarity implies that noncommutativity does not privilege any of
such planes, namely there is a unique noncommutative parameter
$\theta_1=\theta_2=...\ \theta_{n/2}=\theta$.

The other key ingredient of this approach is the interpretation of
conventional coordinates as mean values of coordinate operators
subject to (\ref{ncrules}), to take into account the quantum
geometrical fluctuations of the spacetime manifold. Mean values
are calculated over coherent states, which results eigenstates of
ladder operator, built with noncommutative coordinates only. In
the absence of common eigenstates, the choice of coherent states
is motivated by the fact that  they result the states of minimal
uncertainty and provide the best resolution of the position over a
noncommutative manifold. In other words, the effective outcome is
the loose of the concept of point in favor of smeared position on
the manifold. At actual level, the delocalization of fields is
realized by deforming the source term of their equations of
motion, namely substituting Dirac delta distribution (local
source) with Gaussian distribution (nonlocal source) of width
$\sqrt{\theta}$. As a result, the ultraviolet behavior of
classical and quantum fields  is cured \cite{pnag}.

\section{The noncommutative black hole}

To provide a black hole description by means of a noncommutative
manifold, one should know how to deal with the corresponding
gravity field equations. Fortunately, this nontrivial problem can
be circumvented by a noncommutative deformation of only the matter
source term, leaving unchanged the Einstein tensor. This
procedure, already followed in $(1+1)$ dimensions
\cite{Nicolini:2005de} and $(3+1)$ linearized General Relativity
\cite{Nicolini:2005zi}, is in agreement with the general
prescription to obtain nonlocal field theories from
noncommutativity \cite{ae}. This line of reasoning is supported by
the following motivations. Noncommutativity is an intrinsic
property of a manifold and affects matter and energy distribution,
by smearing point-like objects, also in the absence of curvature.
On the other hand, the metric is a geometrical device defined over
the underlying manifold, while curvature measure the metric
intensity, as a response to the presence of mass and energy
distribution. Being the energy-momentum tensor the tool which
gives the information about the mass and energy distribution, we
conclude that, in General Relativity, noncommutativity can be
taken into account by keeping the standard form of the Einstein
tensor in the l.h.s. of the field equations and introducing a
modified source term in the r.h.s..

Therefore we assume, as mass density of the noncommutative
delocalized particle, the Gaussian function of minimal width
$\sqrt{\theta}$
\begin{equation}
\rho_\theta\left(\,\vec{x}\,\right)=
\frac{M}{\left(\,4\pi\theta\,\right)^{3/2}}\,
\exp\left(-\vec{x}^{\,2}/4\theta\,\right)
 \label{t00}
 \end{equation}
Thus the particle mass $M$ is \textit{diffused} throughout a
region of linear size $\sqrt{\theta}$ taking into account the
intrinsic uncertainty encoded in the coordinate commutator
(\ref{ncx}). The distribution function
$\rho_\theta\left(\,\vec{x}\,\right)$ is static, spherically
symmetric and exponentially vanishing at distances $r>>
\sqrt\theta$. In this limit $\rho_\theta\left(\,\vec{x}\,\right)$
reproduces point-like sources and leads to the conventional
Schwarzschild solution.
On these grounds, we are looking for a static, spherically
symmetric,  asymptotically Schwarzschild solution of Einstein
equations with $T^0_{\,\, 0} = \rho_\theta\left(\,\vec{x}\,\right
)$ as source . There are two further conditions to be taken: the
covariant conservation of the energy-momentum tensor $\nabla_\nu\,
T^{\mu\nu}=0$ and $g_{00}=-g_{rr}^{-1}$ to preserve a
Schwarzschild-like property. Therefore the solution of Einstein
equations is\footnote[1]{We use convenient units $G_N=1$, $c=1$.}:
\begin{equation}
 ds^2 =\left(\, 1- \frac{4M}{r\sqrt{\pi}}\, \gamma \right)\, dt^2 - \left(\,
1-\frac{4M}{r\sqrt{\pi}}\, \gamma \right)^{-1}\, dr^2 -
r^2\,d\Omega^2\label{ncs}
\end{equation}
where $\gamma\equiv\gamma\left(3/2 \ , r^2/4\theta\,
\right)=\int_0^{r^2/4\theta} dt\, t^{1/2} e^{-t} $ is the lower
incomplete Gamma function. This line element describes the
geometry of a noncommutative black hole  and should give us useful
insights about  possible noncommutative effects on Hawking
radiation.
\begin{figure}[h]
\begin{center}
\includegraphics[totalheight=6cm,angle=0]{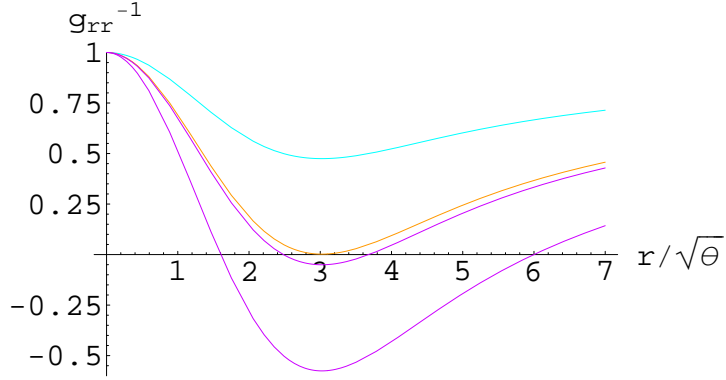}
\caption{\label{hor} $g_{rr}^{-1}$ vs $r$, for various values of
$M/\sqrt\theta$. Intercepts on the horizontal axis give radii of
the event horizons. $M= \sqrt\theta $, ( cyan curve ) no horizon;
$M=1.9\, \sqrt\theta $, (yellow curve) one \textit{degenerate}
horizon $r_0\approx 3.0\, \sqrt\theta$, \textit{extremal}
 black hole;  $M= 3\, \sqrt\theta $ (magenta curve) two horizons.}
\end{center}
\end{figure}
Let's start our analysis from the presence of eventual event
horizons. Since in our case, the equation $g_{00}\left(\,
r_H\,\right)=0$ cannot be solved in closed form, one can
numerically  determine their radius  by plotting $g_{00}$. Figure
(\ref{hor}) shows that noncommutativity introduces a new behavior
with respect to standard Schwarzschild black hole. Instead of a
single event horizon, there are different possibilities: (a) two
distinct horizons  for $M> M_0$ (yellow curve); (b) one degenerate
horizon  in  $r_0=3.0\times \sqrt\theta $, with
 $M= M_0=1.9\times (\sqrt\theta) /G $
 corresponding to \textit{cyan extremal black hole} (cyan curve);
(c) no horizon  for  $M< M_0$ (violet curve).
In view of these results, there can be no black hole if the
original mass is less than the \textit{minimal mass} $M_0$.
Furthermore,  contrary to the usual case, there can be \textit{two
horizons} for large masses. By increasing $M$, i.e. for $M>> M_0$,
the  \textit{inner  horizon} shrinks to zero, while the
\textit{outer} one approaches the Schwarzschild value $r_H=2M$.

For what concerns the covariant conservation of $T^{\mu\nu}$, one
finds that such requirement leads to
\begin{equation}
T^\theta{}_\theta\equiv \partial_r\left(\, r T^r{}_r\,\right)= -
\rho_\theta\left(\,r\,\right)-
r\,\partial_r\rho_\theta\left(\,r\,\right).
\end{equation}
The emerging picture is that of a \textit{self-gravitating},
\textit{droplet } of \textit{anisotropic fluid} of density
$\rho_\theta$, radial pressure $p_r= -\rho_\theta$ and
\textit{tangential pressure} $ p_\perp = -\rho_\theta
-r\,\partial_r\rho_\theta\left(\,r\,\right)$. We are not dealing
with a massive, structure-less point. Thus results reasonable that
a non-vanishing radial pressure balances the inward gravitational
pull, preventing the droplet to collapse into a matter point. This
is the basic physical effect on matter caused by spacetime
noncommutativity and the origin of all new physics at distance
scale of order  $\sqrt\theta$. Regarding the physical
interpretation of the pressure, we underline that it does not
correspond to the inward pressure of outer layers of matter on the
core of a ``star'', but to a totally different quantity of
``quantum'' nature. It is the outward push, which is
conventionally defined to be negative, induced by noncommuting
coordinate quantum fluctuations. In a simplified  picture,  such a
quantum pressure is the relative of the  cosmological constant  in
DeSitter universe. As a consistency check of this interpretation
 we are going to show that line element (\ref{ncs}) is well described near the
 origin by a DeSitter geometry.

\begin{figure}[h]
\begin{center}
 \includegraphics[totalheight=6cm, angle=0 ]{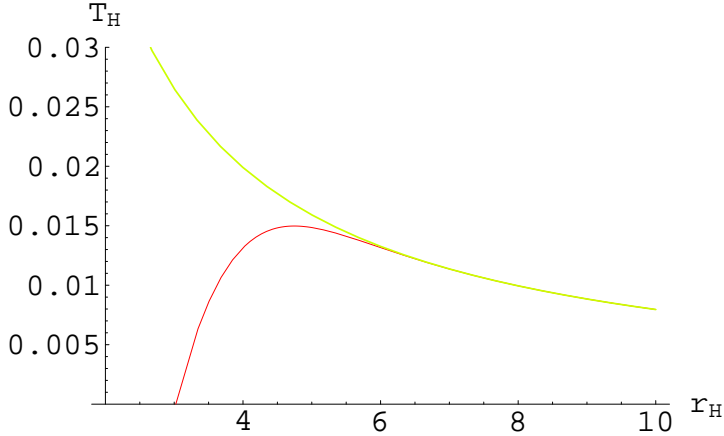}
\end{center}
\caption{\label{T} Plot of $T_H$ vs $r_H$, in $\sqrt\theta$ units.
yellow curve is the plot of (\ref{thnc}): $T_H=0$ for
$r_H=r_0=3.0\sqrt{\theta}$,i.e. for the extremal black hole,
 while the maximum temperature $T_H\simeq 0.015\times
1/\sqrt{\theta}$ corresponds to a mass $M\simeq 2.4
\times\sqrt{\theta}$. For comparison, we plotted in yellow the
standard Hawking temperature.  The two temperatures coincide for
$r_H > 6\,\sqrt\theta$.}
\end{figure}

Let us now consider the black hole temperature $T_H\equiv
\left(\,\frac{1}{4\pi}\, \frac{d g_{00}}{dr}\right)_{r=r_H} $:
\begin{equation}
T_H = \frac{1}{4\pi\,r_H}\left[\, 1
 -\frac{r^3_H}{4\,\theta^{3/2}}\,
  \frac{e^{-r^2_H/4\theta}}{\gamma\left(\, 3/2\ ; r^2_H/4\theta \right)}
\,\right] \label{thnc}
\end{equation}
For large black holes, i.e. $r_H^2/4\theta>>1$, one recovers the
standard result for the Hawking temperature $ T_H= \frac{1}{4\pi\,
r_H} \label{th}$. At the initial state of radiation the black hole
temperature increases while the horizon radius is decreasing. It
is crucial to investigate what happens as $r_H\to \sqrt\theta$. In
the standard (~commutative~) case
  $T_H$ diverges and this puts limit on the
validity of the conventional description of Hawking radiation.
Against this scenario, temperature (\ref{thnc}) includes
noncommutative effects which are relevant at distances comparable
to  $\sqrt\theta$. Behavior of the temperature $T_H$ as a function
of the horizon radius is plotted in Fig.(\ref{T}).
In the region $r_H\simeq \sqrt\theta $, $T_H$ deviates from the
standard hyperbola. Instead of exploding with shrinking $r_H$,
$T_H$ reaches a maximum
in $r_H\simeq 4.7\sqrt\theta $
corresponding to a mass $M\approx 2.4\, \sqrt\theta/ G_N$, then
quickly drops to zero for $r_H=r_0=3.0\sqrt{\theta}$ corresponding
to the radius of the extremal black hole in figure (\ref{hor}). In
the region $r< r_0 $ there is no black hole  and the corresponding
temperature cannot be defined. As a summary of the results, the
emerging picture of non commutative black hole is that for $M >>
M_0$ the temperature is given by the Hawking temperature
(\ref{th}) with negligibly small exponential corrections, and
increases, as the mass is radiated away. $T_H$ reaches a maximum
value at $M= 2.4\, \sqrt\theta$ and then drops  as $M$ approaches
$M_0$. When $M=M_0$, $T_H=0$, event horizon is degenerate, and we
are left with a ``frozen'' extremal black hole.

At this point, important issue of Hawking radiation back-reaction
should be discussed. In commutative case one expects relevant
back-reaction effects during the terminal stage of evaporation
because of huge increase of temperature \cite{swave,backr}. As it
has been shown, the role of noncommutativity is to cool down the
black hole in the final stage. As a consequence, there is a
suppression of quantum back-reaction since the black hole emits
less and less energy. Eventually, back-reaction may be important
during the maximum temperature phase. In order to estimate its
importance in this region, let us look at the thermal energy
$E=T_H\simeq 0.015\, /\sqrt{\theta}$ and the total mass $M\simeq
2.4\,\sqrt{\theta}\,M_{Pl.}^2  $. In order to have significant
back-reaction effect $ T_H^{Max}$ should be of the same order of
magnitude as $M$. This condition leads to the estimate
$\sqrt{\theta}\approx 0.2\, l_{Pl.}\sim 10^{-34}\, cm
\label{stima}$. Expected values of $\sqrt{\theta}$ are well above
the Planck length $l_{Pl.}$, while the back-reaction effects are
suppressed even if $\sqrt{\theta}\approx 10\, l_{Pl.}$ and
$T_H^{Max}\approx 10^{16}\, GeV$.  For this reason we can safely
use unmodified form of the metric (\ref{ncs}) during all the
evaporation process.

 Finally, we would like to clarify what
happens if the starting object has mass smaller than $M_0$, with
particular attention to the eventual presence of  a \textit{naked
singularity}. To this purpose we are going to study the curvature
scalar near $r=0$. The short distance behavior of $R$ is given by
\begin{equation}
 R\left(\, 0\,\right)=\frac{4M}{\sqrt\pi\, \theta^{3/2}}
  \label{ricci0}
\end{equation}
For $r<<\sqrt\theta$ the curvature is actually \textit{constant
and positive}. Thus, an eventual naked singularity is replaced
 by a DeSitter, \textit{regular} geometry around the origin.  Earlier attempts
 to avoid the curvature singularity at the origin of
 Schwarzschild metric have been made by matching DeSitter
 and Schwarzschild geometries both along time-like \cite{olds}, and
space-like matter shells \cite{dscore}, or constructing regular
 black hole geometries by-hand \cite{regbh} or continuosly deforming dilaton 2d models \cite{dejan}. In our approach, it is
 noncommutativity that induces a smooth and continuous transition between
 the two geometries.

 \section{Concluding remarks}
The above results show that the coordinate coherent state approach
to noncommutative effects can cure the singularity problems at the
terminal stage of black hole evaporation.

In particular we have shown that noncommutativity, being an
intrinsic property of the manifold itself, can be introduced in
General Relativity by modifying the matter source. The
Energy-momentum required for this description is of form of the
ideal fluid, although a non-trivial pressure is invoked. In spite
of complicated equation of state it can be studied in the regions
of interest and new black hole behavior is discovered in the
region $r\simeq \sqrt\theta$. Specifically, we have shown that
there is a minimal mass $M_0= 1.9\, \sqrt{\theta}$ to which a
black hole can decay through Hawking radiation.  The reason why it
does not end up into a naked singularity is due to the finiteness
of the  curvature at the origin. The everywhere regular  geometry
and the residual  mass $M_0$ are both manifestations of the
Gaussian delocalization of the source in the noncommutative
spacetime. On the thermodynamic side,  the same kind of
regularization  takes place eliminating the divergent behavior of
Hawking temperature. As a consequence there is a maximum
temperature that the black hole can reach before cooling down to
absolute zero. As already anticipated in the introduction,
noncommutativity regularizes divergent quantities in the final
stage of black hole evaporation in the same way it cured UV
infinities in noncommutative quantum field theory. We have also
estimated that back-reaction does not modify the original metric
in a significant manner. 

\section*{{Acknowledgements}}
The author thanks the ``Dipartimento di Fisica Teorica
dell'Universit\`a di Trieste'', the PRIN 2004 programme ``Modelli
della teoria cinetica matematica nello studio dei sistemi
complessi nelle scienze applicate'' and the CNR-NATO programme for
financial support.

\title{Compactified Time and Likely Entropy\,%
---\,World Inside Time Machine: Closed Time-like Curve\thanks{%
Preprints YITP-05-43; OIQP-05-09}}
\author{H.B. Nielsen${}^1$ and M. Ninomiya${}^2$\thanks{%
Also at
 {Okayama Institute for Quantum Physics, Kyoyama-cho 1-9, 
 Okayama City 700-0015, Japan.}}}
\institute{%
${}^1$ Niels Bohr Institute, Blegdamsvej 17, Copenhagen, Denmark\\
${}^2$ Yukawa Institute for Theoretical Physics,
Kyoto University, Kyoto 606-8502, Japan}

\titlerunning{Compactified Time and Likely Entropy \ldots}
\authorrunning{H.B. Nielsen and M. Ninomiya}
\maketitle

\begin{abstract}
If a macroscopic (random) classical system is put into a random state in phase space, 
it will of course the most likely have an almost maximal entropy
according to second law of thermodynamics.
We will show, however, the following theorem: 
If it is enforced to be periodic with a given period $T$ in advance, 
the distribution of the entropy for the otherwise random state will be much more smoothed out, 
and the entropy could be very likely much smaller than the maximal one. 
Even quantum mechanically we can understand that such a lower than maximal entropy is likely. 
A corollary turns out to be that the entropy in such closed 
time-like loop worlds remain constant.
\end{abstract}

\section{Introduction}
In the present article we shall study a world or a (classical) 
mechanical system 
situated on a closed time-like loop, 
or we can simply consider a model of a universe intrinsically periodic.
This is a study that can be considered a simple exercise playing with ideas 
that could be of relevance for making a (speculative) model 
behind the second law of thermodynamics
\cite{MN1}\cite{MN2}\cite{MN3}\cite{MN3}\cite{MN4}.

One problem with combining the second law with time reversal symmetry, 
even if we hope for global features making $\dot{S}\ge 0$ in some era, 
is that then there is  nothing prevent that in an other era 
-- it is at least logically possible -- we then have $\dot{S}\le 0$.
That in turn means that there are some restrictions known not only about the past 
but also about the future.
Such rules about the future will function in a sense as time machine
\cite{MN5}.

At least we can defend that there is the possibility of ``time machine effects" 
in a model with such regularities about the future 
-- regularities known from a mysterious, presumably not valid law of nature -- 
in the sense that one could
obtain what we should accept as messages from the future.
We could ask:  Why is it 
that we normally do \underline{not} get messages from the future ?

Indeed the point is that a tiny amount of information from or about the past 
can allow us to make great deductions about the past because the past,
which had low entropy, 
is very ordered.
So from a little knowledge we can deduce a lot.
Concerning the future on the other hand we cannot analogously trust the regularities.
So even if we know a tiny amount of photons,
 say to develop into the future 
we cannot count on that they shall be involved with matter.
They can just remain all through the future 
without getting associated with 
any other matters in the future.
We cannot get messages from the future 
because we normally cannot conclude by use of regularity knowledge about the future 
from a tiny amount of information.
For that the future can too easily be irregular.
In our usual world picture (based on second law we might say) 
there is no regularity in the future except for the rudiments 
of regularity
left over from the past.
Therefore we cannot conclude anything great about the future from tiny informations 
like a little bit of light going into future.
Even if we come to know that such a little bit of light runs into the future 
we do not accept that as a message \underline{from the future}, 
because we cannot use that to make further significant deductions about the future.
When we, however, get a letter or a fax, 
sent in the past we can use its content to  
conclude a lot about the past with reasonable reliability.

We want to stress that it is our knowledge about a relatively huge regularity 
in the past that makes it possible for us to consider small letters or light beams 
from the past as messages.

\underline{If} however in some strange violation of the strict second law 
we came to know some regularity about the future, 
then we could begin to accept small pieces of light or paper with text on to give us 
messages about the future.
We would 
then conclude that they would have to go into 
that pattern of regularities, which we had come to know as law of nature.
If we for instance knew that there in some future should exist some very separate 
rather small hot places (time inverted stars) 
that would be  more and more hot by absorbing light, 
then knowing that some light goes out in a certain direction 
will imply such time inverted star(s) in that direction.
Then we could claim it were a message about such a time inverted star.

In any case we argue that because of the conflict 
between time reversal invariance and the second law, 
we can hardly imagine any spontaneous breakdown model 
behind second law providing us with say on era with $\dot{S}\ge 0$ 
unless we also have at least the possibility 
for also messages going the opposite way in time.
So really models that could even have the slightest chance of 
a $T$-invariant model behind second law should have 
at least time machine effect elements in them.

It is therefore reasonable to exercise with as simple 
as possible models with such time machine effect elements in them. 
One of the simplest model of this sort 
for which there can be given several suggestive arguments 
is a world or the mechanical system of the model 
put on a closed time like curve (CTC).
It shall turn out that this model indeed does only a very poor job with respect 
to being a model behind the second law of thermodynamics in 
as far as we shall end up with the conclusion that in such a closed loop time world the entropy 
is almost certainly totally constant, all around the time loop.

Such a time loop model can most simply be taken to mean that 
we postulate the set of moments of time to form an $S^1$-circle 
rather than a straight line 
(or some interval or half time axis in Big Bang theories).
We can simply take it that the general relativity time coordinate 
is of such a nature that going forward in it by a certain constant $T$ (of dimension
of time) 
we come by the coordinate map identifications to
the same time moment 
(or say space-time points) as we started from.
We could call this restriction that the world has got an intrinsic period $T$.

In the usual Big Bang picture which is in simple general relativity enforced 
by the knowledge of the Hubble expansion and suggestive phenomenology 
according to Hawking-Penrose' singularity
 theorem such an intrinsic periodicity 
is of course not possible.

Nevertheless it might be imagined that in a final theory behind general relativity 
and not quite in agreement with the presuppositions of the just mentioned Big Bang 
enforcing theorem, it could e.g. happen that no start or crunch singularities were allowed.

In fact for instance superstring theory is presumably of the type 
without a starting point of time. 
There is possibility that a time parameter would go 
from $-\infty$ to $+\infty$, but it would still mean, in a sense, 
some singularities.
Therefore the most elegant would be a compactified time.

Whatever there might be of essential reasons 
for even nature represented by a compact time or space-time manifold 
it is the purpose of the present article to study a world with closed timelike loops 
-- really with a global forming closed loop -- 
and compactified in space.
As a matter of fact, however, 
we intend to consider so abstract formulation that 
we replace all the fields and their conjugates over the 
whole spatial extension of the compactified universe just by 
one (abstract) phase space.
In other words we do not even consider space explicitly,
 but just a general mechanical system 
 -- i.e. a point moving in phase space according to a Hamilton's equations --. 
The compactness of the time manifold $S^1$ then means 
that this general mechanical system 
is enforced to be strictly periodic with a period $T$ which 
is circumference of the above mentioned $S^1$.

Let us compare what we are concerned with a space-time with holographic principle
\cite{MN6}.
The holographic principle applies to worlds with a periodicity in time 
-- and thus closed timelike loops -- 
but still having a spatial infinity. 
Then what goes on in this space-time 4-volume 
is then determined by the boundary conditions at the spatial ``far away". 
This is holography 
in the sense that the information on the space-time field configurations 
packed into that on boundary surface placed far out in spatial direction.

In the present article we consider a compactified space 
so that there is no surface far out and then even the state of 
the system should be fixed alone by the structure of the system e.g. 
the Hamiltonian $H$ of the system if we think of it as the abstract mechanical system.

It is a very important point to have in mind for the present article that 
when time is compactified 
-- it means
in reality 
that a periodicity with given period is imposed -- 
there are as many periodicity requirements as the dimension of the phase space.
Also the manifold of initial conditions for the system has the same dimension 
since it simply $is$ the phase space.
Unless some of the periodicity conditions are not independent or perhaps contradictory 
the 
expectation is that there are just a discrete set of solutions to enforced periodicity constraint.
By considering a discrete set of solutions as ``essentially" or ``locally" just one solution 
we can say that the periodicity with enforced 
period $T$ fixes an ``essentially" unique solution.
When the Hamiltonian is 
not explicitly time dependent 
and thus conserved there is formally a \underline{one} dimensional manifold of starting points, 
because a start at one of the points in phase space reached along the orbit of the system 
in phase space will just give rise to the same periodic motion just delayed in its development.
But even that can be called an ``essentially" unique solution.
The main point is that compared to a phase space with a huge number of degrees of freedom 
some discrete solutions or even one dimensional sets of solutions 
is little different from a unique solution.
By far most degrees of freedom get settled by the given period with periodicity fixing.

The main purpose of the present article is to deliver a very general estimate of what 
we would consider the entropy of the essentially unique solution as just stressed above, 
to our periodicity constraint with a fixed period.
To put our estimate in perspective one should have in mind that 
if one chooses a random state of a mechanical system with a probability (density) 
given simply by the phase space measure, one almost certainly find it to have maximal entropy.
The reason that 
it is so is because a macro state is associated with a volume of the phase space proportional 
to $e^{\frac{1}{k} S}$ where $k$ is the Boltzmann's constant and $S$ the entropy.
This is certainly true because the entropy $S$ could be defined to be $k \log $ 
(volume of phase space of macro state)
\cite{MN7}.

With the above mentioned consideration in mind that thus the normal situation 
for a state chosen randomly in phase space is to have maximal entropy, 
it would be quite interesting to show that 
for a somewhat differently chosen state or point in phase space, 
the entropy is not maximal. 
This is indeed a remarkable thing.

The main point of the present article is precisely to show that 
with high probability the state obtained 
by the enforcement of the given period in advance for a 
``random system" will $not$ have maximal entropy.

In the following section 2 we set up a general formulation for macro states 
and approximately conserved quantities.
In section 3 we argue for our expectations for the likely entropy of the state determined 
from enforced periodicity under some simplified assumptions.
In section 4 we then take into account more general way of effective macro states 
without having them identified completely by sets of exactly conserved quantities.
In section 5 we investigate the extension of our classical calculation of the likely entropy 
to quantum mechanics.
In section 6 we conclude and present some outlook.


\section{The philosophy of random Hamiltonian}
It is the philosophy -- as is usual -- also in the present article to avoid fine tuning
of any parameters,
quantities involved in a given system.
We can strengthen this by making the general assumption that apart from the symmetry restrictions 
which we impose the Hamiltonian and also other parameters or functions describing the model 
are taken to be random.
This kind of randomness means that we e.g. for the Hamiltonian think 
of a probability distribution over the space of a large class of functions defined over phase space.
That is to say that we have in mind a distribution density $P[H]$ destined to multiply 
a functional measure $\mathcal D H$ 
so that the probability for the Hamiltonian belonging to a certain subset ${\mathcal A}$ 
of ``all" functions is given as
\begin{eqnarray}
P_{\mathcal A} = \int _{\mathcal A} P[H]{\mathcal D} H
~.\end{eqnarray}
We may think of it this way, 
but in reality it is very difficult to find a reasonable functional integral measure 
$\mathcal D H$ so that not almost all functions 
after such a measure become 
very bad functions with respect to continuity and differentiability.
For the purposes for which we want such random functions 
it is, however, not so important that the measure be precisely of such a form 
$P[H]\mathcal{D} H$.
Thus we could as $\mathcal D H$ well instead take, 
some system of parameterized functions $H (\boldsymbol{\xi}, \boldsymbol{q}, \boldsymbol{p})$ 
depending on a set of parameters $\xi$.

Then we can choose -- somewhat arbitrarily, 
but still in a reasonable way a measure over the space of parameters.
By making the parameterization so that it guarantees smooth 
(continuously differentiable) 
functions we can in such a way obtain random differentiable functions.
In as far as the Hamilton equations involve the partial derivatives of $H$ 
with respect to the $p_i$'s and the $q_i$'s 
it is of course strictly speaking needed that $H$ be differentiable.
But in practice we care only for properties 
which are true ``almost certainly" in the mathematical sense.
In this way we mainly ignore null-sets 
(i.e. sets of functions with zero measure).
The point of course is mainly that 
we do not accept our conclusions to be proven wrong 
by just a very specially made up counter-example.
We could take the philosophy that 
it would be exceedingly strange if Nature should just have chosen 
a very special function.
We really rather go for investigating what we shall typically expect.
Really we believe that it is in a way a law of nature that 
unless we have some laws enforcing special feature we shall only 
get the most likely results 
according to such a randomness model.
Really of course if we see something unlikely in this randomness philosophy 
we should make a new law explain so that it would no more be strange.
That is how science works, you must make new laws whenever something is strange 
in the sense of violating the old ones or even just the statistical expectations derived 
using only the old laws.

The philosophy described
in this section  is essentially that of random dynamics, 
taking the laws of nature as random.

\section{Setup of formalism for macro states}

The formalism to make definition of entropy possible which we shall use
here is of the following type: We imagine the phase-space of what we
call the macro system divided up to a large number of ``macro states''
characterized by some ``macroscopic variables''. That is to say we
imagine some functions $\vec{\xi}(p,q)$ 
-- called the ``macroscopic
variables'' -- 
to be defined over the phase-space the coordinates of
which we denote symbolically as $(p,q)$. We then think of the set of all
the points in phase space which within some finite small accuracy has
given values $\vec{\xi}_0$  so that
\begin{eqnarray}
 \left|\vec{\xi}_0 - \vec{\xi}(p,q)\right| &<& \varepsilon_{\rm macro},
\end{eqnarray}
as a macro state. 
Here $\varepsilon_{\rm macro}$ denotes a
small quantity for the macroscopic variables. We can then define entropy
$S(\vec{\xi}_0(\varepsilon))$ for the macro state at
$\vec{\xi}(p,q) \approx \vec{\xi}_0$ as 
\begin{eqnarray}
 S(\vec{\xi}_0(\varepsilon))
  &=&
  k \log {\rm Vol}
  \left\{
   (p,q) \left| | \vec{\xi}_0 - \vec{\xi}(p,q) |\right.
   < \varepsilon_{\rm macro}
  \right\}
\end{eqnarray}
where $k$ denotes the Boltzmann constant.

Often in statistical mechanics one may meet macro states. However, the
macro states appeared in the present paper may not quite seem to be of
this kind. Thus it may be needed for us to define that it is in fact
always allowed to work with our point of view.

Examples of the macro states which is not at first glance of the type we
describe is made by almost all Gibbs ensembles and also by grand
canonical ensembles. 
For a simple canonical ensemble one should think of
the temperature as one of the macro state parameters i.e. one of the
$\vec{\xi}_0$-components.  
Formally, all the states in canonical
ensembles corresponding to two different temperatures $T_1$ and $T_2$
say are the same just with different probability (densities). 
Since it
is well known that one can approximate for macroscopic systems
by a
canonical ensemble,
 we could make the temperature of a state in
phase space $(p,q)$ be assigned as a function of the energy $H(p,q)$ for
that state in phase space.
The ``temperature'' $T(p,q)$ of the state
$(p,q)$ should be made
\begin{eqnarray}
 T(p,q) &=& f(H(p,q))
\end{eqnarray}
where f is the function giving the macroscopic relation
\begin{eqnarray}
 T &=& f(U)
\end{eqnarray}
for the system. Here $U$ denotes the energy $U=\left<H\right>$.

Since the spread in energy $H$ for a canonical ensemble becomes rather
small for highly macroscopic system the error by taking the entropy as
we suggested above with $\varepsilon_{\rm macro}$ sufficiently big to
allow the $H$-spread would be small.

It would be easy to argue similarly for grand canonical ensembles by
approximating it by an ensemble with a fixed number of particles. 
Again
we could use the approximately unique relation between the number of
particles and the chemical potential and thus define a macroscopic
variable among the $\vec{\xi}_0$'s to represent the chemical potential.

A priori it may be a question of our possibilities for finding some
parameters which we can keep track of as our ``macroscopic variables''
$\vec{\xi}$.  
Such keeping track of the variables $\vec{\xi}(p,q)$ is
the easiest if they are reasonably stable under the time development
of the system. 
There will be total stability of the macroscopic
variables $\vec{\xi}(p,q)$ 
if they are indeed conserved quantities such
as changes. 
You could typically imagine that they would be taken to be
conserved quantities such as angular momentum, linear momentum etc.

Since the possibility of choosing the macroscopic variables to be
conserved quantities is such a simple way to ensure the stability of
them we shall take this case as to be used in the next section. More
generally, however, one may imagine a situation in which the macroscopic
variables are not totally conserved but rather suffer some diffusion as
time goes on.

\section{The entropy estimate in the simple case of conserved macroscopic
variables}

Let us now illustrate the main argument for that the system with an
enforced periodicity system will be in a state -- or rather go through
a series of states -- in phase space not having maximal entropy. First
we have, however, to specify in what sense we consider random systems or
rather a system with a random dynamics. First of all it means that one
imagines to put some probability measure over the set of all functions H
defined on the phase space. Then one can assume that the Hamiltonian for
the ``random system'' is obtained by picking a random Hamiltonian form H
with a probability distribution given by the above mentioned probability
measure.

To get an estimate of the probability for finding a periodic orbit with
period $T$ 
beginning at $(q_0, \rho_0)$ we may think like the following:
We imagine a system development by the Hamiltonian equations -- using
the random Hamiltonian H -- which was started at $(q_0, \rho_0)$ at,
say, $t=0$. As the Hamiltonian is ``random'' the development in time
will be ``random'' too, except though for the selection rules or
restrictions from that the macro variables are rather stable. In the
simplest case of macro states characterized by (totally) conserved
quantities $I_i$ say the system remains inside the subset of the phase
space corresponding to the starting set of $I_i$-values. The volume in
phase space of the subset to which the system started at $(q_0, \rho_0)$
is 
\begin{eqnarray}
u_p^{2N} \exp(S(\{I_{i0} \})) = u_p^{2N} \exp(S(\{I_i (q_0, p_0) \}))
\label{6}
\end{eqnarray}
where $S(\{I_i (q_0, p_0) \})$ is the entropy of the macro state 
-- characterized by the conserved quantities
$I_i = I_i (q_0, p_0)$ -- corresponding to $(q_0, p_0)$,
and $U_p$ is a factor inserted for each dimension to make the dimensionality correct 
without assigning the entropy very strange dimensionality.
Thinking of defining $S$ by a quantum mechanical formula like 
(\ref{6}) below we should take
\begin{eqnarray}
u_p = \sqrt{h}
\end{eqnarray}
So the smaller this entropy 
$S(q_0,p_0)$ of the starting state $(q_0, p_0)$, so to speak, the smaller 
the volume to which the system of state will be confined.

Now the main thinking is that it is easier or more likely to reach back
to the starting point $(q_0, p_0)$ by accident if the volume into
which the system can move is smaller. Actually the probability to find
back to the starting point $(q_0, p_0)$ just at time $t=T$ (where $T$
is the imposed period) within a given uncertainty range must be
inversely proportional to the phase space volume which is
$e^{S((q_0, p_0))}$.

We could choose a certain fixes accuracy for which in the phase space
by calling that practically we count all $(q,p)$ with
\begin{eqnarray}
 ||(q,p)-(q_0,p_0)||^2 &<& \varepsilon^2_{\rm p.s.}
\end{eqnarray}
as indistinguishable where $\varepsilon^2_{p.s.}$ denotes a chosen small
accuracy error in phase space. For the purpose of judging if the period
is just $T$, we accept
$||(q_0,p_0)-(q(T),p(T))||^2 < \varepsilon_{\rm p.s.}$
as the criterion for this periodicity condition.  Here the $(q(T),p(T))$
are the time developed ones from $(q_0,p_0)$ during the time $T$. We may
now simply estimate that the probability that $(q_0,p_0)$ with
sufficient accuracy given by $\varepsilon_{\rm p.s.}$ gives rise to a
periodic motion with period $T$ is
\begin{eqnarray}
  &&
  P\left((q_0,p_0) \mbox{ has } T\right) \nonumber\\
  &=&
  P\left[
   (q(T),p(T)),
   \mbox{ is inside $\varepsilon_{p.s.}$ sphere around }
   (q_0,p_0)
  \right] dq_0 dp_0 \nonumber\\
  &\simeq&
  C e^{-S(q_0,p_0)} dq_0 dp_0.
\end{eqnarray}
Here $dp_0 dq_0$ is the (Liouville) phase space measure which is
invariant under canonical transformations. The probability for finding a
period $T$ starting point, i.e. a $(q_0,p_0)$ in a macro state (subset
of phase space) with entropy $S({\rm macro}_1)(=S(q_0,p_0))$ is given
by
\begin{eqnarray}
 C \int_{{\rm macro}_1}e^{-S(q_0,p_0)}dq_0dp_0 
 &=& C e^{S({\rm macro}_1)} \times e^{-S({\rm macro}_1)} \nonumber\\
 &=& C.
\end{eqnarray}
Thus all the possible macro states have the same probability for having a
periodic orbit with period $T$.

This result is remarkable because the different macro states will
typically have wildly different phase space volumes in as far as they
are proportional to $e^{S({\rm macro})}$.

\subsection{How is the realistic situation with respect to $T$-periodic
 orbits?}
 
 In this subsection we should estimate more realistically what the
number of periodic orbits. 
In a macro state with $k$ 
 fixed conserved
quantities in a system with $N$ degrees of freedom so that phase space
has dimension $2N$ this macro state has dimension $2N-k$. By use of
integral invariants of Poincar\'{e}
for small areas extending partly in the direction of a conserved
quantity partly along its conjugate, 
we argue that small line pieces
along these conjugate directions do neither extend nor contract as time
goes on.
This makes $k$ directions of zero scaling. 
The remaining $2N-2k$
dimension will typically scale up or down. 
Let us assume -- we may still have
to prove it -- that half of them i.e. $N-k$ dimensions scale up while
the rest of $N-k$ scales down. 
After long time a little region around
$(q_0,p_0)$ will have scaled up exponentially with ``Lyapunov exponents''
(not only the maximal one, but all positive ones) and will have become
enormous. 
The region obtained from the starting $2N-k$ very small region
will be blown up to a huge $N-k$ dimensional surface, 
still with a bit
of extension in the $k$ directions but strongly contracted in the $N-k$
directions of contraction.

We may imagine reaching such an approximately $N-k$ surface by
propagating the very little region forward from $t=0$ to
$t=\frac{T}{2}$. 
Similarly we should get an effectively $N-k$ dimensional one by
propagating backward from $t=0$ to $t=-\frac{T}{2}$.

To seek periodic solutions for the system we should seek cuts of the two
$N-k$ dimensional surfaces corresponding to $\frac{T}{2}$ and
$-\frac{T}{2}$ respectively both surfaces lying in the $2N-k$
dimensional submanifold in phase space that represents the macro states
characterized by the $k$ fixed conserved quantities ${I_i}$. they will
generically never cut for given ${I_i}$'s because even to just cut in
one point would require generically that the sum of the dimensions of
the surfaces would become equal to the total dimension $2N-k$. Now
$(N-k)+(N-k)= 2N-2k$ and thus the $k$ parameters would have to be tuned
to get a cutting. But $k$ is just the number of the dimensions of the
space of the continuously many macro states and so we expect to find
cutting for discrete values of the macro state characterizing conserved
quantities ${I_i}$.

For the purpose of making an estimate of the number of solutions we
should like to define an effective average of the Lyapunov exponent
$\gamma_{av}$ so as to give the $(N-k)$ dimensional ``area'' of the
$\frac{T}{2}$ surface defined above as 
\begin{eqnarray}
 &
 (\varepsilon_{\rm p.s.})^{N-k}
 \times
 \exp\left[\frac{T}{2} \cdot \gamma_{\rm av} \cdot (N-k)\right]
 &
\end{eqnarray}
Here of course $\gamma_{\rm av} \leq \gamma$ where $\gamma$ is the
maximal Lyapunov exponent. 
We must imagine, if we can consider in
practice the macro state as compact, that the $N-k$ dimensional surfaces
corresponding to $\frac{T}{2}$ and $-\frac{T}{2}$ are much folded back
and forth and essentially cover the whole macro state. 
What we really want to estimate is the typical distances measured 
in the space of the different macro states i.e. in the space with the conserved 
variables $\{I_i\}$ as coordinates.
We ask so to speak how large a $\Delta$ will give us just about one solution 
in the volume $\Delta^k$ in this space with extension $\Delta$.

If we call the distance we imagine to get out in the $\{I_i\}$ from the starting point 
$\Delta$ we cover a space of volume $\Delta^k$.
After such extension we should get generic crossing and could ask how many crossings.

Let us define a ``potential crossing volume" of dimension $2N-k$ 
and imbedded in the macro states as explained of this dimension 
as being the product of the three ``areas": 
\begin{eqnarray}
\varepsilon ^{N-k}_{p, s} \exp(\frac{T}{2}\gamma_{av} (N-k)) \cdot 
\varepsilon ^{N-k}_{p, s} \exp(\frac{T}{2}\gamma_{av} (N-k))\cdot \Delta^k 
~.\end{eqnarray}
The number of solution -- in the $\Delta^k$ volume 
-- is this $(2N-k)$-volume divided by the full volume of the phase
\begin{eqnarray}
u^{2N-k} \exp(S) 
~.
\label{v}
\end{eqnarray}

Now above we though still did not sum over all the cells of size 
$\varepsilon ^{2N-k}_{p, s}$ where the motion could have ``started". 
That of course should be in the phase in question having volume(\ref{v}) 
(see page 19).
So there is place for 
\begin{eqnarray}
\#~\mbox{cells} = \frac{u_{p}^{2N-k}}{\varepsilon_{p, s}^{2N-k}} \cdot \exp(S) 
\end{eqnarray}
small cells.

However we do not really use the extension of the small cells 
in the direction of the $k$ dimensions corresponding to the conjugate variables 
to the conserved quantities, so it is better to think of only counting cells 
in the remaining $(2N-k)-k=2N-2k$ dimensions.
We instead use the $\Delta^k$ volume to count for how far 
we should extend in the $\{I_i\}$-space to get the true crossing.
But that then means that we should count the number of cells being for layers 
in the last $k$ dimensions out of the $2N-k$ but only in the $2N-2k$ ones 
corresponding to the dimensions in which we get the huge extension 
in one or the other time directions.
In this point  of view we should rather say that 
the number of cells we must use is
\begin{eqnarray}
\# (\mbox{cells on surfaces}) &=& 
\frac{u_{s}^{2N-2k}}{\varepsilon_{p, s}^{2N-2k}} 
\exp\left(S \cdot  \frac{2N-2k}{2N-k}\right) \\\nonumber
&=& \left(\frac{u_s}{\varepsilon_{p,s}}\right)^{2N-2k}
\exp\left(S \cdot \left(1- \frac{k}{2N-k}\right)\right) \\\nonumber
&\approx & \left(\frac{u_s}{\varepsilon_{p,s}}\right)^{2N-2k}\cdot \exp (S)
~.\end{eqnarray}
Thus the full number of solutions is 
\begin{eqnarray}\frac{
\varepsilon_{p, s}^{N-k} 
\exp\left( \frac{T}{2} \gamma _{av}(N-k) \right)^2 \Delta ^k
}{u_s^{2N-k}\exp(S)}\cdot 
\frac{u_s^{2N-2k}\exp(S)}{\varepsilon_{p, s}^{2N-2k}
}
~~.\end{eqnarray}

In the light of the number of dimensions $k$ 
corresponding to the conserved quantities being tiny compared 
to the total number of degrees of freedom 
$N\gg k$ we do not consider the factor $\varepsilon_{p, s}^{-k}$ as very important.
If therefore looks (at first) that 
the density of classical periodic solutions with just period $T$ 
is $\frac{1}{\Delta^k}$ evaluated so that  
there is just one solution per k-volume of order $\Delta^k$ which means so that 
\begin{eqnarray}
1\approx 
\varepsilon_{p, s}^{-k}\Delta^k \exp\left(T \gamma _{av}(N-k)\right)
~~.\end{eqnarray}
This density thus becomes 
\begin{eqnarray}
\frac{1}{\Delta^k}\approx
\varepsilon_{p, s}^{-k} \exp \left(T \gamma _{av}(N-k)\right)
~~.\end{eqnarray}
At this stage it thus looks as 
if the density were wildly dependent on the average Lyapunov exponent 
through the presumably hugely varying factor
\begin{eqnarray}
\exp \left(T \gamma _{av}(N-k)\right)
~~.\end{eqnarray}

This would indeed be the result if we took it that 
all the classical solutions had the same probability for being realized.
But as we shall argue in the next subsection this is not realistic.
Rather it will turn out that the bigger 
this huge factor is with the smaller weight should we count the solution in question 
so that actually this huge factor $\exp(T \gamma _{av}(N-k))$ 
gets (essentially) canceled 
and we end up with the result  
-- again -- 
that the density of probability in the space of macro states 
realized 
in the intrinsically periodic world is indeed very smooth,
 slowly varying compared to what huge variation that could have been imaged.

\subsection{Weighting of the different classical solutions}
In this subsection we include a correction
in the sense 
that different classical solutions
\cite{MN8} 
should not a priori be counted as equally likely but that these different tracks rather
obtain a probability weight strongly related to the Lyapunov exponent.

Realistically 
we should always count that there is some uncertainty,
even if small.
If for no other reason, quantum mechanics will provide such a source of uncertainty.
We shall though still postpone quantum mechanics proper to section 6 
and it may be pedagogical here to think of some other source of uncertainty.

Let us here first develop 
-- for self content -- 
the behavior of orbits very close to a given classical solution
by Taylor expanding the Hamiltonian say.
Defining the deviations of the canonical coordinates $q_i$ 
and the conjugate momenta $p_i$ from their values $q_{cl ~i}$, $p_{cl ~i}$ 
along the considered classical motion
\begin{eqnarray}
\Delta q_i &=& q_i - q_{cl ~i} ~,
\\\nonumber
\Delta p_i &=& p_i - p_{cl ~i}
\end{eqnarray}
we easily derive the Hamiltonian for the Hamilton equations for the deviations
to the accuracy of up to second order terms 
in Taylor expansion  
\begin{eqnarray}
\Delta \dot{q}_i 
&=& \dot{q}_i - \dot{q}_{cl ~i} 
\nonumber\\
&=& \sum_{j} \frac{\partial^2 H}{\partial q_j \partial p_i}\Delta q_i
+\sum_{j}\frac{\partial^2 H}{\partial p_j \partial p_i}\Delta p_j
~,\\\nonumber
\Delta \dot{p}_i &=&
-\sum_{j} \frac{\partial^2 H}{\partial q_j \partial q_i}\Delta q_j  
-\sum_{j} \frac{\partial^2 H}{\partial p_j \partial q_i}\Delta p_j ~.
\end{eqnarray}
Equivalently 
in matrix form
they read
\begin{eqnarray}
\left(
  \begin{array}{c}
    \Delta \dot{q}_1    \\
    \Delta \dot{q}_2   \\
    \vdots     \\
    \Delta \dot{q}_N   \\
    \Delta \dot{p}_1   \\
    \Delta \dot{p}_2   \\
    \vdots   \\
    \Delta \dot{p}_N   \\
  \end{array}
\right)
  = \underline{M}
  \left(
  \begin{array}{c}
     \Delta q_1  \\
     \Delta q_2  \\
     \vdots   \\
     \Delta q_N  \\
     \Delta p_1  \\
     \Delta p_2  \\
     \vdots   \\
     \Delta p_N  \\
  \end{array}
\right)
\end{eqnarray}
where $\underline{M}$
 is the $2N\times 2N$ matrix of 
second derivatives of the Hamiltonian
\begin{eqnarray}
\underline{M}=
\left(
  \begin{array}{cc}
  \frac{\partial^2 H}{\partial q_j \partial p_i}     &   
  \frac{\partial^2 H}{\partial p_j \partial p_i} \\ 
 - \frac{\partial^2 H}{\partial q_j \partial q_i}     &  
 - \frac{\partial^2 H}{\partial p_j \partial q_i}  \\
  \end{array}
\right) 
~~.\end{eqnarray}
Here $N$ is the number of degrees of freedom and the four symbols 
$\frac{\partial^2 H}{\partial q_j \partial q_i}$ etc.
stand for the four $N\times N$ submatrices, with $i$ 
enumerating the rows and $j$ the columns.

Introducing the $2N\times 2N$ matrix
\begin{eqnarray}
\underline{J}=
\left(
  \begin{array}{cc}
  \underline{0}      &    \underline{1}\\
  \underline{-1}     &   \underline{0} \\
  \end{array}
\right)
\end{eqnarray}
we note that the time development matrix $\underline{M}$ obeys
\begin{eqnarray}
\underline{J}~ \underline{M}~ \underline{J}=
\left(
  \begin{array}{cc}
  \frac{\partial^2 H}{\partial p_j \partial q_i}     &    
  \frac{\partial^2 H}{\partial q_j \partial q_i}\\
  -\frac{\partial^2 H}{\partial p_j \partial p_i}     &   
  -\frac{\partial^2 H}{\partial q_j \partial p_i} \\
  \end{array}
\right)
=\underline{M}^T
~.\end{eqnarray}

>From this property we immediately obtain that the spectrum 
for frequencies $\omega$  defined from seeking deviations
\begin{eqnarray}
\left(
  \begin{array}{c}
    \Delta q_1    \\
     \vdots     \\
    \Delta q_N   \\
    \Delta p_1   \\
     \vdots   \\
    \Delta p_N   \\
  \end{array}
\right)
  = e^{-i \omega t}
  \left(
  \begin{array}{c}
     \Delta q_{1~0}   \\
      \vdots   \\
     \Delta q_{N~0}  \\
     \Delta p_{1~0}  \\
      \vdots   \\
     \Delta p_{N~0}  \\
  \end{array}
\right)
\end{eqnarray}
with the $\Delta q_{i~0}$ and $\Delta p_{i~0}$'s being constant 
in time is defined as the eigenvalue spectrum for $\omega$ as 
\begin{eqnarray}
-i \omega 
\left(
  \begin{array}{c}
    \Delta q_{1~0}    \\
     \vdots     \\
    \Delta p_{N~0}   \\
  \end{array}
\right)
  = \underline{M}
  \left(
  \begin{array}{c}
     \Delta q_{1~0}   \\
      \vdots   \\
     \Delta p_{N~0}  \\
  \end{array}
\right)
\label{e}
\end{eqnarray}
or from the zeros of the secular equation
\begin{eqnarray}
0=\mathcal{D}et(\underline{M} +i \omega \underline{1})
\label{s}
\end{eqnarray}
from where it is seen to
consist of pairs of opposite eigenvalues.
That is to say if $\omega$ is an eigenvalue then so is also $-\omega$.
In fact equation (\ref{s}) implies
\begin{eqnarray}
0&=&\mathcal{D}et(\underline{M}^T +i \omega \underline{1})
= \mathcal{D}et(\underline{J}~ \underline{M}~ \underline{J} +i 
\omega \underline{1})\\\nonumber
&=& \mathcal{D}et(\underline{-M} +i \omega \underline{1})
\Rightarrow \mathcal{D}et(\underline{M} -i \omega \underline{1})
\end{eqnarray}
and so we see that together with $\omega$ we must also have $-\omega^*$ 
as an eigenfrequency, namely by complex conjugating equation (\ref{e})
which gives that the complex conjugate column 
$
\left(
  \begin{array}{c}
     \Delta q^*_{1~0}   \\
      \vdots   \\
     \Delta p^*_{N~0}  \\
  \end{array}
\right)
$
is an eigenvector corresponding to the frequency  $-\omega^*$.
Provided  
we do not have a whole set of four different related eigenvalues 
$(\omega, -\omega, -\omega^*, \omega^*)$ 
but rather only two related eigenvalues we must have either
\begin{eqnarray}
\label{*}
-\omega^* = \omega 
~~~\mbox{or}~~
-\omega^* = -\omega
~.\end{eqnarray}

The possibility (\ref{*}) only allows that $\omega$ is 
either purely imaginary or purely real.
Both these possibilities are easily seen to be indeed realizable 
respectively the inverted and usual harmonic oscillator used as examples.

In the case of $\omega$  real the paths close 
to a given classical solution circle around the latter, 
much like a harmonic oscillator 
-- or rather several -- in phase space circles around the equilibrium point.
If we however have the purely imaginary $\omega = -\omega^*$ 
then the situation is rather analogous to ``the inverted harmonic oscillator", 
meaning a particles near the top of a hill.
In such an unstable equilibrium situation 
it is well known that the solutions are rather of the form of 
linear combinations of 
exponentially varying solutions in time
of the form $e^{\pm \gamma t}$ 
call 
the coefficients in the exponent,
$\gamma = Im~\omega$,
in time local Lyapunov exponents.
In any case it is very important for the cancellation, 
which we are going to show that averaging over time these eigenvalues
 $\pm \gamma$ are closely related to our averaged Lyapunov exponent $\gamma_{av}$.
In fact we have the relation
\begin{eqnarray}
\int_{\mbox{the period $T$ along the classical path}}
\sum_{\mbox{the different eigenvalues}} 
|\gamma(t)|dt \frac{1}{T} = \gamma_{av}
~.\nonumber\\
{\ \  }\end{eqnarray}

In principle $\gamma_{av}$ depends on the path.

The physical meaning of these imaginary frequencies $\omega=i\gamma$ 
is that the nearby paths soon move away from the given classical path 
$(q_{i~0}(t), p_{i~0}(t))$ exponentially.
If we thus put some cut off -- some accuracy of measurement say 
-- for how far out we still consider a neighboring path connected with 
the given classical path
still in the neighbourhood,
 this border 
will be crossed every unit of time by the fraction of 
all the surrounding neighbouring paths being 
\begin{eqnarray}
\label{l}
\mbox{``Fraction being lost per unit time"} = \sum_{|\gamma|}|\gamma(t)|
~.\end{eqnarray}
Note that the probability for some random neighboring path 
is all the time given by (\ref{l})
independent of the precise size and shape of the cut off chosen to define 
whether a path is still in the (closed) neighbourhood.
It is only required that the cut off is chosen to give 
a small enough neighborhood that the Taylor expansion 
we used is valid as a good approximation.
That we have this loss rate (\ref{l}) 
independent of the details of the cut off is of course a consequence of the spreading of 
the neighboring track/paths is a pure (exponential) scaling up with time, 
a scale invariant operation (i.e. not involving any unit for say $q_i$).

When we now consider a classical periodic path with intrinsic period $T$, 
we should realistically think of it 
as representing a tiny little neighborhood of accompanying paths.
Now following along the classical path 
we have just calculated the loss rate of these accompanying path. 
That means that whatever neighborhood 
we had chosen to represent the ``accompanying paths" 
we loose per unit time the fraction $\sum_{|\gamma|}|\gamma(t)|$ of them.
So the total fraction of the accompanying tracks 
which survive all the period $T$ through is
\begin{eqnarray}
\mbox{``Surviving fraction"}= \exp\left(
-\int_{\mbox{along the period}} \sum_{|\gamma|}|\gamma(t)| dt\right)
~.
\end{eqnarray}
This is actually the
``average Lyapunov".
\begin{eqnarray}
\mbox{``Surviving fraction"}= \exp\left(
-\gamma_{av}(\mbox{for the cl. solution)}T
\right)
~.\end{eqnarray}
It is obvious that starting with a random start very close to one of the
periodic solutions the chance that 
it will return a period later equally very close is only this 
$``survival fraction"= \exp\left(-\gamma_{av}(\mbox{for the track)}T\right)$.
We should therefore not take all the classical periodic 
solutions with period $T$
 as equally likely, 
but rather we should weight them statistically with weight factor
$\exp\left(-\gamma_{av}(\mbox{for the solution)}T\right)$.

Interestingly enough this probability weight 
which a priori could have influenced the relative likelihood for 
different macro states, happens to just cancel the  corresponding factor 
$\exp(\gamma_{av}T)$ occurring in the number of solutions.
That is to say: the effect of loosing the accompanying path 
due to the spreading intervals in time  where some period $\omega$ 
are imaginary is just compensated by the way the number of solutions 
is also sensitive to the Lyapunov exponents.

So we really do not need to calculate Lyapunov exponents to estimate the likelihood 
of the various macro states in our intrinsic periodicity model.
The result that all the macro states have similar probability 
for being realized in the model is still true 
even when the Lyapunov exponents were taken into account 
because their effect is remarkably just canceled out!

\section{More general macro states}

If one makes the assumption used in foregoing section that the
macro states are totally specified by a series ${I_i}$ of conserved
quantum numbers $I_i$, then the macro state cannot change with time at
all. Thus entropy variation with time is completely excluded under such
conditions. Thus any hope of deriving the second law of thermodynamics
in such a model would be excluded
alone for the reason that the macro state could not change.

We therefore must make the assumptions a little more liberal so that
the macro states are able to change the one into the other one. 
These
processes of changing macro state should still be somewhat suppressed 
and
unlikely. Nevertheless it should now be possible.

We wish to argue, however, by an argument similar to that of section 3
that it is unlikely that the periodic orbit with given period T will
have its entropy change, but very little.
Hereby we mean that the
macro state -- if it changes at all -- will only change between
macro states with approximately the same entropy.

The argument runs indeed in the following way: we ask for the
probability of a motion of the system starting at a phase space point
$(q_0,p_0)$ with sufficient accuracy periodic with the period $T$. Let
us say that $(q_0,p_0)$ belongs to a macro state $(q_0,p_0)_M$ with
entropy $S((q_0,p_0)_M)$ where M denotes macro states, but that during
the motion with time $t$ the system runs into other macro states
$M_1,M_2$, etc.. The chance for the system to be at time $t=T$ (by
assuming to at $t=0$) back again at $(q_0,p_0)$ is inversely
proportional to $\sum_i e^{S(M_i)} \approx e^{\max S(M_i)}$. That is to
say that it is inversely proportional to the exponentiated entropy for
that of the passed macro states, which has the biggest entropy,
\begin{eqnarray}
 &\max\left\{S(M_i) \;|\; i \mbox{ corresponding the passed } M\right\}&.
\end{eqnarray}
But the phase space for a starting point $(q_0,p_0)$ in a macro state
$M_k$ is $e^{S(M_k)}=e^{S(q_0,p_0)}$. Thus the chance goes as
\begin{eqnarray}
 e^{S(q_0,p_0) - \max \left\{S(M_i)\right\}} &\leq& 1.
 \label{eq16.eq}
\end{eqnarray}

By having the entropy vary during the periodic motion will -- always --
bring down the relative probability, in fact by the factor given by the
left hand side of (\ref{eq16.eq}).
In this way it gets more and more unlikely the
bigger the spread in the entropy during the passage.

Ignoring the small variation in entropy allowed and it is indeed only a
very small amount allowed, we conclude that the entropy will {\it stay
constant}.

In the above statement we argued completely generally and in an abstract
manner 
with the compactified time manifold of the form $S^1$
that
the initial conditions become essentially fixed and that in such a way
that the entropy becomes completely conserved. 
Note that this means that
in such a machinery 
-- which can be said exists inside a time machine, i.e.
closed time like loop -- 
the entropy becomes constant and there would be
{\it no place for a nontrivial entropy increase}. Therefore the second
law of thermodynamics would only be true in the trivial manner that
entropy stands still. A nontrivial increase would {\it not} be allowed.

Logically it thus seems that since we have certainly the nontrivial
increase in nature we would have to claim that the world would not
possibly turn out to be of type of the compactified time as discussed
in this article.

It could be greatly interesting to remark that as superstring theory is said to
be of the type without the singularities initiating or finalizing the
time-axis, then our way of arguing would imply that superstring theory
would not be compatible with second law of thermodynamics.

We must, however, admit the caveat
that our phase space were taken in the discussion as of finite
volume. Thus if the ``system'' were a field theory in an infinite space,
then perhaps we should reconsider the argument.
Arrow of time axis turn out to infinity
could spoil this argument.

\section{Attempt to extend to quantum mechanics}
 
At first one would have thought that the argument in section 3 that
enforcing periodicity with the period $T$ would lead to essentially all
macro states having equal probability, should easily be extended to work
also for quantum mechanical systems. However, strangely enough we shall
see that at first it does not extend to quantum mechanics.

To repeat the same story quantum mechanically,
in the simplest
case
we should  consider a quantum mechanical system with a series of conserved
operators $I_i$ -- among which $I_i$ is the Hamiltonian -- get that the
imposed periodicity $T$ simply comes to require that 
\begin{eqnarray}
 TH &=& TI_1' \ =\  2\pi n,
 \label{eq17.eq}
\end{eqnarray}
and $I_1'$ the eigenvalue of $I_1=H$
with an integer $n$. 
In this way all the states defined by
\begin{eqnarray}
 &
 \left\{
  \left|\psi\right>
  \left| \; \forall i \;
   [\hat I_i \left|\psi\right> = I_i'\left|\psi\right>]
  \right.
  \right\}
 &
 \label{eq18.eq}
\end{eqnarray}
will be unchanged by the time development operator $e^{-iTH}$ for the
period $T$. 
If we now ask for the macro state specified to have the subspace
characterized by a set of eigenvalues ${I_i}$
the allowed set consists of the set of 
macro
states that satisfy eq (\ref{eq17.eq}).
The dimension of this subspace is obviously to be
identified with $e^{S(M)}$ where $S(M)$ is the entropy of the macro state 
$M$. Here the entropy is given by
\begin{eqnarray}
 S(\{I_i\})
 &=&
 \log \dim
 \left\{
  \left|\psi\right>
  \left| \; \forall i \;
  (\hat I_i \left|\psi\right> = I_i'\left|\psi\right>)
  \right.
 \right\}~.
\end{eqnarray}
Taking seriously that Hamiltonian $H=I_1$, say, is among macro state
specifying parameters, we simply get the condition for the periodic
system with period $T$ as
\begin{eqnarray}
 TH' &=& TI_1' \ =\ 2\pi n
\end{eqnarray}
with an integer $n$. 
It means that the time translation operator through 
the micro states in the macro states in question. 
Actually it is obvious
that on the states in eq (\ref{eq18.eq}), the time translation operator
acts as
\begin{eqnarray}
 e^{-iTH}\left|\psi\right> &=& \left|\psi\right>.
\end{eqnarray}
This simple relation means that the macro states with the imposed period
$T'$ are just those with a series of $I_1'$ values of multiple of
$\frac{2\pi}T$. Such macro states be with a smooth distribution in the
space of $\{I_i'\}$ values. In this manner we reached a result similar
to the one in section 3. The entropy can vary a lot as the $\{I_i'\}$
values are varied and thus we still have the surprise that we get smooth
probability distribution in the manifold of macro state in spite of the
strong variation in entropy $S$ and therefore huge variation of the
phase space volume $e^S$.

\section{Conclusion and outlook}

We have studied the consequences of imposing a given period $T$ in
advance on a ``random'' physical system. What we found were that such a
condition is in general only fulfilled for some or few
macro states. 
The remarkable point is that this macro state is not simply
one with maximal entropy. 
On the contrary we rather get for a random
system same probability for the different macro states
to contain a macro state obeying the
periodicity $T$ requirement.

We got this result both quantum mechanically and classically. It should
be stressed that our quantum mechanical result came by using the
Hamiltonian as one of our macroscopic variables. It should be born in
mind that such a world that has an imposed fixed period $T$ is a world
which can be said to exist inside a time machine. Actually it is obvious
that an enforced periodic time variable running on $S^1$ means a time
machine in the sense that one gets back in time by the periodicity which
simply brings a person -- part of the system we consider -- which
lives into time $T$ back to time $0$. As a consequence of the world
considered -- containing in a time machine, it expected to have ``grand
mother paradoxes'' which have to be resolved by some miracles. We can
consider the surprisingly low entropy resulting for a random model of
the type of the fixed period as an expression of the ``miraculous''
solution to the inconsistencies otherwise easily popping up.

An outlook to an application of the present work could be that we
suppose that for some reason or another -- philosophically that there
be no singularities at which time stops or begins, or superstring theory
-- the time manifold is forced to be compact and thus we are forced to a
periodic world development. 
Then this world would have a promising point
concerning the establishment of a ``deeper understanding'' of the second
law of thermodynamics. However, to get a varying entropy with time is
seemingly not coming, so really it was not a successful model to explain
second law of thermodynamics.

We seek to present our failure to obtain in such a model the genuine 
increase of entropy into a no-go theorem developing the rather trivial point 
that second law is strictly speaking in disagreement 
with the time reversal symmetry principle:
The well known arrow of time problem
\cite{MN9}. 
Could even some time reversal invariant laws allow 
naturally say in time locally in some era, 
an effective second law to be valid.
We hope in near future to see which conditions are enough to make even such models impossible.
 
\section*{Acknowledgments}

This work is supported by Grant-in-Aid for Scientific Research on
Priority Areas, Number of Area 763 ``Dynamics of Strings and Fields'',
from the Ministry of Education of Culture, Sports, Science and
Technology, Japan.
One of us (H.B.N) acknowledges 
for the very helpful discussions
participants of this conference.

\chapter*{Astri Kleppe's Song}
\addcontentsline{toc}{chapter}{Astri Kleppe's Song}

\section*{Wolf light}

\begin{itshape}
\begin{multicols}{2}
\begin{verse}
I

You asked me in my dream\\
what time is. When I answered you turned\\
away\\
But I know, I called out\\
and knew\\
Already was my answer dissolved\\[3mm]

Where does the recent now\\
of a moment pass? My face\\
you just saw,\\
"imagine that the world emerged by a mistake,\\
like when you lose a plexiglass\\
on a stone floor"\\[1cm]

II

I have a picture of you\\
on the wall, black, white\\
and grey\\[3mm]

March twilight, an airplane shadow\\
moves over the roofs\\
What is it the paper attempts to\\
interpret? The granularity\\
of the pigmentation effaces the angles\\
of your eyes, and the mouth keeps moving\\
under my effort to see. What\\
is an other? Air planes\\
plunged in borrowed light,\\[3mm]

on summer mornings, a little child\\
who plays by the seashore (the colours of stones are clearer in water),\\
and the stone which is me\\
is washed over\\
by other rains, by the prospects\\
of other truths\\
Three straight cuts, a giant\\
crosses over. And under the roofs\\
the street space catches hold\\
of the slowly wandering\\
of nearby stars\\
Like us, perhaps they call their neighbours\\
But which homes, which places\\
in this sinking?\\[1cm]

III

It is quiet. I breathe and it is silent,\\
we talked at length and I shut my eyes, lifted\\
by pictures gently passing,\\
the oddness of a touch\\
of other seashores\\[3mm]

It is quiet. Am I asleep?\\
We talked a minute ago about the burned land,\\
the meadow's faithless green. It retains\\
no steps, Earth's file of index cards\\
where exchangeable pass\\
Maybe in the glimpses between sleep and wake,\\
the waves between dead and alive,\\[3mm]

like the sun ray forcing\\
its way through the heavy curtain, a way\\
through the room, in the hair\\
on her shoulders, in the stripes\\
of her lips. She looks up, somehow\\
astonished\\[3mm]

Scent of linen, charlock blossom,\\
an afternoon wind drives across the field\\
And she dips her fingers\\
in the water furrows,\\
perhaps the dead communicate\\
by means of winds, of wind grey\\
aspens\\[1cm]

IV

I was a child\\
by the water side, played\\
with the shells\\
and stones, wrinkles\\
of sand\\
Sometimes we collected mussels\\
But no fire\\[3mm]

Like stars, stones dispersed,\\
glittering in the sun. According to which\\
order are they given room, which clockwork\\
carries stones, the movement of stars?\\
Voids fumbling\\
in each other's space, clamping\\
onto nothing, falling\\
towards nothing\\[3mm]

No, why should anyone\\
be more concerned\\
than the sun by a water surface,\\[3mm]

than a feather by the waves, so light\\
And groping for another Earth, attachments\\
for the weightiness. They said affinities\\
draw limitations through all\\
things, specific ones\\
for every body\\[1cm]

V

"Every man has a part in light", on my wall\\
I have a picture. Of you? These eyes that look at me\\
so questioning, accusingly\\
O, let us chat, let us tell some lies\\
together\\
Breakfast light, let us wake up, perhaps\\
we talk about our dreams, a door\\
that can't be opened and something\\
that approaches, weird\\
and other\\[3mm]

The snow light is so white today\\
I dreamt we walked\\
along the road, a winter day like now\\
Then suddenly I saw the buds, all swelling\\
like in May\\[3mm]

The rooms are much too white\\
You said you are a stranger here, but trees\\
are they not just as strange? Indifferent\\
their frozen time which flows away\\
I looked out of a window\\
in my dream. A yard\\
there under me, enclosure\\
with a tree\\
From which a piece of cloth is hanging, and I strech out\\
to reach it as it flutters back, around\\
the tree and drapes it\\[3mm]

And in the dream you tell me\\
that to carry fruit the tree\\
invokes the darkness\\
where all light is hidden\\[1cm]

VI

The Milky Way hovers\\
in dimmer light,\\
for skaters' moist dance, a city\\[3mm]

where fragments of nightly tales\\
appear. They sang\\
you must tar your back\\
and keep fir twigs\\
in mind, sign secret weeping,\\
weep and tar\\[3mm]

And space grows denser there,\\
your face,\\
your ears spread out, a\\
picture's hunger\\
Thursday, vacuum-cleaning, laundry,\\
and the witch goes to the\\
supermarket, gets\\
her dough for gingersnaps\\
and eggs and matches, tsimtsum, in this\\
best of worlds, laboratory\\
for Paradise studies\\[3mm]

And sure, we have mapped out\\
the trajectories of motion, and you\\
fried eggs for breakfast, they should
be eaten there\\
at once\\[3mm]

you said, and morning\\
struck into the kitchen,\\
pictureless like time,\\
and lemon green\\[1cm]

VII

On my wall\\
I have a portrait\\
Reflected in the wall stains,\\
in shadows long since erased\\[3mm]

"They have taken my face\\
away from me",\\
wrinkles between gold foils, pictures\\
of time, all time, clay eyes\\
out of the dusting earth\\
where the young child plays in the early light\\[3mm]

\ldots perhaps she covered her ears afterwards,\\
when she had ignited\\
(that was the fire)

\end{verse}
\end{multicols}
\end{itshape}



\backmatter


\thispagestyle{empty}
\parindent=0pt
\begin{flushleft}
\mbox{}
\vfill
\vrule height 1pt width \textwidth depth 0pt
{\parskip 6pt

{\sc Blejske Delavnice Iz Fizike, \ \ Letnik~6, \v{s}t. 2,} 
\ \ \ \ ISSN 1580--4992

{\sc Bled Workshops in Physics, \ \  Vol.~6, No.~2}

\bigskip

Zbornik 8. delavnice `What Comes Beyond the Standard Models', 
Bled, 19.~-- 29.~julij 2005

Proceedings to the 8th workshop 'What Comes Beyond the Standard Models', 
Bled, July 19.--29.,  2005

\bigskip

Uredili Norma Manko\v c Bor\v stnik, Holger Bech Nielsen, 
Colin D. Froggatt in Dragan Lukman 

Publikacijo sofinancira Javna agencija za raziskovalno dejavnost Republike Slovenije 

Tehni\v{c}ni urednik Vladimir Bensa

\bigskip

Zalo\v{z}ilo: DMFA -- zalo\v{z}ni\v{s}tvo, Jadranska 19,
1000 Ljubljana, Slovenija

Natisnila Tiskarna MIGRAF v nakladi 100 izvodov

\bigskip

Publikacija DMFA \v{s}tevilka 1618

\vrule height 1pt width \textwidth depth 0pt}
\end{flushleft}



\begin{thebibliography}{99}
\bibitem{imp} Mirman, R., Our Almost Impossible Universe: Why the laws of
nature make the existence of humans extraordinarily unlikely --- to be
published.
\end{thebibliography}
\end{document}